\documentclass[%
 reprint,
superscriptaddress,
 amsmath,amssymb,
 aps,
]{revtex4-1}
\usepackage{amsmath}
\usepackage[export]{adjustbox}
\usepackage{graphicx}
\usepackage{xcolor}
\usepackage{color}
\usepackage{dcolumn}
\usepackage{bm}

\begin{document}

\title{ Sensitivity of third-generation interferometers to extra polarizations\\ in the Stochastic Gravitational Wave Background}

\author{Loris Amalberti}
\email{Electronic address: loris.amalberti@desy.de}
\affiliation{Dipartimento di Fisica e Astronomia ``G. Galilei'', Universit\`a degli Studi di Padova, via Marzolo
8, I-35131 Padova, Italy}
\affiliation{Deutsches Elektronen-Synchrotron DESY, Notkestrasse 85, 22607 Hamburg, Germany}
\affiliation{Humboldt-Universität zu Berlin, Institut für Physik, Newtonstrasse 15, 12489 Berlin, Germany}
\author{Nicola Bartolo}
\email{Electronic address: nicola.bartolo@pd.infn.it}
\affiliation{Dipartimento di Fisica e Astronomia ``G. Galilei'', Universit\`a degli Studi di Padova, via Marzolo
8, I-35131 Padova, Italy}
\affiliation{INFN, Sezione di Padova, via Marzolo 8, I-35131, Padova, Italy}
\author{Angelo Ricciardone}
\email{Electronic address: angelo.ricciardone@pd.infn.it}
\affiliation{Dipartimento di Fisica e Astronomia ``G. Galilei'', Universit\`a degli Studi di Padova, via Marzolo
8, I-35131 Padova, Italy}
\affiliation{INFN, Sezione di Padova, via Marzolo 8, I-35131, Padova, Italy}

\date{\today}

\begin{abstract}
When modified theories of gravity are considered, at most six gravitational wave polarization modes are allowed and classified in tensor modes, the only ones predicted by General Relativity  (GR), along with additional vector and scalar modes. Therefore, gravitational waves represent a powerful tool to test alternative theories of gravitation. In this paper, we forecast the sensitivity  of third-generation ground-based interferometers, Einstein Telescope and Cosmic Explorer, to non-GR polarization modes focusing on the stochastic gravitational wave background.  We consider the latest technical specifications of the two independent detectors and the full network in order to estimate both the optimal signal-to-noise ratio and the detectable energy density limits relative to all polarization modes in the stochastic background for several locations on Earth and orientations of the two observatories. By considering optimal detector configurations, we find that in 5 years of observation the detection limit for tensor and extra polarization modes could reach $h_0^2\Omega^{T,V,S}_{GW} \approx 10^{-12}-10^{-11}$, depending on the network configuration and the stochastic background (i.e. if only one among vector and scalar modes exists or both are present). This means that the network sensitivity to different polarization modes can be approximately improved by a factor $10^3$ with respect to second-generation interferometers. We finally discuss the possibility of breaking the scalar modes degeneracy by considering both detectors angular responses to sufficiently high gravitational wave frequencies.
\end{abstract}

\maketitle

\section{Introduction}
In 2016 it was announced by the LIGO-Virgo collaboration that gravitational waves (GWs) from the merging of a binary black hole system were directly observed for the first time \cite{abbott2016observation, abbott2016tests} and since then, other GWs related to astrophysical processes have been detected (see e.g. \cite{abbott2016gw151226, scientific2017gw170104, abbott2017gw170608, abbott2017gw170814, abbott2017gw170817, abbott2020gw190521}). All these events provided a successful method to test possible deviations from General Relativity (GR) and confirmed that  Einstein's theory is the most accurate and well-tested theory of gravitation we have. Nevertheless the last century saw the rise of a wide number of alternative theories (see e.g. \cite{brans1961mach, sotiriou2010f, de2010f, hellings1973vector, will1972conservation}), also, e.g., in connection with a possible explanation of the late time acceleration of the universe~\cite{DeFelice:2010aj,Capozziello:2011et,Clifton:2011jh} or to model the early universe expansion~\cite{Heisenberg:2018acv, Heisenberg:2018mxx}. When such generic metric theories of gravity are considered, at most six gravitational wave polarization modes are allowed: two tensor modes (usually called tensor-plus and tensor-cross), the only ones permitted by GR,  two scalar modes (scalar-breathing and scalar-longitudinal) and two vector modes (vector-x and vector-y). Not all alternative theories of gravity are excluded by today direct GW observations (see e.g. \cite{will2014confrontation, deRham:2016nuf, will2018theory, Ezquiaga:2018btd} for a detailed overview), therefore the presence or absence of such extra polarization modes provides a useful tool to test and eventually extend the theory of GR, opening the way to other theoretical models.\\

A search for non-GR polarization modes may be pursued through the detection of a stochastic gravitational wave background (SGWB) \cite{maggiore2008gravitational,maggiore2018gravitational, allen1999detecting} which is among the expected targets of future generation laser interferometers \footnote{See also recent constraints on the polarization of GWs from single events \cite{abbott2019tests, abbott2017gw170814, abbott2020tests}}. In order to separate tensor, vector and scalar polarization modes in terms of the corresponding energy density contributions to the SGWB, the correlation analysis \cite{romano2017detection} represents a useful tool: indeed, the polarization modes separation issue was presented in detail by Nishizawa et al. \cite{nishizawa2009probing} for second-generation ground-based interferometers (e.g. the two LIGO observatories, Virgo and KAGRA), where it was shown that energy density contributions from tensor, vector and scalar polarization modes to the SGWB energy density can be isolated from each other considering a network involving at least three GW detectors. Despite the fact that the limited sensitivity of current detectors has not allowed to directly measure any SGWB yet, the LIGO-Virgo collaboration recently provided upper limits for tensor, vector and scalar modes energy densities $\Omega_{GW}^T(25 \text{ Hz}) < 6.4 \times 10^{-9}$, $\Omega_{GW}^V(25 \text{ Hz}) < 7.9 \times 10^{-9}$ and $\Omega_{GW}^S(25 \text{ Hz}) < 2.1 \times 10^{-8}$ \cite{abbott2021upper}. Additionally, the next upgrade LIGO A$+$ for the two observatories in North America \cite{barsotti2018a+} and the addition of the planned IndIGO interferometer in India \cite{unnikrishnan2013indigo} might further improve the current constraints (see also recent work on SGWBs with NANOGrav \cite{arzoumanian2016nanograv, chen2021non, arzoumanian2020nanograv}).\\

The better sensitivity of future third-generation laser interferometers will most probably allow to detect the SGWB and so to possibly probe GW extra polarizations. Both the European project Einstein Telescope \footnote{http://www.et-gw.eu/} and the american project Cosmic Explorer  \footnote{https://cosmicexplorer.org/} are two additional ground-based observatories which are going to be built in the upcoming future. Besides interferometers on Earth, several space-based GW detectors such as LISA \cite{amaro2017laser}, the Japanese project DECIGO \cite{ando2010decigo, kawamura2011japanese} and Chinese projects TianQin \cite{luo2016tianqin} and Taiji \cite{hu2017taiji} are planned to be launched in the near future. Optimistically starting among late 2020s and 2030s, interplays among these detectors are expected to resolve GW sources with incredible precision while exploring different frequency ranges \footnote{See e.g. \cite{barish2020impact} for recent work on ``midband'' detectors which are expected to close the frequency gap between space-based and ground-based interferometers.}, making the following decades a promising period for the study of GWs (see e.g.~\cite{Bartolo:2016ami, Caprini:2019pxz, Baker:2019ync, maggiore2020science, Barausse:2020rsu, Contaldi:2020rht, Orlando:2020oko, sathyaprakash2019cosmology, Flauger:2020qyi, DallArmi:2020dar}).\\

In a recent work by Omiya and Seto \cite{omiya2020searching}, the possibility of detecting vector and scalar polarization modes in a SGWB exploiting the LISA-Taiji network has been explored, leading to the theoretical expectation that ten years of observation could provide much smaller upper limits for the detectable energy density $\Omega_{GW}^V \approx 10^{-12}$ and $\Omega_{GW}^S \approx 10^{-12}$ than the ones provided by \cite{abbott2021upper}. However, due to LISA and Taiji triangular topology and sensitivity, extra polarization modes can be isolated only for a SGWB made of tensor and only one among vector or scalar modes. On the other hand, to date a detailed study for third-generation ground-based interferometers (i.e. Einstein Telescope and Cosmic Explorer) in order to estimate the sensitivity to non-GR polarizations in the SGWB and to distinguish different polarization modes considering the detector latest technical specifications was still missing \footnote{Outside the SGWB context, see \cite{takeda2019prospects} for recent work on ET and CE to test extra polarization modes from compact binary mergers.}.\\

In this work we first investigate the connection between the Einstein Telescope topology (which will most likely consist of three V-shaped interferometers displaced in a triangular way) and extra polarization modes. Indeed, in terms of tensor modes, a well-known result is that the full Einstein Telescope detector would be sensitive to GWs coming from each direction in the sky \cite{regimbau2012mock}, in contrast to single and traditional L-shaped interferometers \cite{romano2017detection}. We further extend this result to vector and scalar modes, showing that more isotropic detector angular responses can be also obtained, with the only ``blind'' direction being the one orthogonal to the detector plane. Next, through correlation analysis \cite{allen1999detecting, romano2017detection}, we address the polarization modes separation problem \cite{nishizawa2009probing, omiya2020searching} with GW interferometers and we investigate different possible SGWBs when scalar, vector or both non-GR polarization classes are present along with usual tensor modes. In particular, we focus on third-generation ground-based detectors and we first carry out an analysis of how different angular separations on Earth between the Einstein Telescope and Cosmic Explorer impact on the SNR and we look for optimal orientations of both observatories for each case. Second, we investigate the possibility of having two Cosmic Explorer-like detectors replacing the two LIGO observatories in North America both in location and orientation and we further assume the Einstein Telescope to be placed either in Sardinia (Italy), or the border region between the Netherlands, Belgium and Germany, which are two possible sites currently under investigation \cite{Note1}. In general, we find that considering the Einstein Telescope-Cosmic Explorer network, detection limits for the detectable energy density $\Omega^M_{GW}$, with $M=T$, $V$ and $S$ (assuming a frequency independent energy density spectrum for the SGWB), would approximately range from $10^{-12}$ to $10^{-11}$ for 5 years of observation depending on the network configuration and the stochastic background considered (i.e. if one or both non-GR polarization classes are present). Finally, we focus on the Einstein Telescope and Cosmic Explorer in order to break the so-called scalar modes degeneracy. The latter implies that for sufficiently low GW frequencies, detector angular responses to breathing and longitudinal polarization modes differ only for a constant factor, making the two scalar modes indistinguishable for a given interferometer (see e.g.  \cite{romano2017detection}). This is no longer true when we consider GWs whose frequencies are higher than a characteristic value inversely proportional to the interferometer arm length: therefore, in analogy e.g. to \cite{liu2020constraining, liang2019frequency} for space-based detectors, we discuss this scenario focusing on the Einstein Telescope and Cosmic Explorer, while obtaining frequency-dependent angular responses and breaking the degeneracy.\\

The paper is structured as follows. In section \ref{GWpolarmodes}, we define all possible GW polarization modes to further introduce the corresponding detector angular responses. In section \ref{3genIFOS}, we consider both Einstein Telescope and Cosmic Explorer angular responses in relation to their different topologies: we recall some differences and similarities between the two and we finally compute and investigate the Einstein Telescope network angular response to extra polarization modes. In section \ref{CorrAn}, we first retrace the fundamental steps of correlation analysis in order to detect a SGWB made of only tensor modes, then we present our forecasts for the detection limit of the SGWB detectable energy density working with the new generation of ground-based interferometers. We further extend these results in sections \ref{2GWB}, \ref{3GWB} and \ref{2CE} to a SGWB made of tensor and $X$-polarization modes (with $X$ being vector or scalar) and to a SGWB made of tensor, vector and scalar polarization modes at the same time: in particular, we explore several network layouts involving the Einstein Telescope and Cosmic Explorer and, for each case considered, we appropriately provide our results for the detection limit of the SGWB detectable energy density contributions for non-GR polarization modes. In section \ref{GWhigh}, we consider detector angular responses to GWs with sufficiently high frequencies in order to discuss the possible breaking of the degeneracy between scalar polarization modes using both the Einstein Telescope and Cosmic Explorer. Finally, in section \ref{CONC}, we summarize our results and we draw out conclusions.\\

We fully focus on the sensitivity of third-generation ground-based interferometers to the SGWB. Extending the result obtained for tensor modes \cite{regimbau2012mock}, we compute and discuss a new expression for the Einstein Telescope network response to vector and scalar polarization modes. Moreover, in comparison to results already available in literature for second-generation ground-based interferometers \cite{allen1999detecting,nishizawa2009probing,abbott2021upper} and space-based ones \cite{omiya2020searching}, for the first time (to our knowledge) we investigate in depth possible setups and proposed locations for the Einstein Telescope and Cosmic Explorer, along with the corresponding detector sensitivities to GWs available to date, in order to carry on a detailed analysis for the detection of the SGWB (in the presence or absence of one or both extra polarization modes) and to provide forecasted limits on the detectable energy density contributions to the SGWB for tensor, vector and scalar modes. We also briefly discuss the possibility of breaking the scalar modes degeneracy given the GW frequency ranges to which the Einstein Telescope and Cosmic Explorer are sensitive, that up to now was only explored for space-based detectors \cite{liu2020constraining, liang2019frequency}.

\section{GW polarization modes \label{GWpolarmodes}}
In this section we briefly discuss the detector response to each possible GW polarization \cite{nishizawa2009probing, romano2017detection}. We assume the existence of both tensor and extra (i.e. vector and scalar) GW polarization modes and we further decompose the spatial metric perturbation as follows~\cite{maggiore2008gravitational,maggiore2018gravitational}
\begin{eqnarray}
h_{ij}(t,\mathbf{\bar{x}})=&&\sum\limits_{P}\int_{-\infty}^{+\infty}df\int_{S^2}d\Omega\left[\mathbf{\tilde{e}}_{ij}^P(\mathbf{\hat{\Omega}})h_P(f,\mathbf{\hat{\Omega}})\right]\nonumber\\
&&\times e^{i2\pi f\left(t-\mathbf{\hat{\Omega}}\cdot\frac{\mathbf{\bar{x}}}{c}\right)},
\label{eq:1}
\end{eqnarray}
where $h_P(f,\mathbf{\hat{\Omega}})$ and $\mathbf{\tilde{e}}_{ij}^P(\mathbf{\hat{\Omega}})$ are the GW amplitude and corresponding polarization tensor respectively for $P$ = $+$ (tensor-plus), $\times$ (tensor-cross), $x$ (vector-x), $y$ (vector-y), $b$ (scalar-breathing) and $l$ (scalar-longitudinal). The unit vector $\mathbf{\hat{\Omega}}$ is defined on the 2-sphere $S^2$ and denotes the general GW direction, which we further assume to be traveling at the speed of light \cite{liu2020measuring}. Let us consider the following orthonormal coordinate system
\begin{eqnarray}
\begin{cases}
\mathbf{\hat{x}}=(1,0,0)\nonumber\\
\mathbf{\hat{y}}=(0,1,0)\nonumber\\ 
\mathbf{\hat{z}}=(0,0,1)\nonumber\\
\end{cases},
\end{eqnarray}
where unit vectors $\mathbf{\hat{x}}$ and $\mathbf{\hat{y}}$ identify the plane where the GW detector lies, while $\mathbf{\hat{z}}$  denotes the orthogonal direction to such plane. Let us further consider a second orthonormal coordinate system rotated by angles $(\theta,\phi)$ given by
\begin{eqnarray}
\begin{cases}
\mathbf{\hat{u}}=(\cos\theta\cos\phi,\cos\theta\sin\phi,-\sin\theta)\nonumber\\
\mathbf{\hat{v}}=(-\sin\phi,\cos\phi,0)\nonumber\\ 
\mathbf{\hat{\Omega}}=(\sin\theta\cos\phi,\sin\theta\sin\phi,\cos\theta)\nonumber\\
\end{cases}.
\end{eqnarray}
Since we wish to consider the most general choice of coordinates, we perform a plane-rotation by an angle $\psi$, the so-called polarization angle, around the axis identified by the unit vector $\mathbf{\hat{\Omega}}$, thus we get
\begin{eqnarray}
\begin{cases}
\mathbf{\hat{m}}=\mathbf{\hat{u}}\cos\psi+\mathbf{\hat{v}}\sin\psi\nonumber\\
\mathbf{\hat{n}}=-\mathbf{\hat{u}}\sin\psi+\mathbf{\hat{v}}\cos\psi\nonumber\\ 
\mathbf{\hat{\Omega}}=\mathbf{\hat{\Omega}}\nonumber\\
\end{cases}. 
\end{eqnarray}
This allows us to choose the orthonormal basis $(\mathbf{\hat{m}},\mathbf{\hat{n}},\mathbf{\hat{\Omega}})$ to give a proper expression to polarization tensors
\begin{itemize}
\item Tensor modes: \begin{eqnarray}
&&\mathbf{\tilde{e}}_+=\mathbf{\hat{m}}\otimes\mathbf{\hat{m}}-\mathbf{\hat{n}}\otimes\mathbf{\hat{n}}, \nonumber\\
&&\mathbf{\tilde{e}}_{\times}=\mathbf{\hat{m}}\otimes\mathbf{\hat{n}}+\mathbf{\hat{n}}\otimes\mathbf{\hat{m}}.
\label{eq:2}
\end{eqnarray}
\item Vector modes: \begin{eqnarray}
&&\mathbf{\tilde{e}}_x=\mathbf{\hat{m}}\otimes\mathbf{\hat{\Omega}}+\mathbf{\hat{\Omega}}\otimes\mathbf{\hat{m}}, \nonumber\\
&&\mathbf{\tilde{e}}_y=\mathbf{\hat{n}}\otimes\mathbf{\hat{\Omega}}+\mathbf{\hat{\Omega}}\otimes\mathbf{\hat{n}}.
\label{eq:3}
\end{eqnarray}
\item Scalar modes: \begin{eqnarray}
&&\mathbf{\tilde{e}}_b=\mathbf{\hat{m}}\otimes\mathbf{\hat{m}}+\mathbf{\hat{n}}\otimes\mathbf{\hat{n}}, \nonumber\\
&&\mathbf{\tilde{e}}_l=\sqrt{2}\mathbf{\hat{\Omega}}\otimes\mathbf{\hat{\Omega}}.
\label{eq:4}
\end{eqnarray}
\end{itemize}
The detector response to incoming GWs is represented by its angular pattern functions (APFs), which are given by
\begin{eqnarray}
F^P(\mathbf{\hat{\Omega}})=D^{ij}e_{ij}^P(\mathbf{\hat{\Omega}}),
\label{eq:5}
\end{eqnarray}
where $\mathbf{D}$ is the so-called detector tensor containing information on the detector geometry. For a given interferometer 
\begin{eqnarray}
\mathbf{D}=\frac{1}{2}\bigl\{\mathbf{\hat{e}_1}\otimes\mathbf{\hat{e}_1} - \mathbf{\hat{e}_2}\otimes\mathbf{\hat{e}_2}\bigr\},
\label{eq:6}
\end{eqnarray}
where $\mathbf{\hat{e}_{1}}$ and $\mathbf{\hat{e}_{2}}$ are the unit vectors directed along each interferometer arm. Note that the expression given in Eq.\eqref{eq:6} becomes valid only while considering GW frequencies lower than a characteristic value, which is inversely proportional to the interferometer arm length ($f \ll f_* = c/2\pi L$). Although this assumption is fairly good in the cross-correlation scenario, third-generation ground-based interferometers are expected to be sensitive to GW frequencies higher than their respective characteristic frequency $f_*$: whenever the low-frequency limit is no longer valid, the detector tensor expression is modified and a detector transfer functions need to be taken into account (see e.g. \cite{schilling1997angular}). In section \ref{GWhigh} we discuss the detector response to GW frequencies $f \gtrsim f_*$, while in the rest of this paper we always take the low-frequency limit to be valid: we shall briefly mention why this assumption is possible along the way.

\section{Third-Generation ground-based Interferometers \label{3genIFOS}}
In this section, we quickly retrace the well-known results available in literature for detector angular responses (see e.g. \cite{romano2017detection, cutler1998angular}) to each polarization mode ($P$ = $+$, $\times$, $x$, $y$, $b$ and $l$), then we discuss the detector angular response to joined tensor, vector and scalar polarization modes \cite{schutz2011networks} with the third-generation of ground-based interferometers, which consists of two upcoming detectors: the European project Einstein Telescope \cite{maggiore2020science, punturo2010einstein, Note1} and the American project Cosmic Explorer \cite{reitze2019cosmic, Note2}. Some well-known results for tensor modes are finally extended to extra polarization modes for the first time.  

\subsection{The Einstein Telescope}
\begin{figure}
\includegraphics[width=0.35\textwidth]{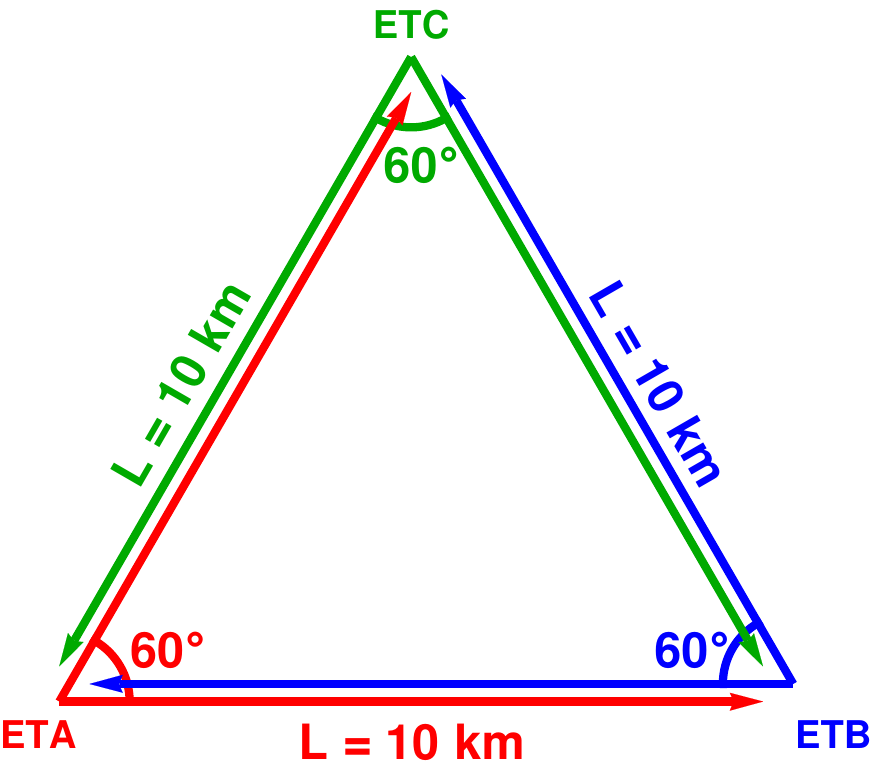}
\caption{ET detector topology: three V-shaped interferometers ETA, ETB and ETC with opening angle of $60^{\circ}$ and arm length of $10$ km displaced in a triangular way.}
\label{fig:3}
\end{figure}
\begin{figure*}
\includegraphics[width=0.3\textwidth]{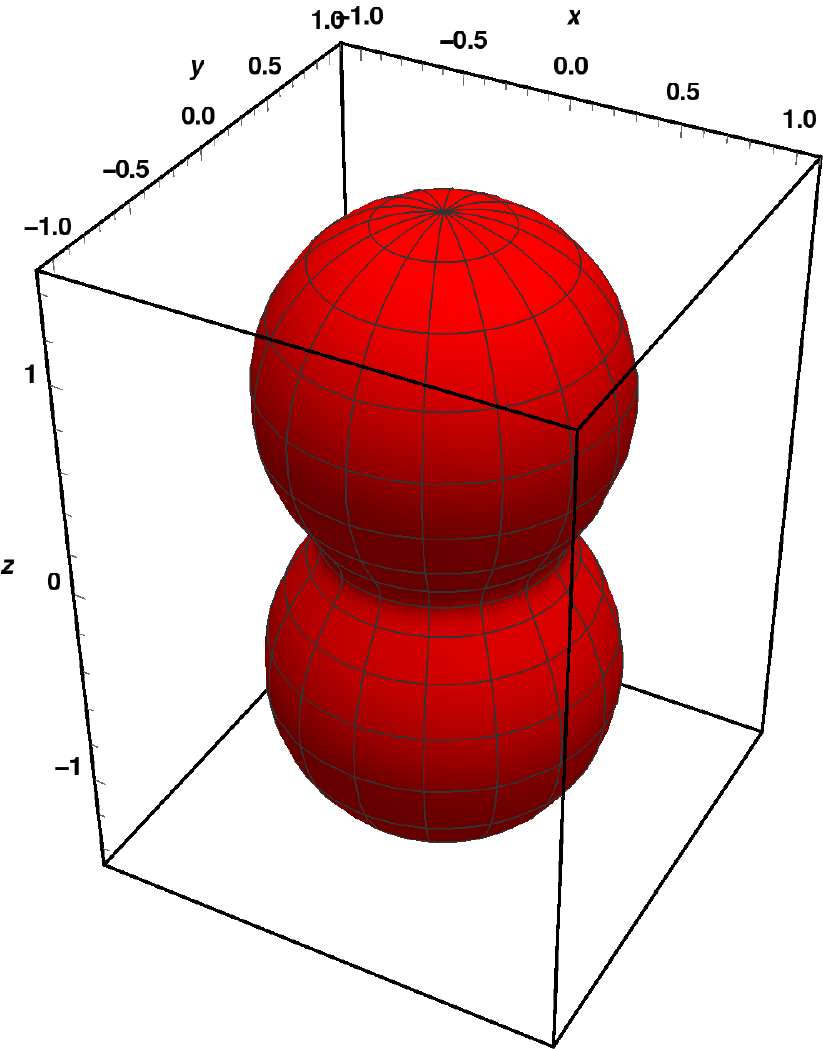}
\includegraphics[width=0.3\textwidth]{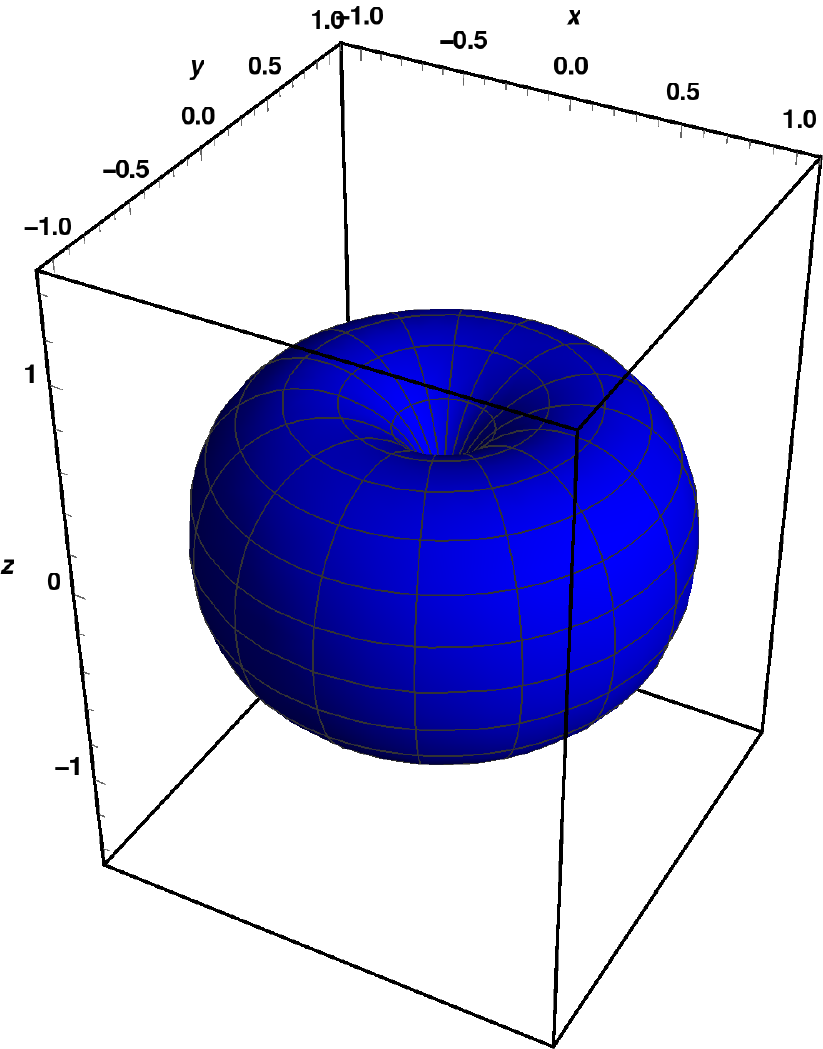}
\includegraphics[width=0.3\textwidth]{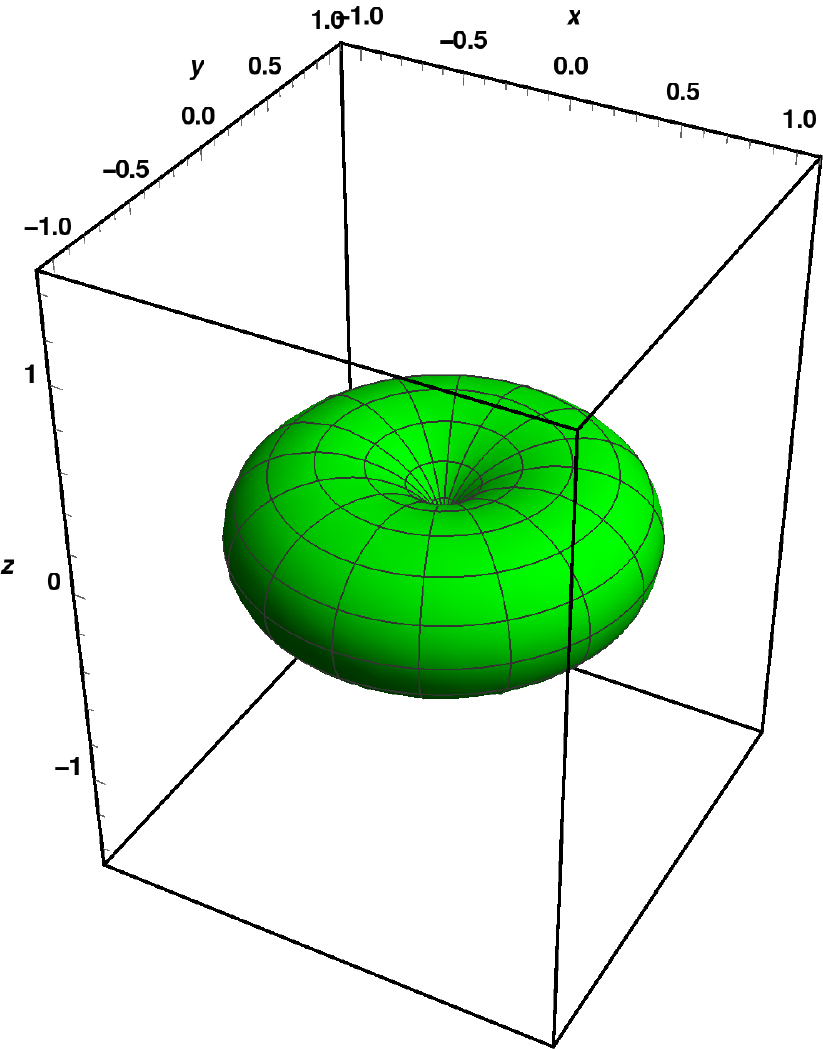}
\caption{(Left to right) Einstein Telescope square root joint responses for tensor ($\sqrt{F_{ET}^T}$, red), vector ($\sqrt{F_{ET}^V}$, blue) and scalar ($\sqrt{F_{ET}^S}$, green) polarization modes: in terms of tensor modes ET has full sky-coverage, while in terms of vector and scalar modes the only GW insensitive direction is the one orthogonal to the detector plane.}
\label{fig:4}
\end{figure*}

The Einstein Telescope (ET) observatory will most likely consist of three V-shaped interferometers (throughout the paper, we refer to ET three interferometers as ETA, ETB and ETC) with arm length $L=10$ km (meaning $f_* \approx 4774$ Hz) and opening angle of $60^{\circ}$ displaced in a triangular way \cite{Note1} as shown in Fig.\ref{fig:3}. In contrast to L-shaped interferometers, a triangular layout is preferred since it would be equally sensitive to plus and cross polarization modes and, as we shall see, would provide a more isotropic angular response in terms of tensor polarization modes. Moreover, in order to compute the detector angular response to each GW polarization mode, in this case we need to set unit vectors directed along each detector arm for every possible interferometer
\begin{eqnarray}
&&\mathbf{\hat{e}_{A1}}=\left(1,0,0\right), \hspace{0.3cm} \mathbf{\hat{e}_{A2}}=\left(\frac{1}{2},\frac{\sqrt{3}}{2},0\right),\nonumber\\
&&\mathbf{\hat{e}_{B1}}=\left(-\frac{1}{2},\frac{\sqrt{3}}{2},0\right), \hspace{0.3cm} \mathbf{\hat{e}_{B2}}=\left(-1,0,0\right),\nonumber\\
&&\mathbf{\hat{e}_{C1}}=\left(-\frac{1}{2},-\frac{\sqrt{3}}{2},0\right), \hspace{0.3cm} \mathbf{\hat{e}_{C2}}=\left(\frac{1}{2},-\frac{\sqrt{3}}{2},0\right).\nonumber
\end{eqnarray}
We further consider Eq.\eqref{eq:5} to compute APFs \cite{romano2017detection} relative to the single detector ETA, finding the following expressions
\begin{itemize}
\item Tensor modes: \begin{eqnarray}
F_A^+(\mathbf{\hat{\Omega}},\psi)=&&\frac{\sqrt{3}}{8}\left[\left(3+\cos 2 \theta \right) \cos 2 \psi  \sin \left(\frac{\pi}{3} -2 \phi \right)\right. \nonumber\\
&&\left.-4 \cos \theta  \sin 2 \psi  \cos \left(\frac{\pi}{3} -2 \phi \right)\right],
\end{eqnarray}
\begin{eqnarray}
F_A^{\times}(\mathbf{\hat{\Omega}},\psi)=&&-\frac{\sqrt{3}}{8} \left[4 \cos \theta  \cos 2 \psi  \cos \left(\frac{\pi}{3} -2 \phi \right)\right. \nonumber\\
&&\left.+(3+\cos 2 \theta) \sin 2 \psi  \sin \left(\frac{\pi}{3} -2 \phi \right)\right],
\end{eqnarray}
\item Vector modes: \begin{eqnarray}
F_A^x(\mathbf{\hat{\Omega}},\psi)=&&\frac{\sqrt{3}}{2} \sin \theta  \left[\cos \theta  \cos \psi  \sin \left(\frac{\pi}{3} -2 \phi \right)\right. \nonumber\\
&&\left.-\sin \psi  \cos \left(\frac{\pi}{3} -2 \phi \right)\right],
\end{eqnarray}
\begin{eqnarray}
F_A^y(\mathbf{\hat{\Omega}},\psi)=&&-\frac{\sqrt{3}}{2} \sin \theta  \left[\cos \theta  \sin \psi  \sin \left(\frac{\pi}{3} -2 \phi \right)\right.\nonumber\\
&&\left.+\cos \psi  \cos \left(\frac{\pi}{3} -2 \phi \right)\right],
\end{eqnarray}
\item Scalar modes: \begin{eqnarray}
F_A^b(\mathbf{\hat{\Omega}})=-\frac{\sqrt{3}}{4} \sin ^2\theta  \sin \left(\frac{\pi}{3} -2 \phi \right),
\label{eq:breathET}
\end{eqnarray}
\begin{eqnarray}
F_A^l(\mathbf{\hat{\Omega}})=\frac{\sqrt{3} \sin ^2\theta  \sin \left(\frac{\pi}{3} -2 \phi \right)}{2\sqrt{2}}.
\label{eq:longET}
\end{eqnarray}
\end{itemize}
Note that breathing and longitudinal APFs differ only for a multiplicative constant factor, thus making the two scalar polarization modes degenerate and indistinguishable for a single ET detecor. In order to compute the APFs relative to ETB and ETC for each polarization mode we can simply exploit the following identities (given ET triangular topology)
\begin{eqnarray}
&&F_B^P(\theta,\phi,\psi)=F_A^P\left(\theta,\phi-\frac{2\pi}{3},\psi\right)\nonumber\\
&&F_C^P(\theta,\phi,\psi)=F_A^P\left(\theta,\phi+\frac{2\pi}{3},\psi\right),
\end{eqnarray}
with $P$ = $+$, $\times$, $x$, $y$, $b$ and $l$. Since we have three interferometers, we can consider ET as a network to compute its joint angular response \cite{schutz2011networks} to tensor and, for the first time, to extra polarization modes
\begin{eqnarray}
F_{ET}^T(\theta)&&=\sum_{k=A,B,C}\left[ \left(F^+_k\right)^2+\left(F^{\times}_k\right)^2 \right]\nonumber\\
&&=\frac{9}{256}\bigl(35+28\cos2\theta+\cos4\theta \bigr),
\end{eqnarray}
\begin{eqnarray}
F_{ET}^V(\theta)&&=\sum_{k=A,B,C}\left[ \left(F^x_k\right)^2+\left(F^y_k\right)^2 \right]\nonumber\\
&&=\frac{9}{16}\bigl(3+\cos2\theta \bigr)\sin^{2}\theta,
\label{respV}
\end{eqnarray}
\begin{eqnarray}
F_{ET}^S(\theta)&&=\sum_{k=A,B,C}\left[ \left(F^b_k\right)^2+\left(F^l_k\right)^2 \right]\nonumber\\
&&=\frac{27\sin^{4}\theta}{32}.
\label{respS}
\end{eqnarray}
Plots of square roots of ET joint responses for tensor, vector and scalar polarization modes are shown in Fig.\ref{fig:4}.
In terms of tensor modes, ET sensitivity has already been investigated in literature \cite{regimbau2012mock}: the network optimal response is given by $F^T_{ET}(\theta=0)=3/2$, while the minimum value is given by $F^T_{ET}(\theta=\pi/2)=3/(2\sqrt{8})$. This also means that ET presents no blind directions, therefore the network has full sky-coverage. Additionally, ET averaged response over the solid angle $\mathbf{\Omega}$ is given by $\sqrt{\langle \left(F^T_{ET}\right)^2 \rangle}$ = $3/(\sqrt{10})$. Here, we extend these results to vector and scalar modes: optimal responses are given by $F^V_{ET}(\theta=\pi/2)=3/(2\sqrt{2})$ and $F^S_{ET}(\theta=\pi/2)=3\sqrt{3}/(4\sqrt{2})$, where the only blind direction is the one orthogonal to the detector plane identified by
\begin{eqnarray}
&&F_{ET}^V(\theta=0 \text{ mod }\pi)=0, \nonumber
\end{eqnarray}
\begin{eqnarray}
&&F_{ET}^S(\theta=0 \text{ mod }\pi)=0.\nonumber
\end{eqnarray}
Finally, ET averaged responses to extra polarization modes are given by $\sqrt{\langle \left(F^V_{ET}\right)^2 \rangle}$ = $3/(\sqrt{10})$ and $\sqrt{\langle \left(F^S_{ET}\right)^2 \rangle}$ = $3/(2\sqrt{5})$. Note that ET angular joint response to tensor modes is $\psi$-independent, but it presents cylindrical symmetry with respect to the orthogonal direction to the detector plane. These properties are a consequence of ET triangular topology and, considering Eqs.\eqref{respV} and \eqref{respS}, it is straightforward to extend them to extra polarization modes.

\subsection{The Cosmic Explorer}
\begin{figure}
\includegraphics[width=0.27\textwidth]{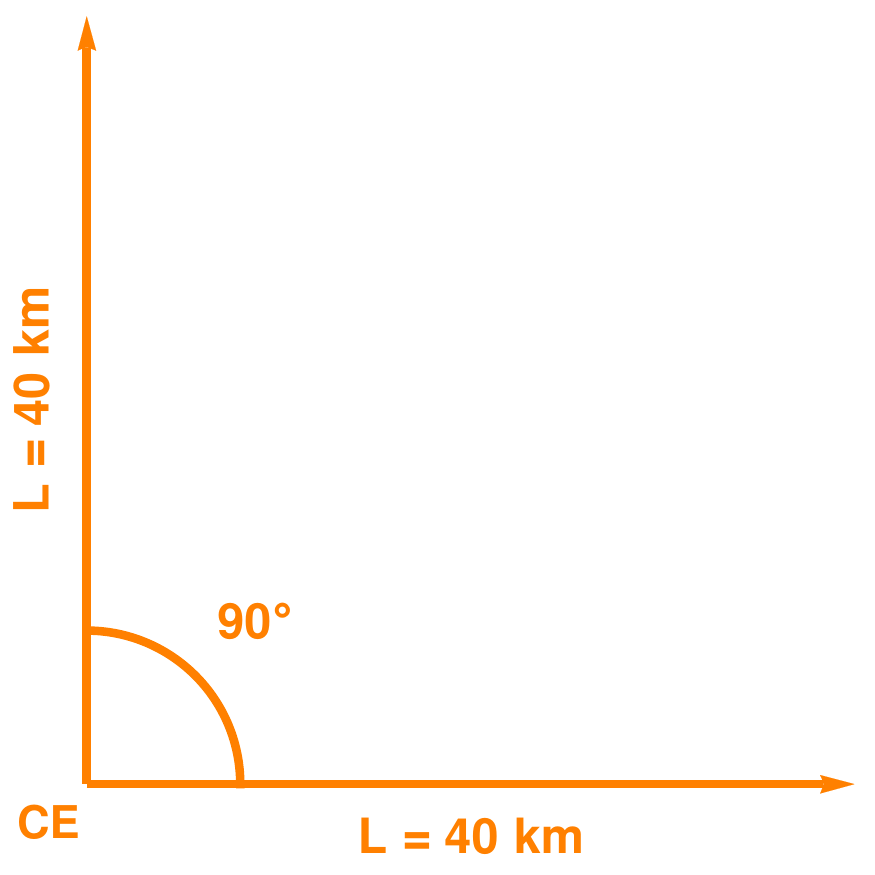}
\caption{Cosmic Explorer detector topology: the observatory will be a single L-shaped interferometer with an opening angle of $90^{\circ}$ and arm length of $40$ km.}
\label{fig:1}
\end{figure}
\begin{figure*}
\includegraphics[width=0.3\textwidth]{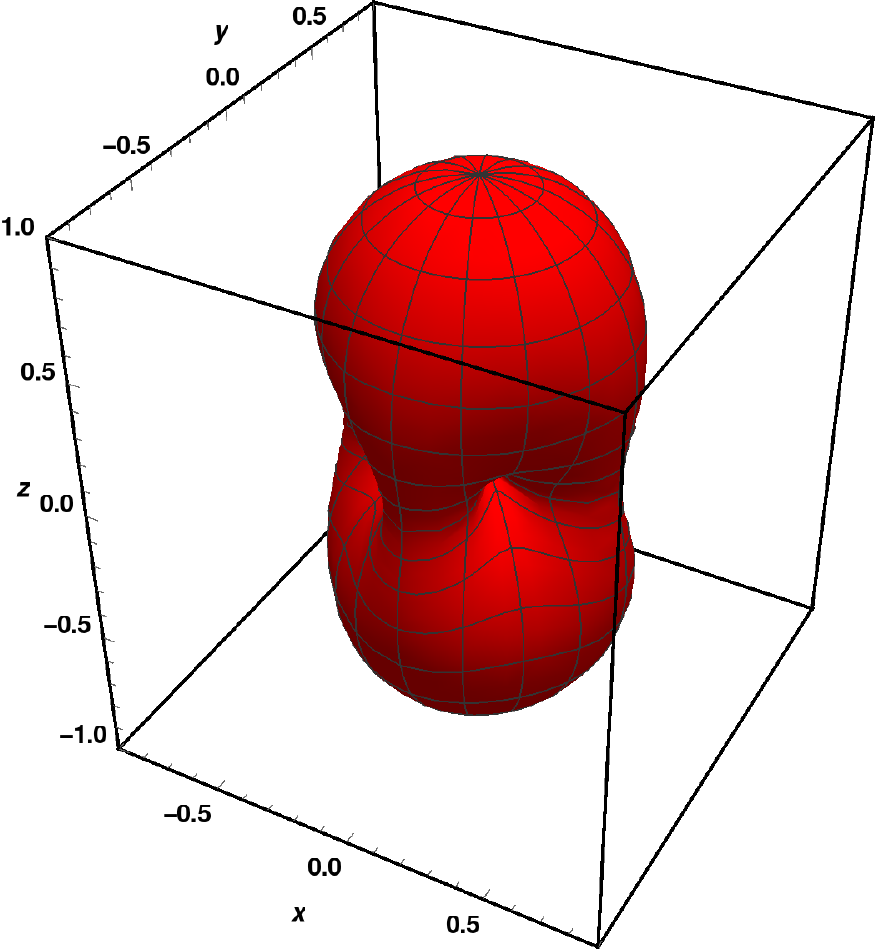}
\includegraphics[width=0.3\textwidth]{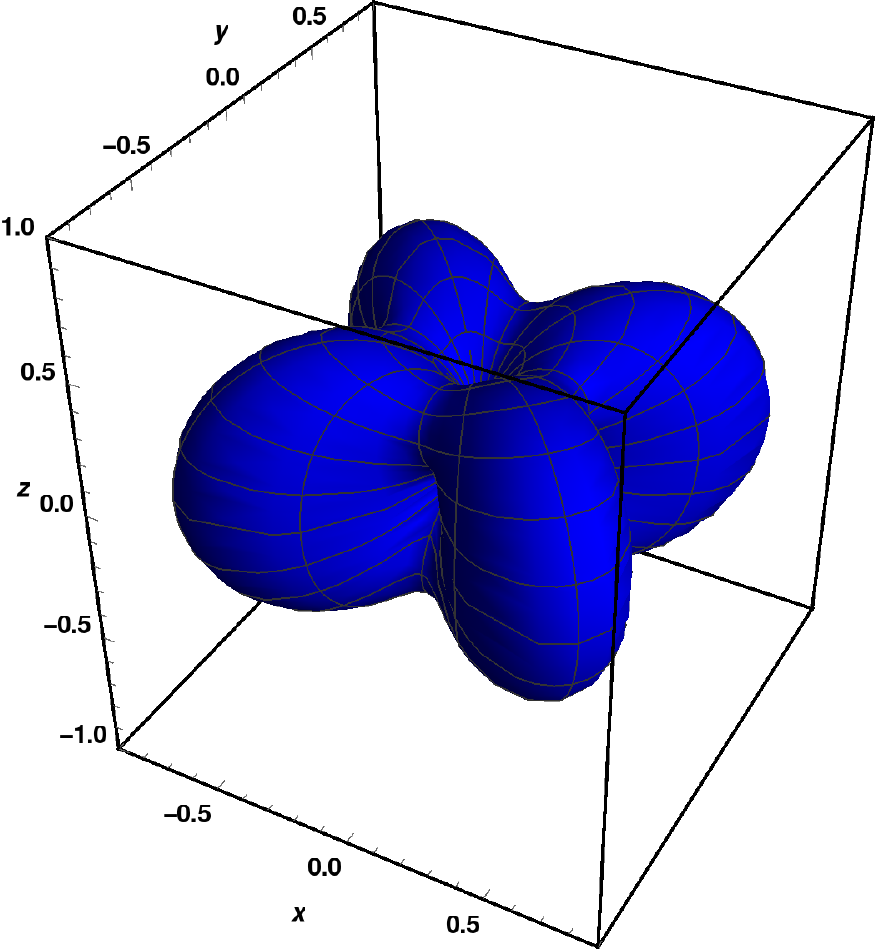}
\includegraphics[width=0.3\textwidth]{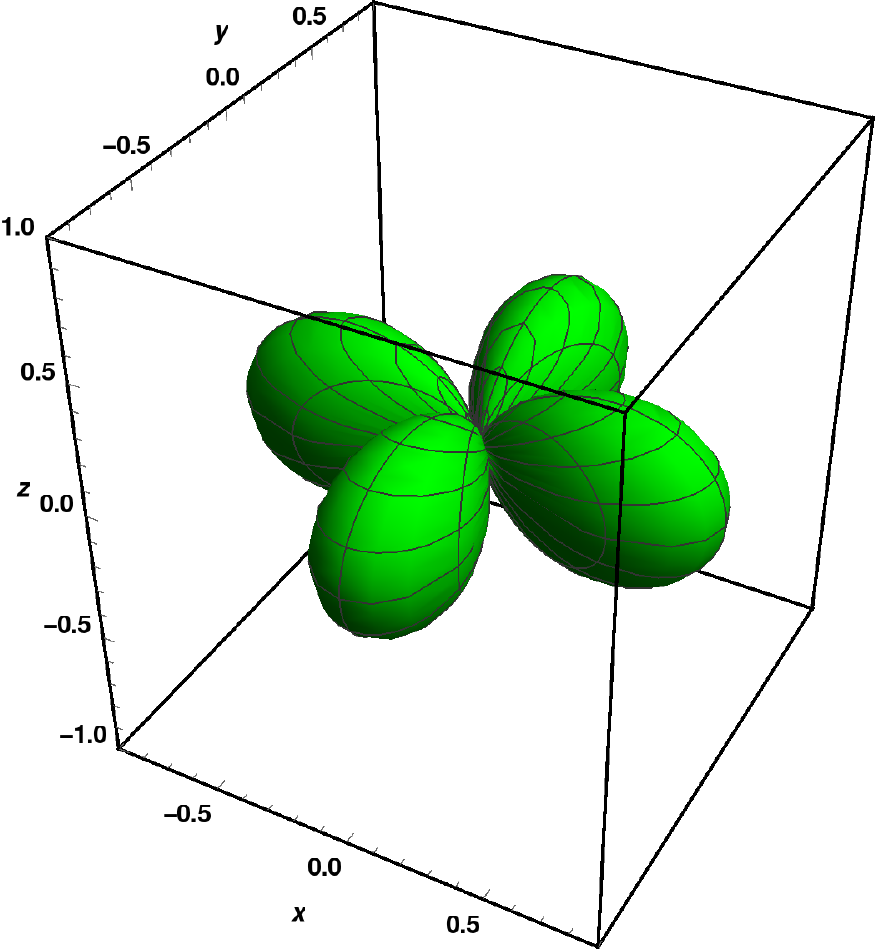}
\caption{(Left to right) Cosmic Explorer square root antenna power pattern functions for tensor ($\sqrt{F_{CE}^T}$, red), vector ($\sqrt{F_{CE}^V}$, blue) and scalar ($\sqrt{F_{CE}^S}$, green) polarization modes: each response presents different insensitive directions to incoming GWs.}
\label{fig:2}
\end{figure*}
\begin{figure*}
\includegraphics[width=0.4\textwidth]{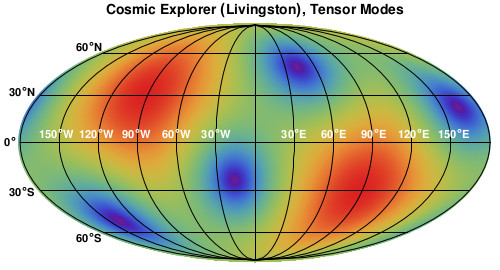}
\includegraphics[width=0.6cm,height=3.5cm]{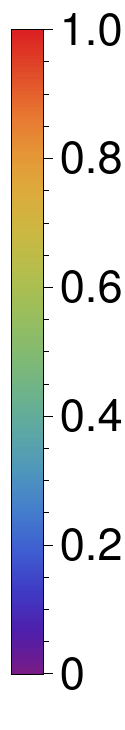}
\hspace{0.3cm}
\includegraphics[width=0.4\textwidth]{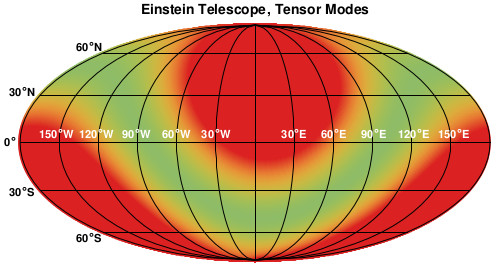}
\includegraphics[width=0.6cm,height=3.5cm]{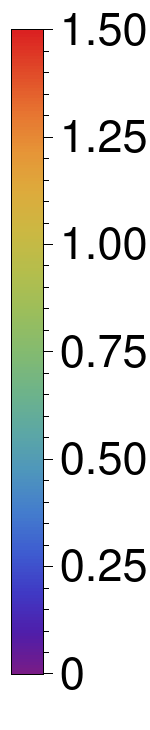}\\
\vspace{0.8cm}
\includegraphics[width=0.4\textwidth]{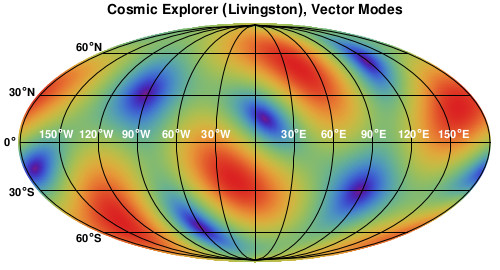}
\includegraphics[width=0.6cm,height=3.5cm]{figs/Bar2genTV.pdf}
\hspace{0.3cm}
\includegraphics[width=0.4\textwidth]{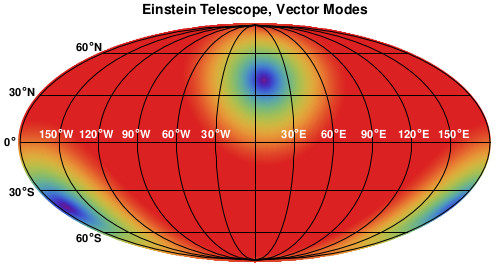}
\includegraphics[width=0.6cm,height=3.5cm]{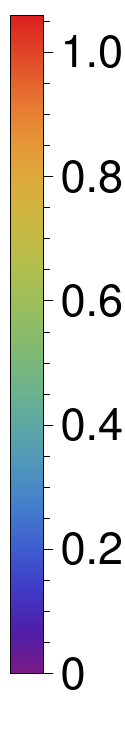}\\
\vspace{0.8cm}
\includegraphics[width=0.4\textwidth]{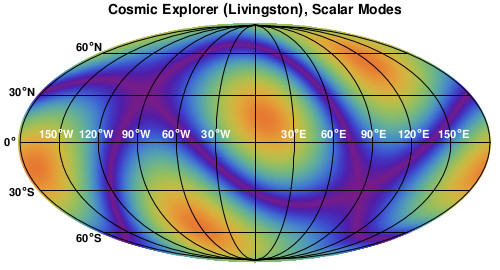}
\includegraphics[width=0.6cm,height=3.5cm]{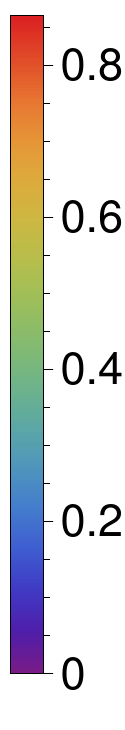}
\hspace{0.3cm}
\includegraphics[width=0.4\textwidth]{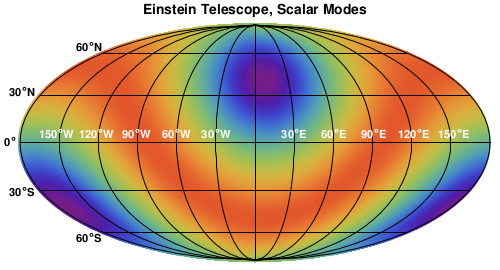}
\includegraphics[width=0.6cm,height=3.5cm]{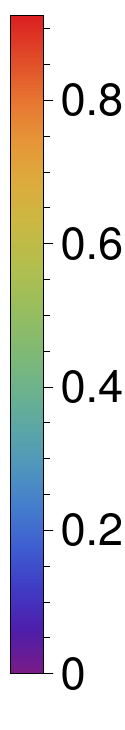}
\caption{(Left) Mollweide projection of CE APPFs for tensor, vector and scalar polarization modes, where CE was assumed to replace the LIGO Livingston observatory both in location and orientation. (Right) Mollweide projection of ET network joint responses to tensor, vector and scalar polarization modes, where ET was assumed to be located in the Sardinia island, Italy. In contrast to an L-shaped interferometer, ET triangular topology provides more isotropic angular responses to both tensor and non-GR polarization modes.}
\label{fig:5}
\end{figure*}

The Cosmic Explorer (CE) observatory will consist of one single L-shaped interferometer with arm length $L=40$ km ($10$ times the length of LIGO and $4$ times the length of one Einstein Telescope interferometer) and opening angle of $90^{\circ}$ as shown in Fig.\ref{fig:1} and a characteristic frequency $f_* \approx 1194$ Hz. In order to compute the detector angular response to each GW polarization mode we need to set the unit vectors directed to each detector arm 
\[\mathbf{\hat{e}_{1}}=(1,0,0), \hspace{0.3cm} \mathbf{\hat{e}_{2}}=(0,1,0). \]
Then, from Eq.\eqref{eq:5}, we get the following APFs \cite{romano2017detection}
\begin{itemize}
\item Tensor modes: \begin{eqnarray}
F^+(\mathbf{\hat{\Omega}},\psi)=&&\frac{1}{2}\left(1+\cos^{2}\theta\right)\cos2\phi\cos2\psi \nonumber\\
&&-\cos\theta\sin2\phi\sin2\psi,
\end{eqnarray}
\begin{eqnarray}
F^{\times}(\mathbf{\hat{\Omega}},\psi)=&&-\frac{1}{2}\left(1+\cos^{2}\theta\right)\cos2\phi\sin2\psi \nonumber\\
&&-\cos\theta\sin2\phi\cos2\psi,
\end{eqnarray}
\item Vector modes: \begin{eqnarray}
F^x(\mathbf{\hat{\Omega}},\psi)=&&\sin\theta(\cos\theta\cos2\phi\cos\psi\nonumber\\
&&-\sin2\phi\sin\psi),
\end{eqnarray}
\begin{eqnarray}
F^y(\mathbf{\hat{\Omega}},\psi)=&&-\sin\theta(\cos\theta\cos2\phi\sin\psi\nonumber\\
&&+\sin2\phi\cos\psi),
\end{eqnarray}
\item Scalar modes: \begin{eqnarray}
F^b(\mathbf{\hat{\Omega}})=-\frac{1}{2}\sin^{2}\theta\cos2\phi,
\label{eq:breathCE}
\end{eqnarray}
\begin{eqnarray}
F^l(\mathbf{\hat{\Omega}})=\frac{1}{\sqrt{2}}\sin^{2}\theta\cos2\phi.
\label{eq:longCE}
\end{eqnarray}
\end{itemize}
Despite the fact we are considering a GW interferometer with a bigger opening angle with respect to ET, once again the two scalar modes differ only for a constant factor and they are still degenerate. We shall see in section \ref{GWhigh} that the choice of the interferometer opening angle does not break the degeneracy in the low-frequency limit. We further compute the CE angular response to joined tensor, vector and scalar polarization modes in the following way  
\begin{eqnarray}
F_{CE}^T(\theta,\phi)&&\equiv\left(F^+\right)^2+\left(F^{\times}\right)^2\nonumber\\
&&=\frac{1}{4} \left(1+\cos ^2\theta \right)^2 \cos ^22 \phi +\cos ^2\theta  \sin^2 2 \phi, \nonumber\\
\end{eqnarray}
\begin{eqnarray}
F_{CE}^V(\theta,\phi)&&\equiv\left(F^x\right)^2+\left(F^y\right)^2\nonumber\\
&&=\sin ^2\theta  \left(\cos ^2\theta  \cos ^22 \phi +\sin ^22 \phi \right), \nonumber\\
\end{eqnarray}
\begin{eqnarray}
F_{CE}^S(\theta,\phi)&&\equiv\left(F^b\right)^2+\left(F^l\right)^2\nonumber\\
&&=\frac{3}{4} \sin ^4\theta  \cos ^22 \phi. 
\end{eqnarray}
These responses are also called detector antenna power pattern functions (APPFs) for tensor, vector and scalar modes \cite{schutz2011networks} and plots of their corresponding square root are shown in Fig.\ref{fig:2}.
Note that also CE angular responses to tensor, vector and scalar polarization modes are independent from the polarization angle $\psi$. Optimal values for APPFs are given by $F^T_{CE}(\theta = 0 \text{ mod } \pi,\phi) = 1$, $F^V_{CE}(\theta = \pi/2,\phi = \pi/4 \text{ mod } \pi/4) = 1$ and $F^S_{CE}(\theta = \pi/2,\phi = 0 \text{ mod } \pi/2) = \sqrt{3}/2$. However, there are clearly some directions to which the detector results utterly insensitive \cite{romano2017detection}: these blind spots are given by 
\begin{eqnarray}
F_{CE}^T\left(\theta=\frac{\pi}{2},\phi=\frac{\pi}{4} \text{ mod }\frac{\pi}{2}\right)=0,
\end{eqnarray}
\begin{eqnarray}
&&F_{CE}^V\left(\theta=\frac{\pi}{2},\phi=0 \text{ mod }\frac{\pi}{2}\right)=0,\nonumber\\
&&F_{CE}^V\left(\theta=0 \text{ mod }\pi,\phi\right)=0,
\end{eqnarray}
\begin{eqnarray}
&&F_{CE}^S\left(\theta=\frac{\pi}{2},\phi=\frac{\pi}{4}\text{ mod }\frac{\pi}{2}\right)=0,\nonumber\\
&&F_{CE}^S\left(\theta=0\text{ mod }\pi,\phi\right)=0.
\end{eqnarray}
Similarly to what was done for ET, we compute the values of the averaged angular responses over the solid angle $\mathbf{\Omega}$, which are given by $\sqrt{\langle \left(F^T_{CE}\right)^2 \rangle}$ = $\sqrt{2/5}$, $\sqrt{\langle \left(F^V_{CE}\right)^2 \rangle}$ = $\sqrt{2/5}$ and $\sqrt{\langle \left(F^S_{CE}\right)^2 \rangle}$ = $1/\sqrt{5}$.
In terms of tensor modes, if we compare the average response of CE to the one we obtained for ET, we see the latter is smaller by a factor $\sin \pi/3$ = $\sqrt{3}/2$, but its three detectors enhance its response by a factor $\sqrt{3}$ \cite{regimbau2012mock}. Moreover, this enhancement is due to the detector geometry, rather than the polarization modes considered: indeed, our results show this is also true for vector and scalar polarization modes, therefore we have
\begin{eqnarray}
\sqrt{\langle \left(F^M_{ET}\right)^2 \rangle}=\frac{3}{2}\sqrt{\langle \left(F^M_{CE}\right)^2 \rangle},
\end{eqnarray}
for $M=T,V$ and $S$.
At the time of writing, both Einstein Telescope and CE projects are still being discussed and we have very little or no information on their definitive location and orientation, although we know that two preferred sites for ET are the Sardinia island in Italy and the border region between the Netherlands, Belgium and Germany \cite{Note1}. Therefore, we further consider an Earth-based coordinate system and we show the plot relative to CE (which we assume to replace the LIGO Livingston observatory both in location and orientation) APPFs and to ET (located in the Sardinia island in Italy) joint angular responses to tensor, vector and scalar polarization modes in Fig.\ref{fig:5}: note how ET triangular topology visibly provides a more isotropic angular response to tensor modes, in contrast to a single L-shaped interferometer. As we discussed, indeed this also applies to extra polarization modes, where the number of insensitive directions to incoming GWs is reduced to one (i.e. the orthogonal direction to the detector plane).

\section{Correlation Analysis \label{CorrAn}}
GWs have tensorial nature, though we expect the detector output to be given by a scalar quantity expressed as the sum of two terms \cite{maggiore2008gravitational, romano2017detection}: the first one given by $h(t)$, representing the true GW signal, and the second one given by $n(t)$, representing the detector noise
\begin{eqnarray}
s(t)=h(t)+n(t).
\end{eqnarray}
In particular, for a detector located in $\mathbf{\bar{x}}$, the GW signal is expressed as
\begin{eqnarray}
h(t)&&\equiv D^{ij}h_{ij}(t,\mathbf{\bar{x}})\nonumber\\
&&=\sum\limits_{P}\int_{-\infty}^{+\infty}df\int_{S^2}d\Omega\left[F^P(\mathbf{\hat{\Omega}})h_P(f,\mathbf{\hat{\Omega}})\right]\nonumber\\
&&\hspace{0.35cm}\times e^{i2\pi f\left(t-\mathbf{\hat{\Omega}}\cdot\frac{\mathbf{\bar{x}}}{c}\right)},
\label{eq:7}
\end{eqnarray}
where in the second equality we used Eq.\eqref{eq:5}. It is usually more convenient to switch to the frequency domain, thus expressing the signal as
\begin{eqnarray}
s(f)=h(f)+n(f), 
\end{eqnarray}
where $h(f)$ is the Fourier transform of Eq.\eqref{eq:7} given by
\begin{eqnarray}
h(f)=\sum\limits_{P}\int_{S^2}d\Omega\left[F^P(\mathbf{\hat{\Omega}})h_P(f,\mathbf{\hat{\Omega}})\right]e^{-i2\pi f\mathbf{\hat{\Omega}}\cdot\frac{\mathbf{\bar{x}}}{c}},
\end{eqnarray}
and $n(f)$ is the Fourier transform of the noise term. In the rest of this section we want to understand how GWs can be detected through the correlation analysis technique while focusing on a SGWB \cite{romano2017detection} which we assume to be stationary, unpolarized and in first approximation both gaussian and isotropic \footnote{for some recent works about non-Gaussianities and anisotropies of the SGWB, see e.g. \cite{bartolo2018probing, Bartolo:2019oiq, bartolo2020characterizing, bartolo2020gravitational, DallArmi:2020dar}.}. Whenever these four assumptions can be taken to be valid, all SGWB statistical properties are characterized by the so-called two-point correlator \cite{romano2017detection}
\begin{eqnarray}
\langle h^*_P(f,\mathbf{\hat{\Omega}}) h_{P'}(f',\mathbf{\hat{\Omega}}')  \rangle =&&\frac{1}{4 \pi}\delta(\mathbf{\hat{\Omega}},\mathbf{\hat{\Omega}}') \delta(f-f') \nonumber\\
&&\times\delta_{PP'} \frac{1}{2}S^P(f),
\label{eq:8}
\end{eqnarray}
where $\langle \cdot \rangle$ denotes the ensemble average, while $S^P(f )$ is a real function called power spectral density, it is defined for each polarization mode and it has dimensions Hz$^{-1}$. In order to characterize the SGWB energy
density, the energy density per logarithmic frequency bin normalized by the critical energy density of the Universe is introduced for each polarization mode \cite{maggiore2008gravitational}
\begin{eqnarray}
\Omega^P_{GW}(f)\equiv\frac{1}{\rho_c}\frac{d \rho^P_{GW}}{d \ln f},
\label{eq:spectrum}
\end{eqnarray}
where $\rho_c=3H_0^2/8\pi G$ and $H_0$ is the Hubble constant. Moreover, there exists a precise relation between the two functions $S^P(f)$ and $\Omega^P_{GW}(f)$ given by
\begin{eqnarray}
\Omega^P_{GW}(f)=\frac{2 \pi^2}{3 H_0^2}f^3S^P(f).
\end{eqnarray}
Since we are considering an unpolarized SGWB, this also means that the energy density related to tensor, vector and scalar modes is given by
\begin{eqnarray}
&&\Omega^T_{GW}(f)=\Omega^+_{GW}(f)+\Omega^{\times}_{GW}(f),\nonumber \\
&&\Omega^V_{GW}(f)=\Omega^x_{GW}(f)+\Omega^y_{GW}(f), \nonumber \\
&&\Omega^S_{GW}(f)=(1+ \kappa)\Omega^b_{GW}(f),
\end{eqnarray}
where plus and cross polarization modes equally contribute to the tensor modes energy density ($\Omega^+_{GW}=\Omega^{\times}_{GW}$), x and y equally contribute to the vector modes energy density ($\Omega^x_{GW}=\Omega^y_{GW}$) and we set the longitudinal polarization mode energy density as a fraction of the breathing one as a consequence of the two scalar modes being indistinguishable for a ground-based GW-interferometer in the low-frequency limit. Throughout this paper we assume the detector noise to be stationary, which means that given two detectors $I$ and $J$ we have \cite{romano2017detection}
\begin{eqnarray}
\langle n_I^*(f) n_J(f') \rangle = \delta_{IJ}\delta(f-f') \frac{1}{2}P(f),
\label{eq:9}
\end{eqnarray}
where $P(f)$ is called noise power spectral density (PSD), it has dimensions Hz$^{-1}$ and it describes the detector noise statistical properties. It is worth noting that the $\delta_{IJ}$ factor in Eq.\eqref{eq:9} indicates that we are taking different detector noises to be uncorrelated. This might be no longer the case while considering ET, since interferometers are expected to be approximately colocated and the noise contributions might be correlated. Therefore, one extra correlation term could appear in Eq.\eqref{eq:9} and it would need a proper treatment to be canceled \footnote{Techniques to identify noise correlation terms relative to a pair of co-located detectors have been developed e.g. for the two LIGO Hanford interferometers \cite{fotopoulos2008searching} and could be generalized to ET}. Moreover, we further assume
\[P_{ETA}(f)=P_{ETB}(f)=P_{ETC}(f)\equiv P_{ET}(f).\]
Sensitivity curves for ground-based interferometers are usually represented by the so-called amplitude spectral density and they are provided for both ET and CE.
\begin{figure*}
\includegraphics[width=0.45\textwidth]{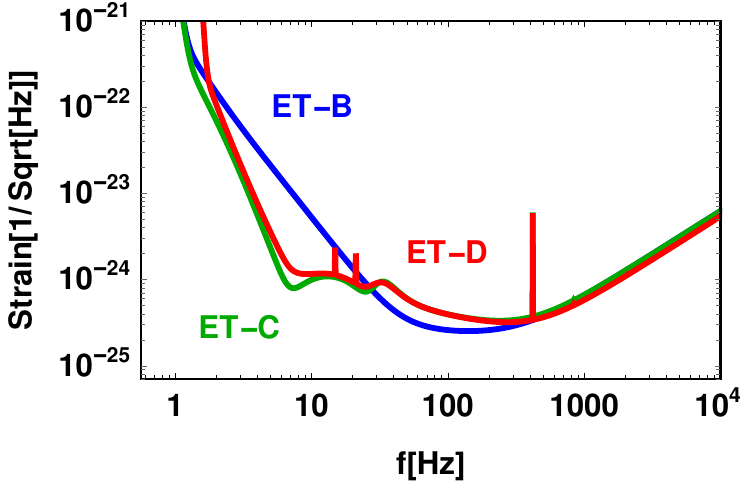}
\hspace{0.3cm}
\includegraphics[width=0.45\textwidth]{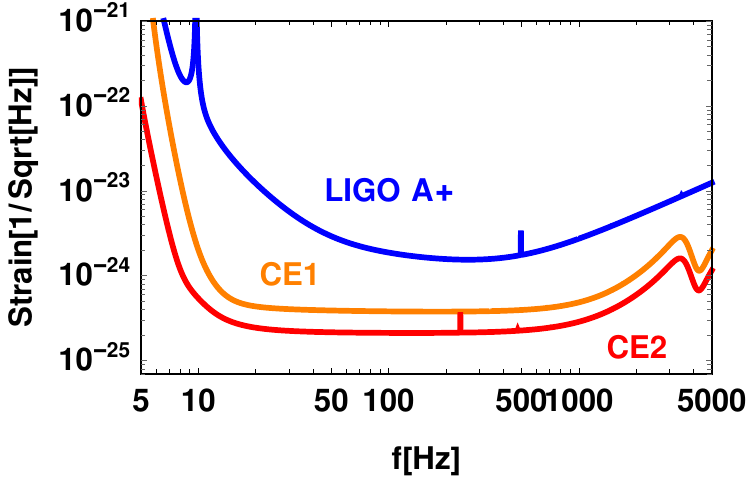}\\
\caption{(Left) Einstein Telescope sensitivity curves for three discussed configurations ET-B (single interferometer sensitive to the whole frequency range expected for ET \cite{hild2008pushing}), ET-C and ET-D (xylophone configuration \cite{hild2009xylophone, hild2011sensitivity}). (Right) Cosmic Explorer projected sensitivity curves for Stage 1 ($2030$s) and Stage 2 ($2040$s) \cite{reitze2019us} compared to LIGO A+ design sensitivity \cite{barsotti2018a+}.}
\label{fig:NOISE}
\end{figure*}
ET sensitivity curves denoted with ET-B, ET-C and ET-D are shown in the left panel of Fig.\ref{fig:NOISE}. In particular, ET-B \cite{hild2008pushing} refers to a single ET interferometer which is sensitive to the whole Einstein Telescope frequency range ($1$-$10000$ Hz), while ET-C \cite{hild2009xylophone} refers to the so-called xylophone configuration where each of the three ET interferometers is in turn composed of two more interferometers, the first one being more sensitive to low GW frequencies ($1$-$250$ Hz) and the second one specialized for higher GW frequencies ($10$-$10000$ Hz) instead. Finally, ET-D represents an upgraded  version of the xylophone configuration with respect to ET-C accounting for more noise sources \cite{hild2011sensitivity}. On the other hand, sensitivity curves for CE Stage 1 (CE1) and Stage 2 (CE2) \cite{reitze2019us} are shown in the right panel of Fig.\ref{fig:NOISE} along with the LIGO A+ design sensitivity \cite{barsotti2018a+} for a direct comparison. The first configuration denoted by CE1 is assumed to be operative in the late $2030$s, while the configuration CE2, which will upgrade the detector performance, is expected to begin its operations in the mid $2040$s; both CE1 and CE2 are currently defined between $5$-$5000$ Hz. Throughout this paper, we mostly focus on the possibility of working with three ET interferometers in their proposed xylophone configuration along with the Stage 1 CE: in our analysis, whenever we consider a detector pair or network involving both ET and CE characterized by ET-D and CE1 sensitivity curves, we refer to the corresponding combined configuration as ``D1''. Similarly, when we consider each ET interferometer characterized by ET-B along with the Stage 1 CE, we refer to the relative combined configuration as ``B1''. While investigating SGWBs, a direct comparison between B1 and D1 will allow us to understand how different ET sensitivities may produce different forecasts, thus where the xylophone configuration would be the optimal one. We shall not consider the Stage 2 CE: clearly, this would further improve the detector sensitivity to incoming GWs, though this upgrade will only be available much later in time with respect to Stage 1.

\subsection{Overlap reduction functions}
We now define the cross-correlation of two detector outputs by taking the following ensemble average \cite{maggiore2008gravitational, romano2017detection}
\begin{eqnarray}
G_{IJ}(f,f')&&=\langle s_I^*(f) s_J(f') \rangle = \langle h_I^*(f) h_J(f') \rangle,
\label{eq:10}
\end{eqnarray}
where we used Eq.\eqref{eq:9} to get the second equality as long as $I \neq J$. We assume the existence of all extra polarization modes so we can write the previous result in a more compact way as
\begin{eqnarray}
G_{IJ}(f,f')&&=G_{IJ}^T(f,f')+G_{IJ}^V(f,f')+G_{IJ}^S(f,f')\nonumber\\
&&=\frac{1}{10}\bigl[ \gamma^T_{IJ}(f)S^T(f)+\gamma^V_{IJ}(f)S^V(f)\nonumber\\
&&\hspace{0.35cm}+\xi\gamma^S_{IJ}(f)S^S(f) \bigr] \times \delta(f-f')\nonumber \\
&&=\frac{3H_0^2}{20 \pi^2}f^{-3}\bigl[ \gamma^T_{IJ}(f)\Omega^T_{GW}(f)+\gamma^V_{IJ}(f)\Omega^V_{GW}(f)\nonumber\\
&&\hspace{0.35cm}+\xi\gamma^S_{IJ}(f)\Omega^S_{GW}(f) \bigr]\times \delta(f-f'),
\label{eq:11}
\end{eqnarray}
where we have defined 
\[G_{IJ}^M(f,f') \equiv \frac{1}{10}[ \gamma^M_{IJ}(f)S^M(f)]\delta(f-f'),\]
for $M$ = $T$, $V$ and $S$. We also introduced the so-called normalized overlap reduction functions (ORFs), which are defined separately for tensor, vector and scalar modes
\begin{eqnarray}
\gamma^{T}_{IJ}(f)=&&\frac{5}{2}\int_{S^2}\frac{d\Omega}{4\pi}e^{i2\pi f\mathbf{\hat{\Omega}}\cdot\frac{\Delta\overline{\mathbf{X}}}{c}}\left(\sum\limits_{P=+, \times}F^P_I(\mathbf{\hat{\Omega}})F^P_J(\mathbf{\hat{\Omega}})\right),\nonumber\\
\end{eqnarray}
\begin{eqnarray}
\gamma^{V}_{IJ}(f)=\frac{5}{2}\int_{S^2}\frac{d\Omega}{4\pi}e^{i2\pi f\mathbf{\hat{\Omega}}\cdot\frac{\Delta\overline{\mathbf{X}}}{c}}\left(\sum\limits_{P=x, y}F^P_I(\mathbf{\hat{\Omega}})F^P_J(\mathbf{\hat{\Omega}})\right),\nonumber\\
\end{eqnarray}
\begin{eqnarray}
\gamma^{S}_{IJ}(f)=&&\frac{15}{1+2\kappa}\int_{S^2}\frac{d\Omega}{4\pi}e^{i2\pi f\mathbf{\hat{\Omega}}\cdot\frac{\Delta\overline{\mathbf{X}}}{c}}\left( F^b_I(\mathbf{\hat{\Omega}})F^b_J(\mathbf{\hat{\Omega}})\right.\nonumber\\
&&\left.+\kappa F^l_I(\mathbf{\hat{\Omega}})F^l_J(\mathbf{\hat{\Omega}})\right),
\end{eqnarray}
where we defined the spatial separation between the detector pair
\begin{equation}
\Delta\overline{\mathbf{x}}=\overline{\mathbf{x}}_I-\overline{\mathbf{x}}_J \,,
\end{equation}
and the parameter
\begin{equation}
\xi=\frac{1}{3}\biggl( \frac{1+2\kappa}{1+\kappa}\biggr)\,,
\end{equation}
which ranges from $\xi=1/3$ (no longitudinal polarization mode present) to $\xi=2/3$ (no breathing polarization mode present). ORFs start to oscillate when $f\approx f_c=c/2 \pi |\Delta\overline{\mathbf{x}}|$ causing a loss of sensitivity of the detector pair correlated responses to the SGWB signal. In literature, the ORF analytic expressions for tensor, vector and scalar polarization modes are provided in \cite{nishizawa2009probing} along with some related interesting properties, we have
\begin{itemize}
\item Tensor modes:
\begin{eqnarray}
\gamma^T_{IJ}(\alpha,\beta,\sigma_1,\sigma_2)=&&\sin\nu_I\sin\nu_J\bigl[\Theta_T^+(\alpha,\beta)\cos2(\sigma_1+\sigma_2)\nonumber\\
&&+\Theta_T^-(\alpha,\beta)\cos2(\sigma_1-\sigma_2)\bigr],
\label{eq:ORFT}
\end{eqnarray}
\begin{eqnarray}
\Theta_T^+(\alpha,\beta)=&&-\biggl(\frac{3}{8}j_0-\frac{45}{56}j_2+\frac{169}{896}j_4 \biggr)\nonumber\\
&&+\biggl(\frac{1}{2}j_0-\frac{5}{7}j_2-\frac{27}{224}j_4 \biggr)\cos\beta\nonumber\\
&&-\biggl(\frac{1}{8}j_0+\frac{5}{56}j_2+\frac{3}{896}j_4 \biggr)\cos2\beta,
\label{eq:ORFT1}
\end{eqnarray}
\begin{eqnarray}
\Theta_T^-(\alpha,\beta)=&&\biggl(j_0+\frac{5}{7}j_2+\frac{3}{112}j_4 \biggr)\cos^{4}\biggl(\frac{\beta}{2}\biggr),
\label{eq:ORFT2}
\end{eqnarray}

\item Vector modes:
\begin{eqnarray}
\gamma^V_{IJ}(\alpha,\beta,\sigma_1,\sigma_2)=&&\sin\nu_I\sin\nu_J\bigl[\Theta_V^+(\alpha,\beta)\cos2(\sigma_1+\sigma_2) \nonumber\\
&&+\Theta_V^-(\alpha,\beta)\cos2(\sigma_1-\sigma_2)\bigr],
\label{eq:ORFV}
\end{eqnarray}
\begin{eqnarray}
\Theta_V^+(\alpha,\beta)=&&-\biggl(\frac{3}{8}j_0+\frac{45}{112}j_2-\frac{169}{896}j_4 \biggr)\nonumber\\
&&+\biggl(\frac{1}{2}j_0+\frac{5}{14}j_2+\frac{27}{56}j_4 \biggr)\cos\beta\nonumber\\
&&-\biggl(\frac{1}{8}j_0-\frac{5}{112}j_2-\frac{3}{224}j_4 \biggr)\cos2\beta,
\label{eq:ORFV1}
\end{eqnarray}
\begin{eqnarray}
\Theta_V^-(\alpha,\beta)=&&\biggl(j_0-\frac{5}{14}j_2-\frac{3}{28}j_4 \biggr)\cos^{4}\biggl(\frac{\beta}{2}\biggr),
\label{eq:ORFV2}
\end{eqnarray}

\item Scalar modes:
\begin{eqnarray}
\gamma^S_{IJ}(\alpha,\beta,\sigma_1,\sigma_2)=&&\sin\nu_I\sin\nu_J\bigl[\Theta_S^+(\alpha,\beta)\cos2(\sigma_1+\sigma_2) \nonumber\\
&&+\Theta_S^-(\alpha,\beta)\cos2(\sigma_1-\sigma_2)\bigr],
\label{eq:ORFS}
\end{eqnarray}
\begin{eqnarray}
\Theta_S^+(\alpha,\beta)=&&-\biggl(\frac{3}{8}j_0+\frac{45}{56}j_2+\frac{507}{448}j_4 \biggr)\nonumber\\
&&+\biggl(\frac{1}{2}j_0+\frac{5}{7}j_2-\frac{81}{112}j_4 \biggr)\cos\beta\nonumber\\
&&-\biggl(\frac{1}{8}j_0-\frac{5}{56}j_2+\frac{9}{448}j_4 \biggr)\cos2\beta,
\label{eq:ORFS1}
\end{eqnarray}
\begin{eqnarray}
\Theta_S^-(\alpha,\beta)=&&\biggl(j_0-\frac{5}{7}j_2+\frac{9}{56}j_4 \biggr)\cos^{4}\biggl(\frac{\beta}{2}\biggr).
\label{eq:ORFS2}
\end{eqnarray}
\end{itemize}
Here, $\beta$ is the separation angle between the two detectors (with internal opening angles $\nu_I$ and $\nu_J$) with respect to the center of the Earth, while $\sigma_1$ and $\sigma_2$ are the two bisector orientation angles measured in a counterclockwise manner with respect to the great circle connecting the pair and wrapping the planet. Moreover, $j_n(\alpha)$ are spherical Bessel functions, where \[\alpha\equiv \frac{f}{f_c}=\frac{2\pi f|\Delta\overline{\mathbf{x}}|}{c}.\]

\begin{figure}
\includegraphics[width=0.45\textwidth]{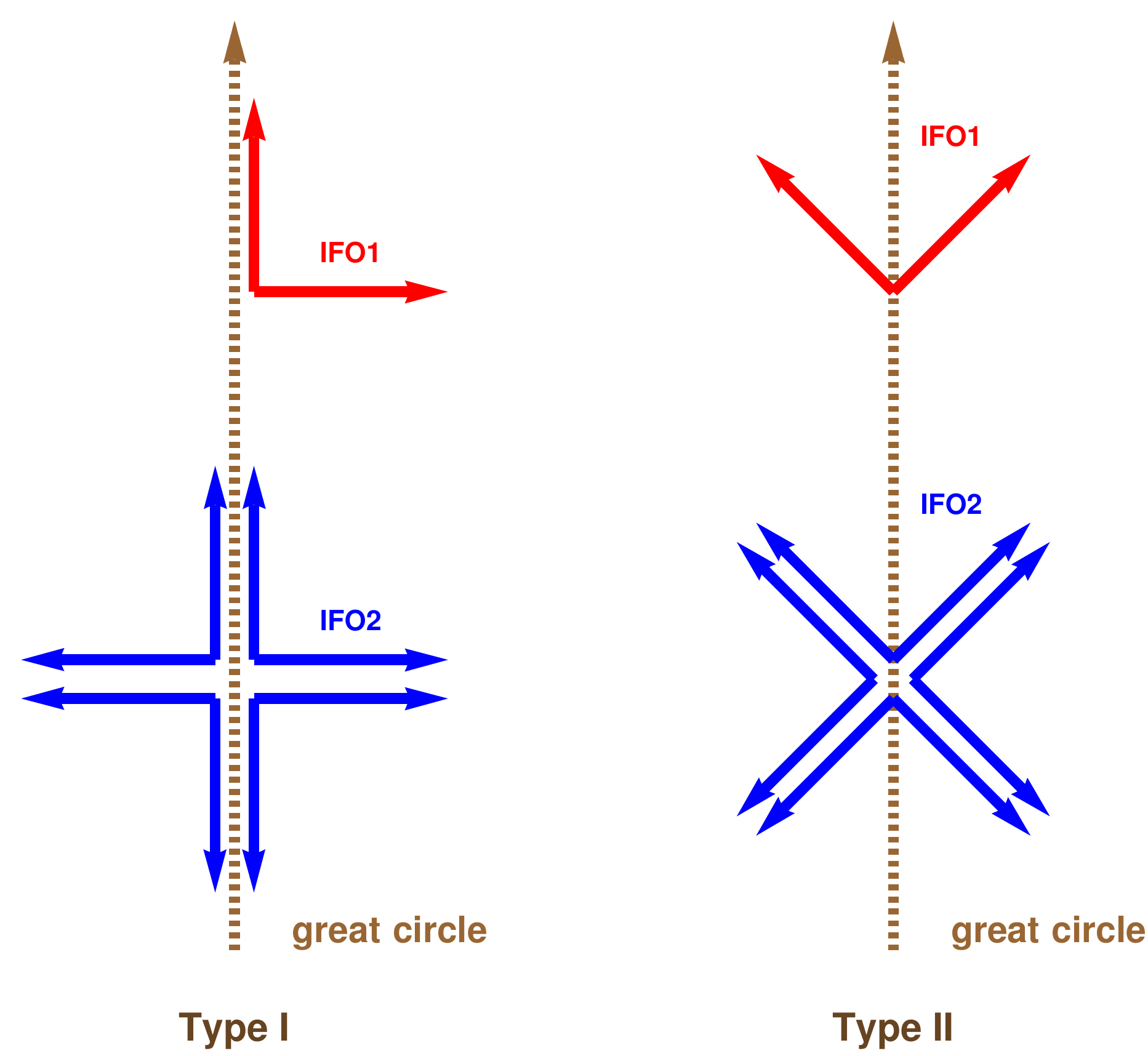}
\caption{Possible optimal configurations for a detector pair in the presence of only one between tensor, vector and scalar polarization modes \cite{nishizawa2009probing}. For a fixed orientation of one interferometer (IFO1) there are four possible orientations for the second one (IFO2).}
\label{fig:optimal}
\end{figure}
\begin{figure*}
\includegraphics[width=0.45\textwidth]{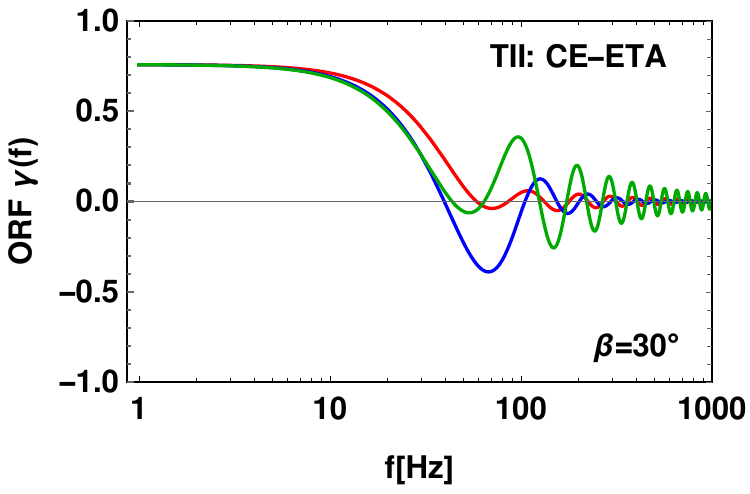}
\hspace{0.3cm}
\includegraphics[width=0.45\textwidth]{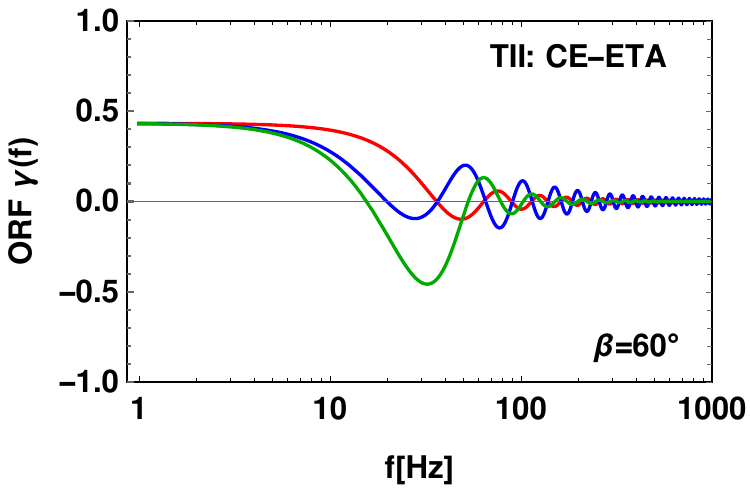}\\
\vspace{0.3cm}
\includegraphics[width=0.45\textwidth]{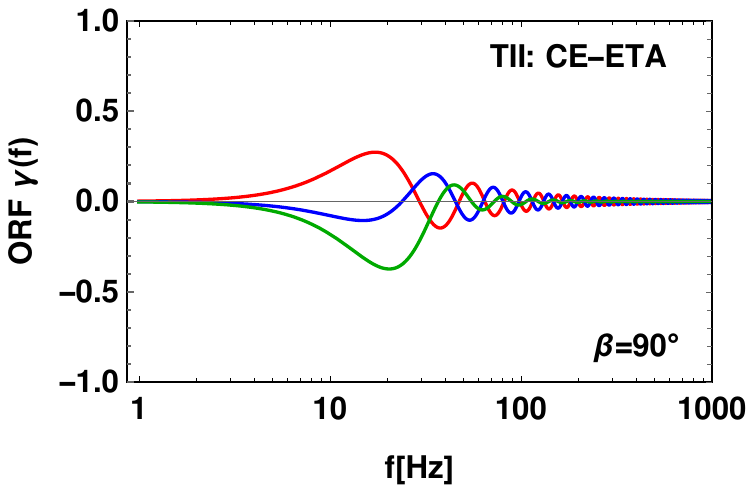}
\hspace{0.3cm}
\includegraphics[width=0.45\textwidth]{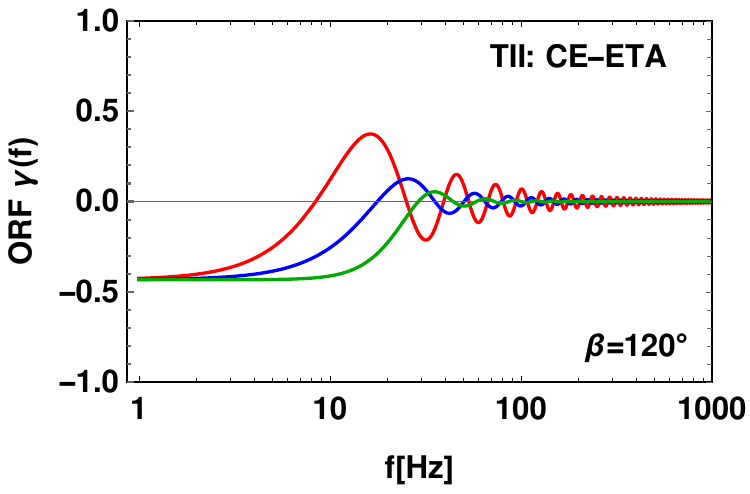}\\
\vspace{0.3cm}
\includegraphics[width=0.45\textwidth]{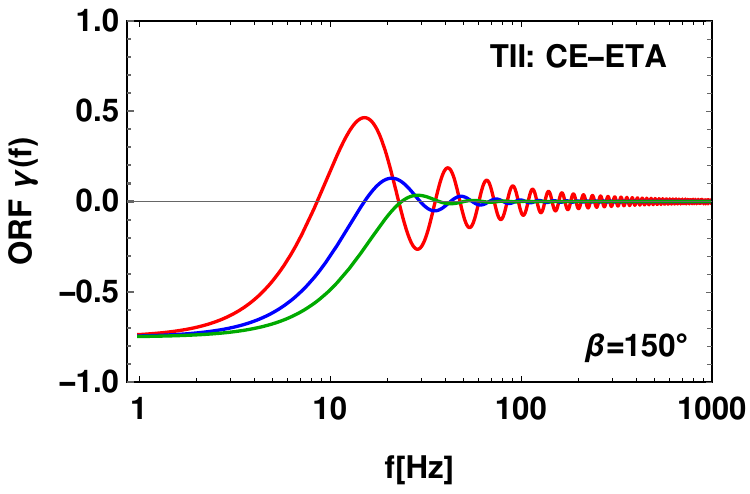}
\hspace{0.3cm}
\includegraphics[width=0.45\textwidth]{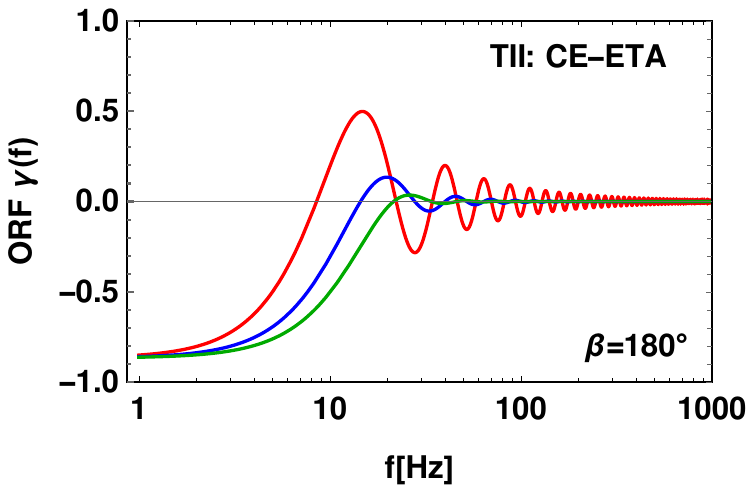}\\
\caption{Overlap reduction functions (ORFs) relative to the Type II optimal configuration for different angular separations and for tensor (red), vector (blue) and scalar (green) polarization modes. The detector pair is given by CE and one among three ET interferometers (i.e. ETA)}
\label{fig:ORFTYPEII}
\end{figure*}

\subsection{SGWB made of tensor modes}
\begin{figure*}
\includegraphics[width=0.27\textwidth,valign=m]{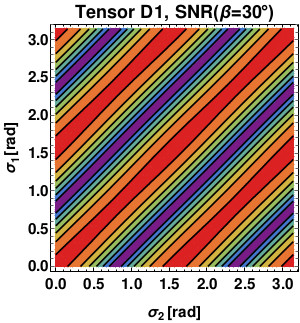}
\includegraphics[width=0.4cm,height=3.85cm,valign=m]{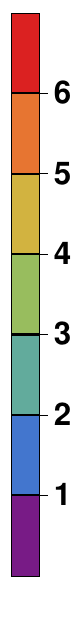}
\hspace{0.2cm}
\includegraphics[width=0.27\textwidth,valign=m]{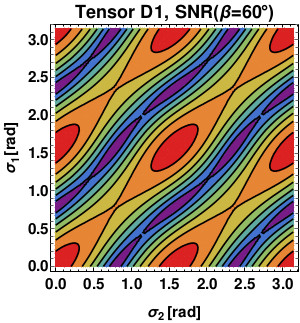}
\includegraphics[width=0.5cm,height=3.85cm,valign=m]{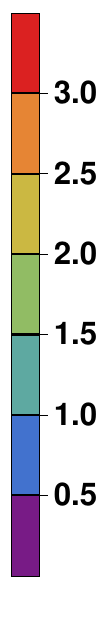}
\hspace{0.19cm}
\includegraphics[width=0.27\textwidth,valign=m]{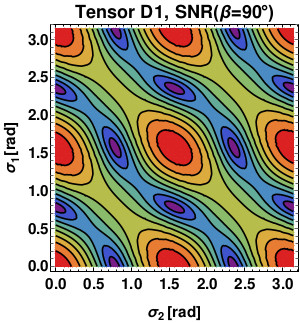}
\includegraphics[width=0.6cm,height=3.85cm,valign=m]{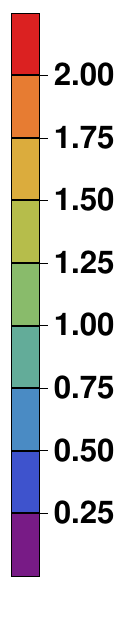}\\
\vspace{0.3cm}
\includegraphics[width=0.27\textwidth,valign=m]{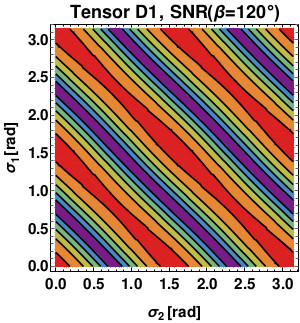}
\includegraphics[width=0.5cm,height=3.85cm,valign=m]{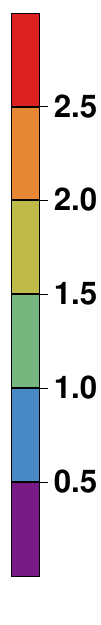}
\hspace{0.19cm}
\includegraphics[width=0.27\textwidth,valign=m]{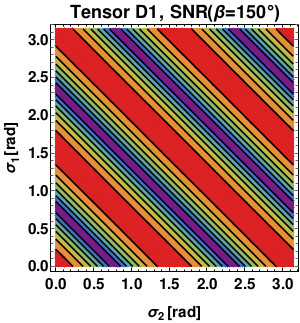}
\includegraphics[width=0.5cm,height=3.85cm,valign=m]{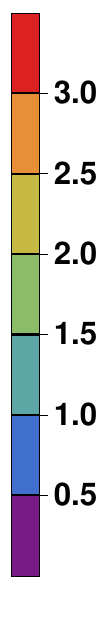}
\hspace{0.19cm}
\includegraphics[width=0.27\textwidth,valign=m]{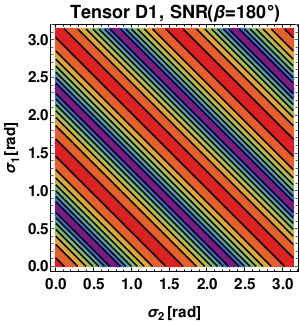}
\includegraphics[width=0.5cm,height=3.85cm,valign=m]{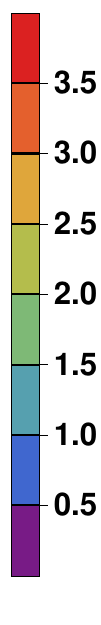}\\
\caption{Expected SNR ($\Omega^T_{ref}$ = $10^{-12}$ and $T=5$ yrs) for a SGWB made of tensor polarization modes as a function of both ET ($\sigma_1$) and CE ($\sigma_2$) orientations for fixed angular separations ($\beta$) between the pair. The configuration D1 refers to two Einstein Telescope interferometers (e.g. ETA and ETB) in the xylophone layout and the Stage 1 CE.}
\label{fig:1TD12}
\end{figure*}

\begin{figure*}
\includegraphics[width=0.45\textwidth]{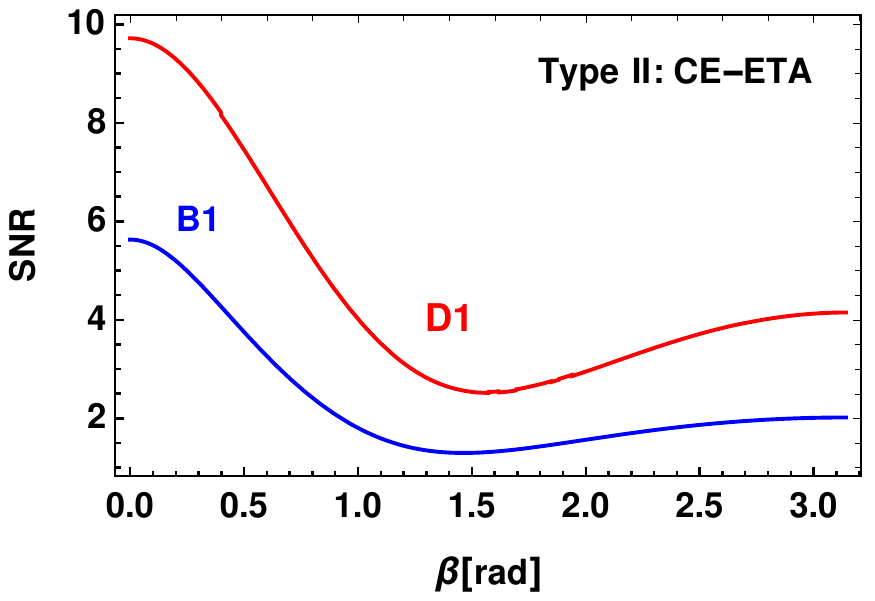}
\hspace{0.3cm}
\includegraphics[width=0.45\textwidth]{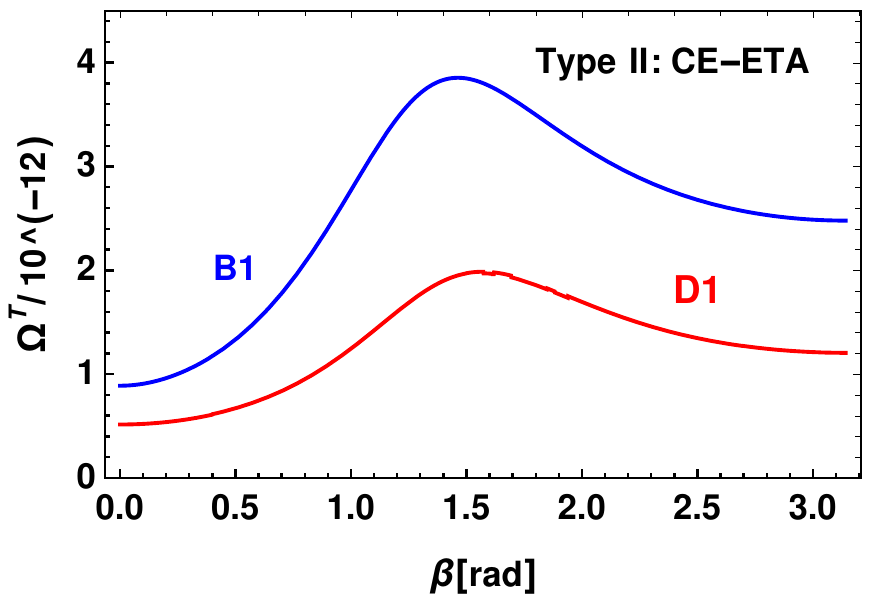}\\
\caption{(Left) Expected SNR ($\Omega^T_{ref}=10^{-12}$ and $T=5$ yrs) for the ETA-CE detector pair in Type II optimal configuration and assuming a SGWB made of tensor modes and (right) relative detectable energy density (SNR$=5$ and $T=5$ yrs) for tensor modes normalized by the reference value $\Omega^T_{ref}=10^{-12}$. Both normalized SNR and detectable energy density are shown for configurations B1 (blue) and D1 (red).}
\label{fig:NORMT}
\end{figure*}

The cross-correlation analysis for a detector pair in the presence of only tensor modes has already been well developed in the literature for second-generation ground-based interferometers (i.e. LIGO observatories, Virgo and KAGRA) \cite{allen1999detecting, nishizawa2009probing}. Here, after recalling some general and useful results, our aim is to carry out a cross-correlation analysis focusing on the third-generation of ground-based interferometers (i.e. ET and CE) instead.
If only tensor modes are present, the appropriate SNR expression (see e.g. \cite{romano2017detection, nishizawa2009probing}) is derived in Appendix A and we have
\begin{eqnarray}
\text{SNR$_T$} =\frac{3H_0^2}{10\pi^2}\sqrt{T}\biggl[2\int_{0}^{+\infty}df\frac{(\gamma^T(f)\Omega^T_{GW}(f))^2}{f^3P_{I}(f)P_{J}(f)}\biggr]^{1/2}.\nonumber\\
\label{eq:11.4}
\end{eqnarray}
More generally, if we replace $\gamma^T\Omega^T_{GW}$ with the term $\gamma^M\Omega^M_{GW}$, $M$ = $T$, $V$ and $S$, Eq.\eqref{eq:11.4} provides the SNR formula for a SGWB made of only one polarization class. Then, from Eqs.\eqref{eq:ORFT}, \eqref{eq:ORFV} and \eqref{eq:ORFS}, it is possible to extract an ideal layout for the detector pair, in terms of their orientations, which maximizes the SNR in Eq.\eqref{eq:11.4}. Indeed, depending on the sign of functions $\Theta_M^+$ and $\Theta_M^-$, with $M=T,V$ and $S$, there exists only one among two optimal configurations classified as \cite{allen1999detecting, nishizawa2009probing}
\begin{eqnarray}
&&\text{Type I:}\hspace{0.43cm} \cos2(\sigma_1+\sigma_2)=-\cos2(\sigma_1-\sigma_2)=\pm 1, \nonumber\\
&&\text{Type II:}\hspace{0.3cm} \cos2(\sigma_1+\sigma_2)=\cos2(\sigma_1-\sigma_2)= \pm 1, \nonumber\\
\label{eq:type}
\end{eqnarray}
and shown in Fig.\ref{fig:optimal} for a general detector pair. In particular, we further consider one ET interferometer (e.g. ETA) and CE in Type II optimal configurations: the corresponding ORFs for different values of $\beta$ are shown in Fig.\ref{fig:ORFTYPEII} for tensor, vector and scalar polarization modes. Moreover, note how ORFs relative to the ETA-CE pair mediate to zero way before reaching characteristic values $f_*$ of both ET and CE: this means we are safe to work in the low-frequency limit. In the rest of this section, we apply all previous tools to investigate a SGWB made of tensor modes using third-generation ground-based detectors in order to list our results. Since we aim to estimate the SGWB detectable energy, we assume a frequency-independent energy density spectrum for tensor modes. Then, the expected SNR can be found using Eq.\eqref{eq:11.4} once we fix the observation time to $5$ yrs and we set a reference value $\Omega^T_{ref}=10^{-12}$ for the detectable energy density. Let us consider again the ETA-CE pair: in Fig.\ref{fig:1TD12} we show the expected SNR using the D1 configuration for several fixed values of $\beta$ while leaving the orientation of both detectors ($\sigma_1$ and $\sigma_2$ respectively) free to vary. 
All these cases present the common Type II optimal configuration, although the more the separation angle approaches $\pi$, the more we get some narrow bands of ideal orientations which maximize the SNR. This can be understood considering the limit $\Theta_T^-(\beta \rightarrow \pi) \rightarrow 0$, thus the ORF in Eq.\eqref{eq:ORFT} is only affected by the sum of the two detector orientations (and not anymore by their difference) and both Type I and Type II configurations are optimal, along with all configurations lying on the straight lines. We should mention that analogous results can be found for B1. Since the optimal Type II configuration is shared among all possible distances between the two detectors, in the left panel of Fig.\ref{fig:NORMT} we show the expected SNR as a function of $\beta$ using B1 and D1 configurations for a direct comparison. 
If only tensor modes are present, we find that D1 is the best configuration for every angular separation considered between the pair, with the latter approximately doubling its sensitivity to the SGWB with respect to B1. Note how in both scenarios the SNR rapidly decreases for relatively small distances between the two detectors until it reaches a minimum value, then slowly starts to increase again for higher relative distances between members of the pair. This behavior is analogous to the one discussed in \cite{nishizawa2009probing} for the second-generation of ground-based interferometers. It is also possible to understand how the detectable SGWB energy density varies as a function of the spatial distance between the two interferometers, as shown in the right panel of Fig.\ref{fig:NORMT}. Here we assumed again a frequency independent energy density spectrum for tensor modes in Eq.\eqref{eq:11.4}, though this time we fix the value of the SNR$=5$ (which we use to claim detection) and set $T=5$ yrs. Both expected SNR and detectable energy density values are listed in Tab.\ref{tab:TENSORONLY} for both B1 and D1 configurations using different fixed angular separations: in terms of detectable energy density, we generally find that the ETA-CE pair sensitivity to a SGWB made of tensor modes only is approximately improved by a factor $10^3$ with respect to current limits (see recent limits provided by \cite{abbott2021upper}).\\
\begin{table}
\begin{ruledtabular}
\begin{tabular}{ccccc}
 &\multicolumn{2}{c}{SNR$_T$}&\multicolumn{2}{c}{$h_0^2\Omega^T_{GW}$}\\
$\beta$ & B1 & D1 & B1 & D1\\ \hline
$30^{\circ}$ & $3.14$ & $6.30$  & $1.59\times 10^{-12}$ & $7.93\times 10^{-13}$  \\
$60^{\circ}$ & $1.46$ & $3.26$ & $3.40\times 10^{-12}$ & $1.53\times 10^{-12}$ \\

$90^{\circ}$ & $1.14$ & $2.18$ & $4.39\times 10^{-12}$ & $2.29\times 10^{-12}$ \\

$120^{\circ}$ & $1.41$ & $2.68$ & $3.54\times 10^{-12}$ & $1.86\times 10^{-12}$ \\

$150^{\circ}$ & $1.66$ & $3.34$ & $3.01\times 10^{-12}$ & $1.50\times 10^{-12}$ \\

$180^{\circ}$& $1.75$ & $3.59$  & $2.86\times 10^{-12}$ & $1.39\times 10^{-12}$ \\
\end{tabular}
\end{ruledtabular}
\caption{Expected SNR ($\Omega^T_{ref}=10^{-12}$ and $T=5$ yrs) and detectable energy density (SNR$=5$ and $T=5$ yrs) for a SGWB made of tensor modes with one ET interferometer (e.g. ETA) and CE in Type II optimal configuration for a fixed angular separation ($\beta$). Numerical results are shown using B1 and D1 configurations for the detector pair.}
\label{tab:TENSORONLY}
\end{table}
\begin{figure}
\includegraphics[width=0.45\textwidth]{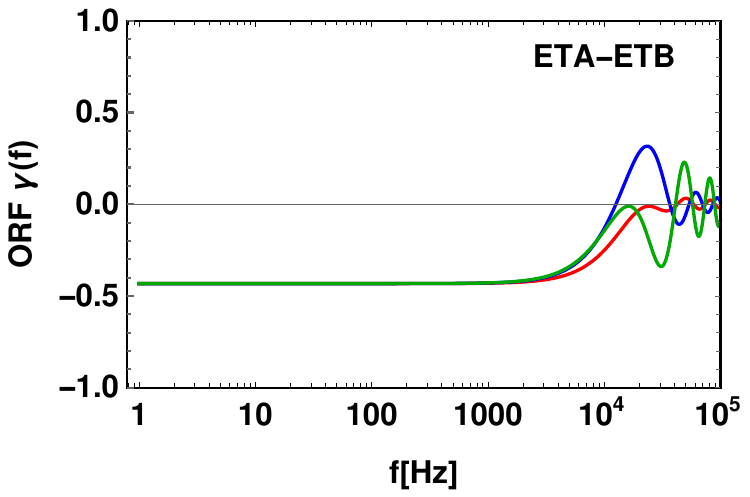}
\caption{ORFs relative to two ET interferometers (e.g. ETA-ETB) for tensor (red), vector (blue) and scalar (green) polarization modes. Due to the proximity between the detector pair, note how ORFs stay approximately constant up to $f \approx 5000$ Hz.}
\label{fig:ETETORF}
\end{figure}
We now focus on two ET detectors (e.g. ETA and ETB) at our disposal and we consider the corresponding ORF shown in Fig.\ref{fig:ETETORF}. Note how due to the two detectors relative small distance ($|\Delta\overline{\mathbf{x}}|=L=10$ km), the ORF starts to oscillate at higher GW frequencies. Moreover, given the frequency range to which ET is expected to be sensitive ($1$-$10000$ Hz), we can assume $\gamma_{IJ}^T$ = $\gamma_{IJ}^V$ = $\gamma_{IJ}^S \approx -3/8$ with $I$ and $J$ being one among ETA, ETB and ETC. In real situations, whenever $f \approx f_*$ Hz we no longer are in the low frequency limit and we should consider how ORFs are affected by detector transfer functions: however, as a first approximation, neglecting the latter does not have a significant impact on the final results. This time we have two possible configurations: either both interferometer sensitivity curves are represented by ET-B, or both interferometers are in the xylophone configuration with sensitivity curves given by ET-D. We refer to the first ad second configuration as ``BB'' and ``DD'' respectively.  Similarly to what we previously did, we assume a frequency-independent energy density spectrum for the tensor modes and we show the results for both expected SNR ($\Omega^T_{ref}=10^{-12}$ and $T=5$ yrs) and detectable energy density (SNR$=5$ and $T=5$ yrs) in Tab.\ref{tab:ETETdata}.
\begin{table}
\begin{ruledtabular}
\begin{tabular}{ccc}
 & SNR$_T$ & $h_0^2\Omega^T_{GW}$ \\ \hline
BB & $1.14$  &$4.39 \times 10^{-12}$\\

DD & $6.29$  & $7.94 \times 10^{-13}$\\

\end{tabular}
\end{ruledtabular}
\caption{Expected SNR ($\Omega^T_{ref}=10^{-12}$ and $T=5$ yrs) and detectable energy density (SNR$=5$ and $T=5$ yrs) for a SGWB made of tensor modes using two ET interferometers (e.g. ETA and ETB). Numerical results are shown using BB and DD configurations for the detector pair.}
\label{tab:ETETdata}
\end{table}
If both interferometers are in a xylophone configuration, we find that the pair sensitivity to the SGWB is approximately doubled with respect to the BB one and results for the detectable energy density can be further compared to the ones obtained for the ETA-CE pair. Moreover, considering Tab.\ref{tab:TENSORONLY} and Tab.\ref{tab:ETETdata}, configurations DD and D1 with $\beta=\pi/6$ provide very similar results, while taking higher angular separations between ET and CE, the DD layout is always more sensitive to the SGWB with respect to D1. Therefore, if only tensor modes are present, in first approximation we find that an ET pair is generally slightly more sensitive to the SGWB with respect to a ET-CE pair \footnote{For a more detailed treatment one needs to consider some ET technical specifications in order to account for the presence of transfer functions \cite{schilling1997angular}.}.

\section{SGWB made of Tensor and Vector or Scalar polarization modes \label{2GWB}} 
In this section we want to extend the previous results obtained for a SGWB involving only tensor modes to a SGWB made of both tensor and one other class of extra polarization modes $X$, with $X$ = $V$ or $S$. Once again, our results refer to the third-generation of ground-based interferometers. In realistic situations, we need to consider the possibility where we dispose of two ET interferometers along with CE: we mentioned that ET comes with three detectors, though we shall see in this section a first example showing how only two of them can be taken to work independently for our considerations (see \cite{philippoz2018gravitational} for a detailed discussion on this matter). Moreover, after we understand how to remove tensor modes contributions to the SNR for the SGWB \cite{omiya2020searching}, we will separately focus on a SGWB made of tensor along with vector or scalar polarization modes to show some differences and similarities between the two cases. Considering ET and CE, we finally present our forecasts for vector and scalar polarization modes detectable energy density contributions to the SGWB. 
\subsection{Tensor modes deletion}

Let us consider a SGWB made of both tensor and extra $X$-polarization modes (i.e. tensor and vector or scalar polarization modes): drawing from \cite{omiya2020searching} and considering a generic triad of detectors $1$, $2$ and $3$, the SNR expression for the extra polarization modes is derived in Appendix A and it is given by
\begin{widetext}
\begin{eqnarray}
\text{SNR$_X$}&&=\frac{3 H_0^2}{10 \pi^2}\sqrt{T}\left\{2\int_{0}^{+\infty}df \frac{\bigl[\bigl(\gamma^T_{12}(f)\gamma^X_{13}(f)-\gamma^T_{13}(f)\gamma^X_{12}(f)\bigr)\Omega^X_{GW}(f)\bigr]^2}{f^6\bigl[(\gamma^T_{12}(f))^2P_1(f)P_3(f)+(\gamma^T_{13}(f))^2P_1(f)P_2(f) \bigr]} \right\}^{1/2}\, . 
\label{eq:SNREXTRA}
\end{eqnarray}
\end{widetext}
As a consistency check, note that if only tensor modes exist, then $X=T$ and the integrand numerator collapses to zero, making the SNR null for every GW-frequency range considered. Indeed this cancellation technique allow us to reduce to zero the tensor modes contribution to the SNR defined for the SGWB, thus singling out the additional $X$-polarization modes. In order to study the latter, we also require the integrand numerator to be different from zero, meaning that the condition
\[\gamma^T_{12}(f)\gamma^X_{13}(f)-\gamma^T_{13}(f)\gamma^X_{12}(f) \neq 0,\]
needs to be satisfied. While working with ET and CE, we shall see that this happens only for finite frequency ranges whose values are much smaller than both the ET and CE characteristic frequency $f_*$, meaning we can safely work in the low-frequency limit. Note how the interferometer $1$ plays the role of a ``dominant'' detector, since it affects all ORFs in Eq.\eqref{eq:SNREXTRA}. We will show how its choice (whether it is identified with one ET interferometer or CE) finally influences the optimal configurations and SNR formula. Additionally, once we specify the nature of the ``dominant'' detector, we find that Eq.\eqref{eq:SNREXTRA} can be written in a more compact way, therefore we now show our results for different possible scenarios while having at our disposal three ET interferometers and CE.
\subsubsection{``Dominant'' detector represented by the Cosmic Explorer}
We first consider the case where the ``dominant'' detector is given by CE, while detectors $2$ and $3$ are ET interferometers (e.g. ETA and ETB). The SNR for X-extra polarization modes can be written as 
\begin{eqnarray}
\text{SNR$_X$}=\frac{3 H_0^2}{10 \pi^2}\sqrt{T}\left[2\int_{0}^{+\infty}df \frac{(\Gamma^X(f)\Omega^X_{GW}(f))^2}{f^6P_{CE}(f)P_{ET}(f)} \right]^{1/2} \,,\nonumber\\
\label{eq:16.0}
\end{eqnarray}
where we have introduced the effective overlap reduction function (EORF) for X-polarization modes
\begin{eqnarray}
\Gamma^X(f) = \frac{\gamma^T_{12}(f)\gamma^X_{13}(f) - \gamma^T_{13}(f)\gamma^X_{12}(f)}{\sqrt{\bigl(\gamma^T_{12}(f)\bigr)^2+\bigl(\gamma^T_{13}(f)\bigr)^2}}.
\label{eq:EORFX}
\end{eqnarray}
When $\alpha \equiv f/f_c \ll 1$ EORFs are null, reflecting the fact that $\gamma^T$ and $\gamma^X$ share approximately the same value in this limit \cite{allen1999detecting}. Recalling Eqs.\eqref{eq:ORFT}, \eqref{eq:ORFV}, \eqref{eq:ORFS}, let $\sigma_2$ be the orientation of the CE bisector and $\sigma_1$ the orientation of the ETB one: due to ET topology the orientation of the ETA interferometer bisector will always be $\sigma_1-2\pi/3$. Finally, let $\beta$ be the angular separation between ET and CE. Under these assumptions, for general X-polarization modes we get
\[\Gamma^X(f) \propto \cos^4\left( \frac{\beta}{2} \right)\sin(4\sigma_2)j_2(f)j_4(f).\]
Clearly, the CE orientation filters the sensitivity to GWs: indeed whenever $\sin(4\sigma_2)=0$, meaning $\sigma_2=\pi/4 \text{ mod }\pi/4$, the signal is null for every frequency range. If $\beta=0$, then $j_2=j_4=0$ and it is straightforward to show that the EORF numerator is zero, while if $\beta=\pi$, then $\cos(\pi/2)=0$ and the EORF is zero again: these two angular separation values correspond to a null signal, meaning that in order to detect X-polarization modes the detector plane normal vector cannot be the same or the opposite for both ET and CE.

\subsubsection{``Dominant'' detector represented by the Einstein Telescope}
Let us now consider the case where ETB is the ``dominant'' detector, with the second and third interferometers being ETA and CE respectively, therefore we further assume $\gamma^T_{12}=\gamma^X_{12} \approx -3/8$ (as we shall see, we are safe to work in the low-frequency limit). The SNR  is given by
\begin{eqnarray}
\text{SNR$_X$}=\frac{3 H_0^2}{10 \pi^2}\sqrt{T}\left[2\int_{0}^{+\infty}df \frac{(\tilde{\Gamma}^X(f)\Omega^X_{GW}(f))^2}{f^6P_{CE}(f)P_{ET}(f)} \right]^{1/2},\nonumber\\
\end{eqnarray}
where we have introduced the noise-affected EORF for X-polarization modes
\begin{eqnarray}
\tilde{\Gamma}^X(f) = \frac{\gamma^T_{13}(f) - \gamma^X_{13}(f)}{\sqrt{1+\tilde{g}_{13}(f)}},
\label{eq:ETEORFX}
\end{eqnarray}
with $\tilde{g}_{IJ}$ a frequency-dependent function defined as
\[\tilde{g}_{IJ}(f) \equiv \left(\frac{8\gamma^T_{IJ}(f)}{3}\right)^2\frac{P_{ET}(f)}{P_{CE}(f)}.\]
Noise-affected EORFs tell us we only need to worry about ORFs relative to the single ETB-CE detector pair, therefore we shall see how the SNR is then maximized by one among Type I or Type II layouts introduced in Eq.\eqref{eq:type}, depending on the angular separation between ET and CE.

\subsubsection{Three Einstein Telescope interferometers}
Let us now consider the case where we have three ET detectors at our disposal. Due to ET triangular topology, once we select tensor, vector or scalar polarization modes, the respective ORFs (shown in Fig.\ref{fig:ETETORF}) for each possible pair involving detectors ETA, ETB and ETC are the same. More explicitly, this means that $\gamma^M_{AB}$ = $\gamma^M_{BC}$ = $\gamma^M_{AC}$ for $M$ = $T$, $V$ and $S$, thus the SNR numerator in Eq.\eqref{eq:SNREXTRA} is always equal to zero. Unfortunately, this means that only two ET interferometers may be taken to work independently while working with a network of detectors \cite{philippoz2018gravitational}, thus it is mandatory to consider CE (or another ground-based detector) in order to distinguish between different polarization modes.

\subsection{SGWB made of Tensor and X-polarization modes}

\begin{figure*}
\includegraphics[width=0.27\textwidth,valign=m]{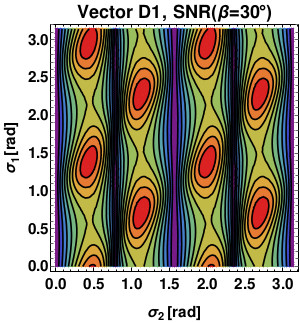}
\includegraphics[width=0.6cm,height=3.85cm,valign=m]{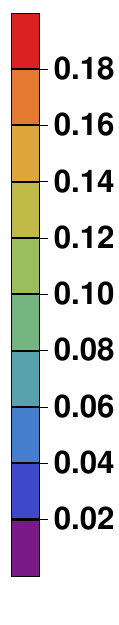}
\hspace{0.18cm}
\includegraphics[width=0.27\textwidth,valign=m]{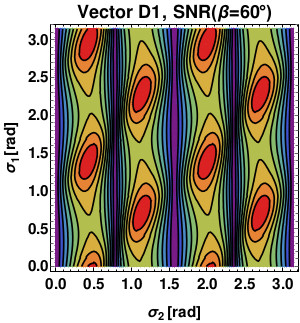}
\includegraphics[width=0.5cm,height=3.85cm,valign=m]{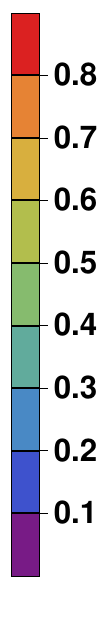}
\hspace{0.19cm}
\includegraphics[width=0.27\textwidth,valign=m]{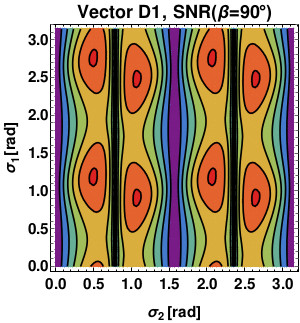}
\includegraphics[width=0.5cm,height=3.85cm,valign=m]{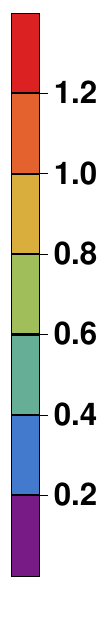}\\
\vspace{0.3cm}
\includegraphics[width=0.27\textwidth,valign=m]{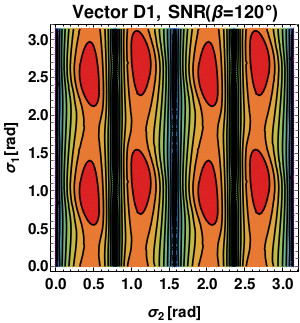}
\includegraphics[width=0.6cm,height=3.85cm,valign=m]{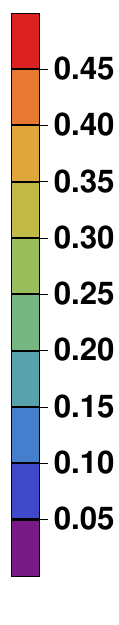}
\hspace{0.18cm}
\includegraphics[width=0.27\textwidth,valign=m]{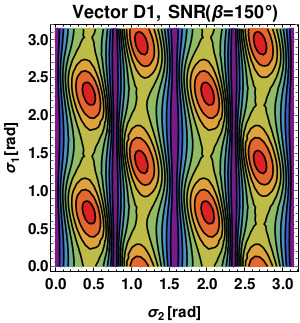}
\includegraphics[width=0.6cm,height=3.85cm,valign=m]{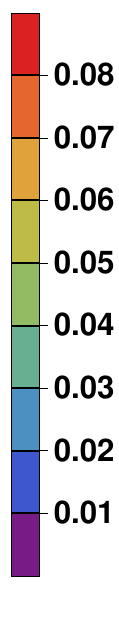}\\
\caption{Expected SNR$_V$ ($\Omega^V_{ref}$ = $10^{-12}$ and $T=5$ yrs) for a SGWB made of tensor and vector polarization modes as a function of both ET ($\sigma_1$) and CE ($\sigma_2$) orientations for fixed angular separations ($\beta$) between the pair with the ``dominant'' detector represented by CE. The configuration D1 refers to two Einstein Telescope interferometers (e.g. ETA and ETB) in the xylophone layout and the Stage 1 CE.}
\label{fig:2TVD12}
\end{figure*}

\begin{figure*}
\includegraphics[width=0.27\textwidth,valign=m]{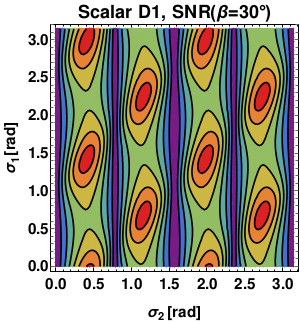}
\includegraphics[width=0.6cm,height=3.85cm,valign=m]{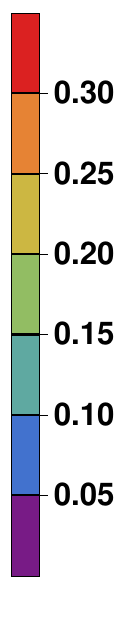}
\hspace{0.18cm}
\includegraphics[width=0.27\textwidth,valign=m]{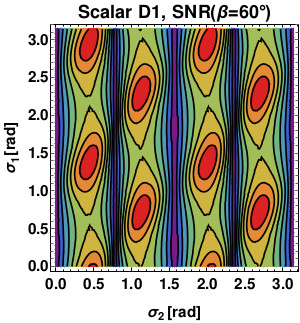}
\includegraphics[width=0.5cm,height=3.85cm,valign=m]{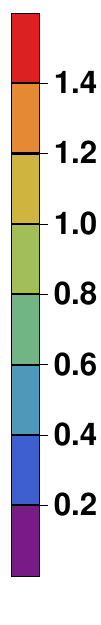}
\hspace{0.19cm}
\includegraphics[width=0.27\textwidth,valign=m]{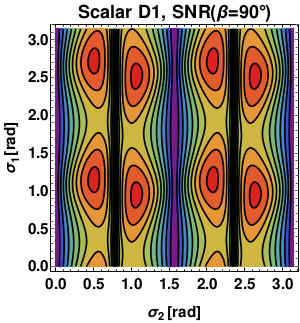}
\includegraphics[width=0.5cm,height=3.85cm,valign=m]{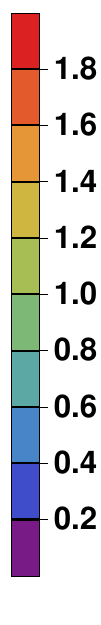}\\
\vspace{0.3cm}
\includegraphics[width=0.27\textwidth,valign=m]{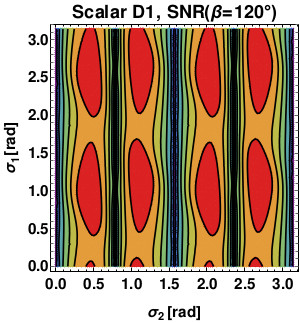}
\includegraphics[width=0.5cm,height=3.85cm,valign=m]{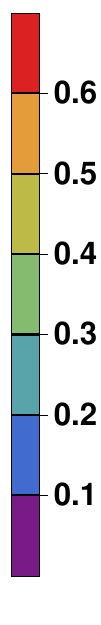}
\hspace{0.19cm}
\includegraphics[width=0.27\textwidth,valign=m]{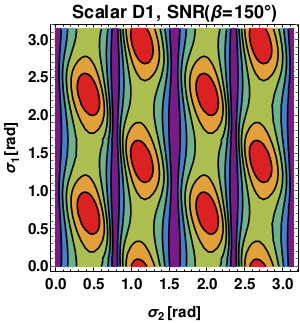}
\includegraphics[width=0.6cm,height=3.85cm,valign=m]{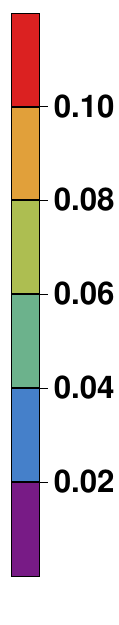}\\
\caption{Expected SNR$_S$ ($\Omega^S_{ref}$ = $10^{-12}$ and $T=5$ yrs) for a SGWB made of tensor and scalar polarization modes as a function of both ET ($\sigma_1$) and CE ($\sigma_2$) orientations for fixed angular separations ($\beta$) between the pair with the ``dominant'' detector represented by CE. The configuration D1 refers to two Einstein Telescope interferometers (e.g. ETA and ETB) in the xylophone layout and the Stage 1 CE.}
\label{fig:2TSD12}
\end{figure*}

\begin{figure*}
\includegraphics[width=0.45\textwidth]{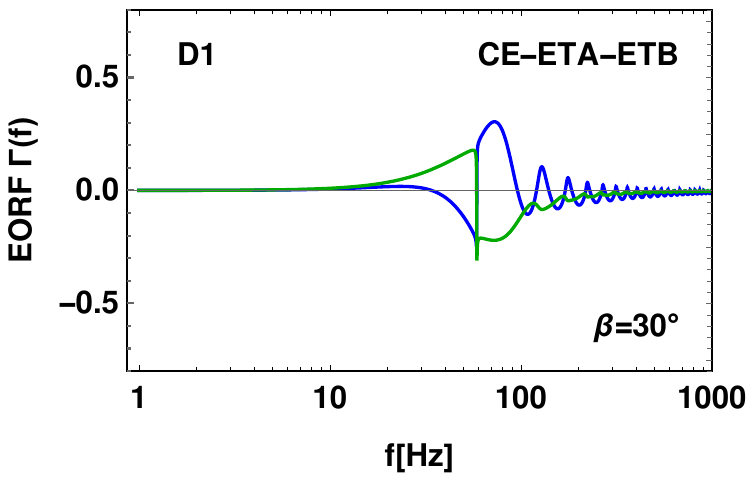}
\hspace{1cm}
\includegraphics[width=0.45\textwidth]{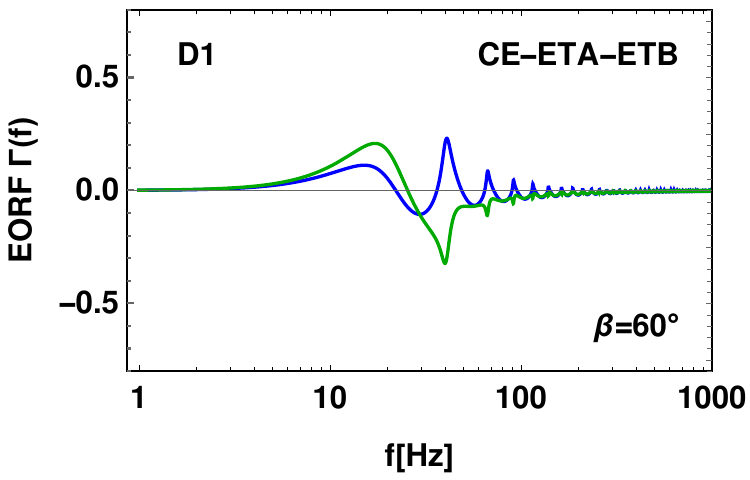}\\
\vspace{0.3cm}
\includegraphics[width=0.45\textwidth]{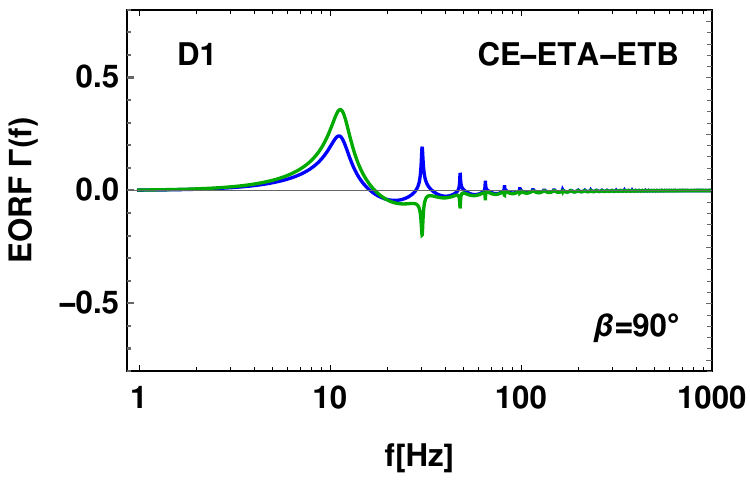}
\hspace{1cm}
\includegraphics[width=0.45\textwidth]{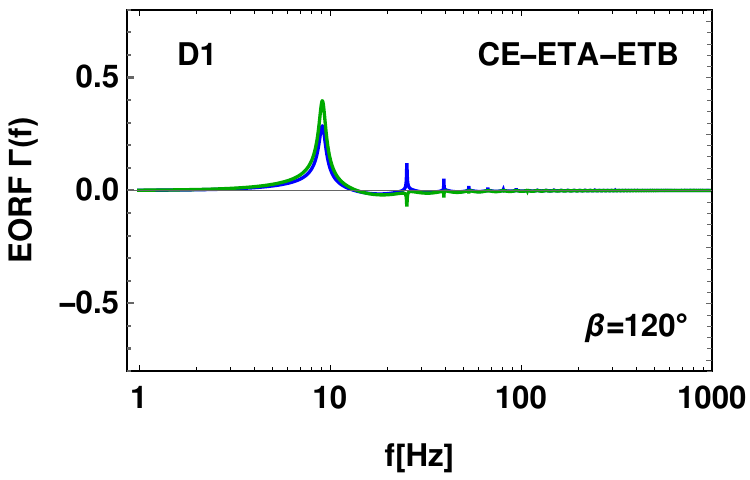}\\
\vspace{0.3cm}
\includegraphics[width=0.45\textwidth]{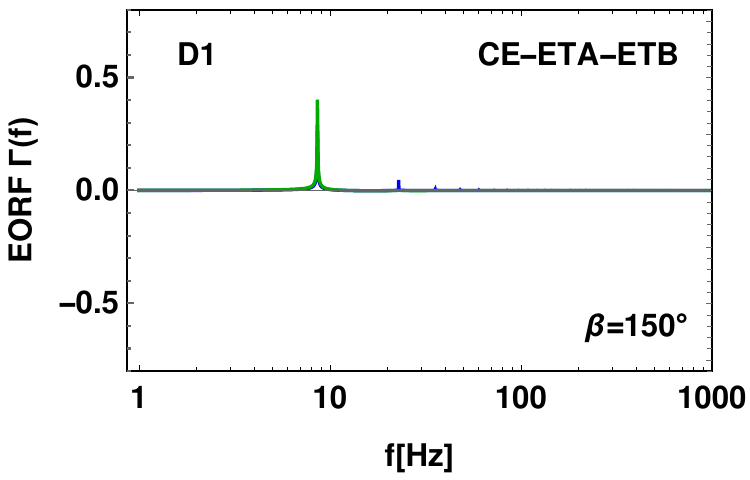}
\caption{Vector (blue) and scalar (green) EORFs (see Eq.\eqref{eq:EORFX}) corresponding to optimal configurations listed in Figs.\ref{fig:2TVD12} and \ref{fig:2TSD12} respectively. The ``dominant'' detector is represented by CE and the configuration used is D1, which refers to the ETA and ETB interferometers in the xylophone configuration along with CE in Stage 1.}
\label{fig:EORFVS}
\end{figure*}

We now consider a SGWB made of tensor and X-polarization modes, with $X$=$V$ or $S$ (see e.g. ~\cite{Jimenez:2008sq,Kimura:2016rzw,Heisenberg:2017hwb} and ~\cite{ fujii_maeda_2003,Alonso:2016suf,Heisenberg:2018mxx} for vector-tensor and scalar-tensor theories of gravity respectively).  Given two possible scenarios, we begin by considering CE as the ``dominant'' detector and we assume a frequency-independent energy density spectrum for both tensor and vector polarization modes. The expected value for the SNR can be found using Eq.\eqref{eq:16.0} once we set both the observation time $T=5$ yrs and a reference value $\Omega^X_{ref}$ = $10^{-12}$ for the detectable energy density. Let now $\sigma_2$ be the orientation of the CE bisector and $\sigma_1$ the orientation of the ETB one: similarly to what we did for a SGWB made of tensor modes, in Figs.\ref{fig:2TVD12} (vector modes) and \ref{fig:2TSD12} (scalar modes) we show the results for the expected SNR using the configuration D1 for several fixed values of $\beta$ while leaving the orientation of both detectors free to vary. 
As previously anticipated while discussing EORFs, there are four ``bands'' separated by vertical straight lines of SNR null values corresponding to the case where $\sin (4\sigma_2) = 0$. We find that the CE optimal orientation is approximately the same for every value of $\beta$ for both vector and scalar polarization modes, though this is no longer true for ET and we do not have same optimal orientations shared among all angular separations between ET and CE. Considering Figs.\ref{fig:2TVD12} and \ref{fig:2TSD12}, a set (given ORFs and EORFs periodicity) of ideal values for $\sigma_1$ and $\sigma_2$ which maximize the SNR for each value of $\beta$ can be extrapolated: we further show the corresponding optimal EORFs for vector and scalar polarization modes in Fig.\ref{fig:EORFVS}. Note how these functions filter the frequency range where the isolation of extra polarization modes becomes possible. The peaks appearing come from the almost simultaneous zeros of the ORFs in the denominator of $\Gamma$. Finally, in Tabs.\ref{tab:VECTORONLY} and \ref{tab:SCALARONLY} we show results for both expected SNR ($\Omega^X_{ref}=10^{-12}$ and $T$ = $5$ yrs) and detectable energy density (SNR $= 5$ and $T=5$ yrs) using configurations B1 and D1 with optimal orientations for ET and CE. We find that the D1 configuration is generally the best one, which in particular for $\beta$ = $\pi/3$ and $\beta$ = $\pi/2$ approximately doubles and triples its sensitivity to vector and scalar polarization modes in terms of detectable energy density with respect to the B1 configuration, while for larger angles, the process reverses, leading to completely worse expected forecasts (i.e. for $\beta = 5\pi/6$). An exception is made by small spatial distances between the pair ($\beta=\pi/6$) where B1 results the best layout (note this will not be the case if final locations for ET and CE are Europe and North America respectively).
\begin{table}
\begin{ruledtabular}
\begin{tabular}{ccccc}
 &\multicolumn{2}{c}{SNR$_V$}&\multicolumn{2}{c}{$h_0^2\Omega^V_{GW}$}\\
$\beta$ & B1 & D1 & B1 & D1\\ \hline
$30^{\circ}$ & $0.28$ &  $0.20$  & $1.81\times 10^{-11}$ & $2.54\times 10^{-11}$ \\

$60^{\circ}$ & $0.51$ & $0.88$ & $9.82\times 10^{-12}$ & $5.65\times 10^{-12}$ \\

$90^{\circ}$ & $0.41$ & $1.22$ & $1.21\times 10^{-11}$ & $4.10\times 10^{-12}$ \\

$120^{\circ}$ & $0.12$ & $0.49$ & $4.04\times 10^{-11}$ & $1.02\times 10^{-11}$  \\

$150^{\circ}$ & $0.02$  & $0.08$ & $3.41\times 10^{-10}$ & $5.96\times 10^{-11}$  \\
\end{tabular}
\end{ruledtabular}
\caption{Expected SNR$_V$ ($\Omega^V_{ref}=10^{-12}$ and $T$ = $5$ yrs) and detectable energy density (SNR$=5$ and $T=5$ yrs) for a SGWB made of tensor and vector modes with ETB  and CE (``dominant'' detector) with optimal orientations for D1 shown in Fig.\ref{fig:2TVD12}) and for a fixed angular separation ($\beta$). Numerical results are shown using the B1 and D1 configurations for the detector pair.}
\label{tab:VECTORONLY}
\end{table}
\begin{table}
\begin{ruledtabular}
\begin{tabular}{ccccc}
 &\multicolumn{2}{c}{SNR$_S$}&\multicolumn{2}{c}{$\xi h_0^2\Omega^S_{GW}$}\\
$\beta$ & B1 & D1 & B1 & D1 \\ \hline
$30^{\circ}$ & $0.40$ &  $0.33$ & $1.24\times 10^{-11}$ & $1.54\times 10^{-11}$ \\

$60^{\circ}$ & $0.92$ & $1.56$ & $5.46\times 10^{-12}$ & $3.21\times 10^{-12}$ \\

$90^{\circ}$ & $0.63$ & $1.86$ & $7.96\times 10^{-12}$ & $2.68\times 10^{-12}$ \\

$120^{\circ}$ & $0.16$ & $0.69$  & $3.06\times 10^{-11}$ & $7.21\times 10^{-12}$ \\

$150^{\circ}$ & $0.02$ & $0.11$ & $2.51\times 10^{-10}$ & $4.33\times 10^{-11}$ \\
\end{tabular}
\end{ruledtabular}
\caption{Expected SNR$_S$ ($\Omega^S_{ref}=10^{-12}$ and $T$ = $5$ yrs) and detectable energy density (SNR$=5$ and $T=5$ yrs) for a SGWB made of tensor and scalar modes with ETB  and CE (``dominant'' detector) with optimal orientations for D1 shown in Fig.\ref{fig:2TSD12}) and for a fixed angular separation ($\beta$). Numerical results are shown using the B1 and D1 configurations for the detector pair.}
\label{tab:SCALARONLY}
\end{table}

\begin{figure*}
\includegraphics[width=0.27\textwidth,valign=m]{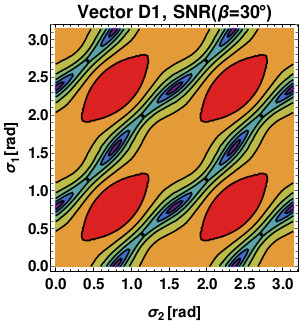}
\includegraphics[width=0.6cm,height=3.85cm,valign=m]{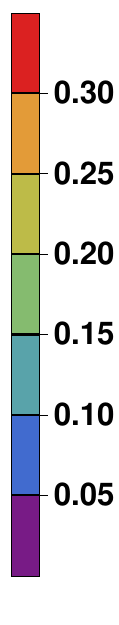}
\hspace{0.18cm}
\includegraphics[width=0.27\textwidth,valign=m]{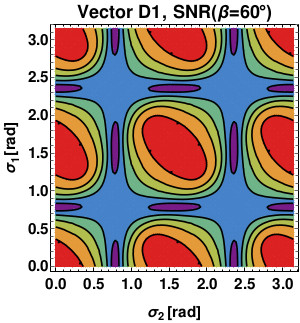}
\includegraphics[width=0.5cm,height=3.85cm,valign=m]{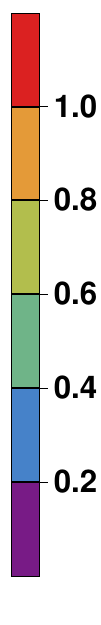}
\hspace{0.19cm}
\includegraphics[width=0.27\textwidth,valign=m]{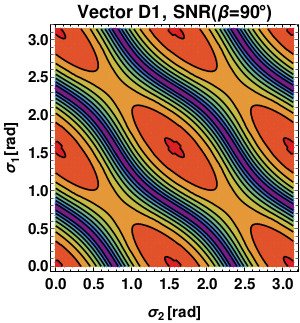}
\includegraphics[width=0.6cm,height=3.85cm,valign=m]{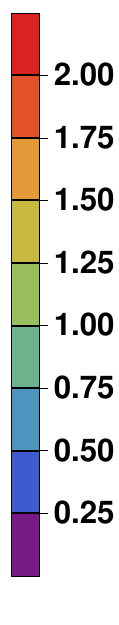}\\
\vspace{0.3cm}
\includegraphics[width=0.27\textwidth,valign=m]{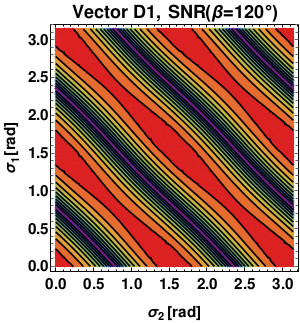}
\includegraphics[width=0.6cm,height=3.85cm,valign=m]{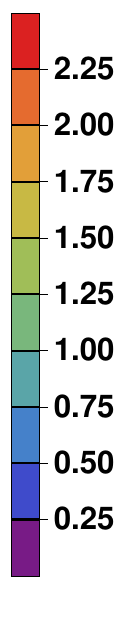}
\hspace{0.18cm}
\includegraphics[width=0.27\textwidth,valign=m]{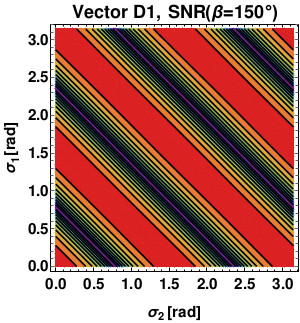}
\includegraphics[width=0.6cm,height=3.85cm,valign=m]{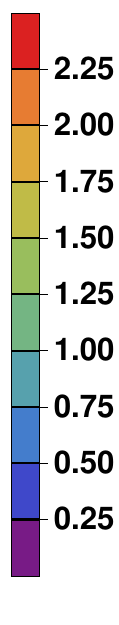}
\hspace{0.18cm}
\includegraphics[width=0.27\textwidth,valign=m]{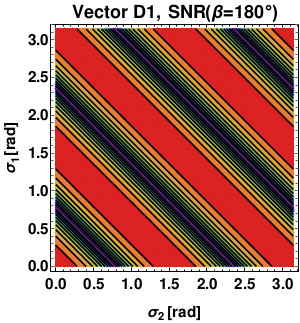}
\includegraphics[width=0.6cm,height=3.85cm,valign=m]{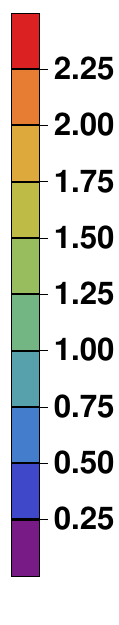}\\
\caption{Expected SNR$_V$ ($\Omega^V_{ref}$ = $10^{-12}$ and $T=5$ yrs) for a SGWB made of tensor and vector polarization modes as a function of both ET ($\sigma_1$) and CE ($\sigma_2$) orientations for fixed angular separations ($\beta$) between the pair with the ``dominant'' detector represented by ET. The configuration D1 refers to two Einstein Telescope interferometers (e.g. ETA and ETB) in the xylophone layout and the Stage 1 CE.}
\label{fig:2TVD12ETET}
\end{figure*}

\begin{figure*}
\includegraphics[width=0.27\textwidth,valign=m]{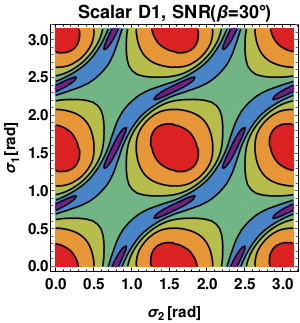}
\includegraphics[width=0.5cm,height=3.85cm,valign=m]{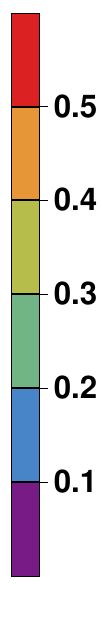}
\hspace{0.19cm}
\includegraphics[width=0.27\textwidth,valign=m]{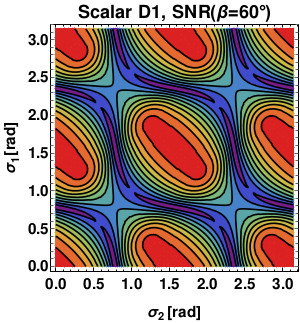}
\includegraphics[width=0.5cm,height=3.85cm,valign=m]{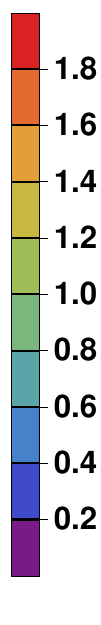}
\hspace{0.19cm}
\includegraphics[width=0.27\textwidth,valign=m]{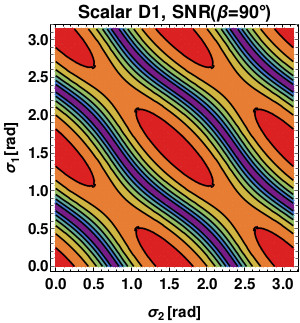}
\includegraphics[width=0.5cm,height=3.85cm,valign=m]{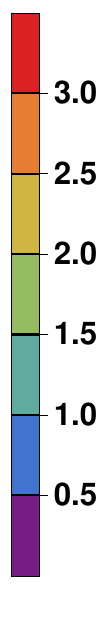}\\
\vspace{0.3cm}
\includegraphics[width=0.27\textwidth,valign=m]{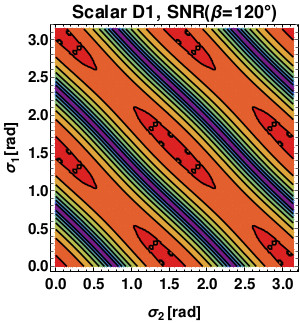}
\includegraphics[width=0.5cm,height=3.85cm,valign=m]{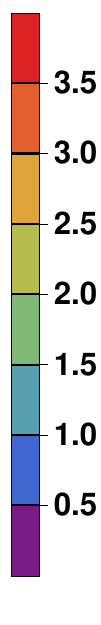}
\hspace{0.19cm}
\includegraphics[width=0.27\textwidth,valign=m]{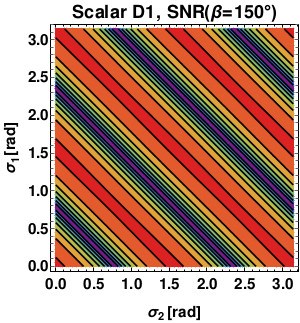}
\includegraphics[width=0.5cm,height=3.85cm,valign=m]{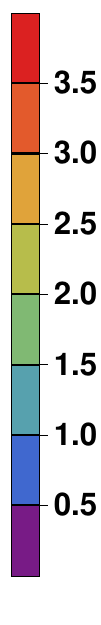}
\hspace{0.19cm}
\includegraphics[width=0.27\textwidth,valign=m]{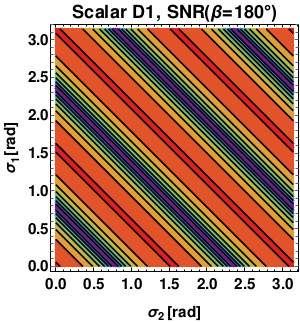}
\includegraphics[width=0.5cm,height=3.85cm,valign=m]{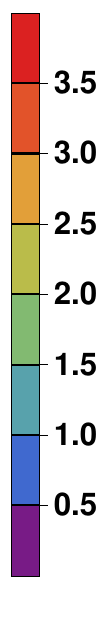}\\
\caption{Expected SNR$_S$ ($\Omega^S_{ref}$ = $10^{-12}$ and $T=5$ yrs) for a SGWB made of tensor and scalar polarization modes as a function of both ET ($\sigma_1$) and CE ($\sigma_2$) orientations for fixed angular separations ($\beta$) between the pair with the ``dominant'' detector represented by ET. The configuration D1 refers to two Einstein Telescope interferometers (e.g. ETA and ETB) in the xylophone layout and the Stage 1 CE.}
\label{fig:2TSD12ETET}
\end{figure*}

\begin{figure*}
\includegraphics[width=0.45\textwidth]{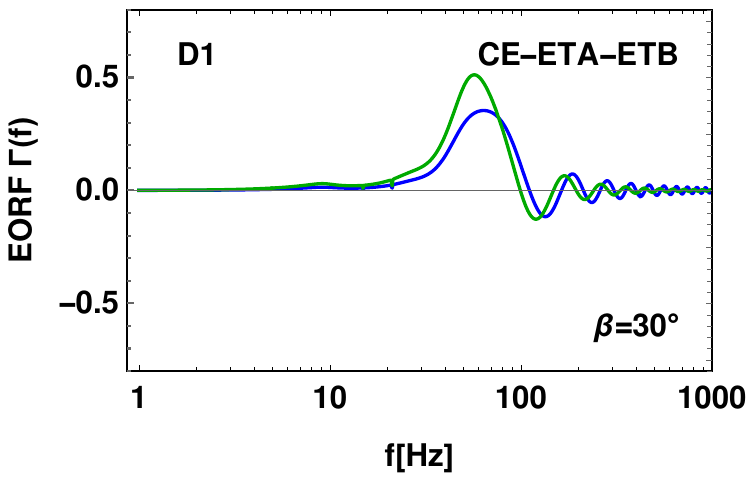}
\hspace{1cm}
\includegraphics[width=0.45\textwidth]{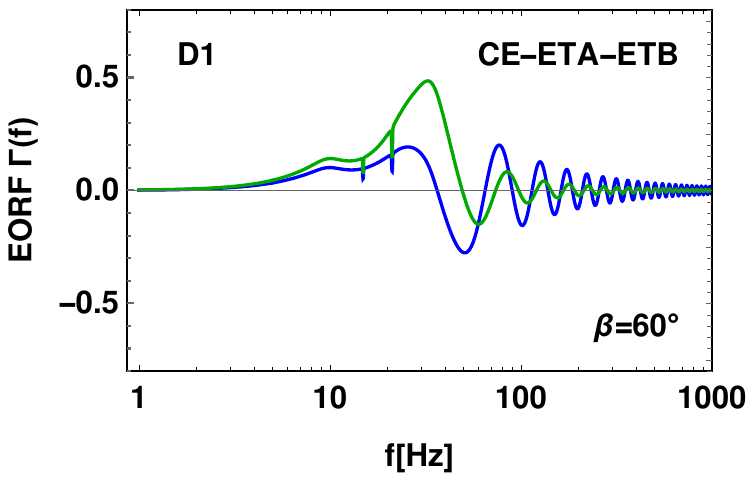}\\
\vspace{0.3cm}
\includegraphics[width=0.45\textwidth]{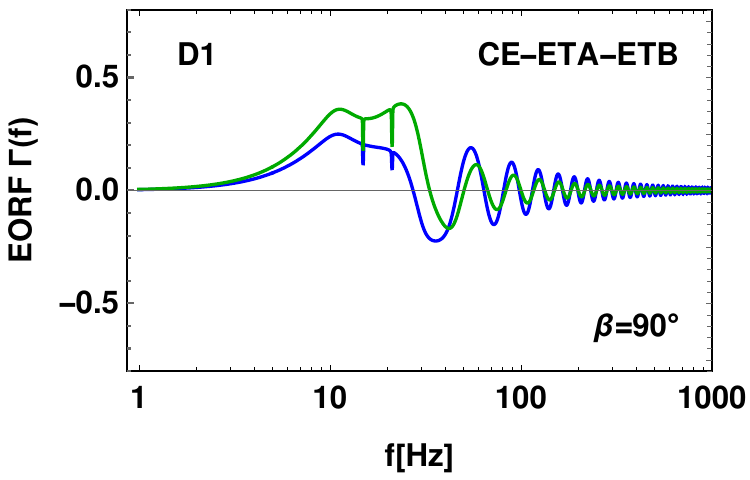}
\hspace{1cm}
\includegraphics[width=0.45\textwidth]{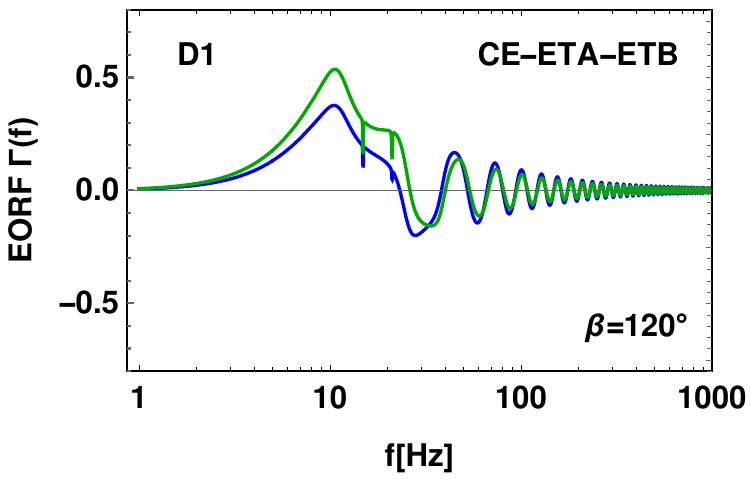}\\
\vspace{0.3cm}
\includegraphics[width=0.45\textwidth]{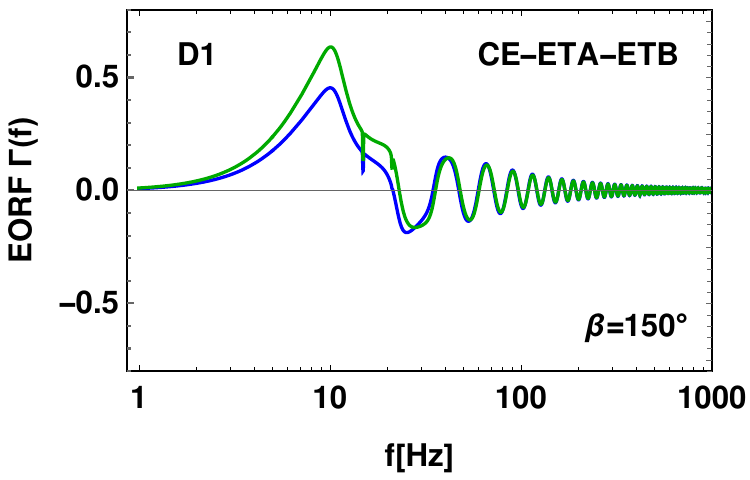}
\hspace{1cm}
\includegraphics[width=0.45\textwidth]{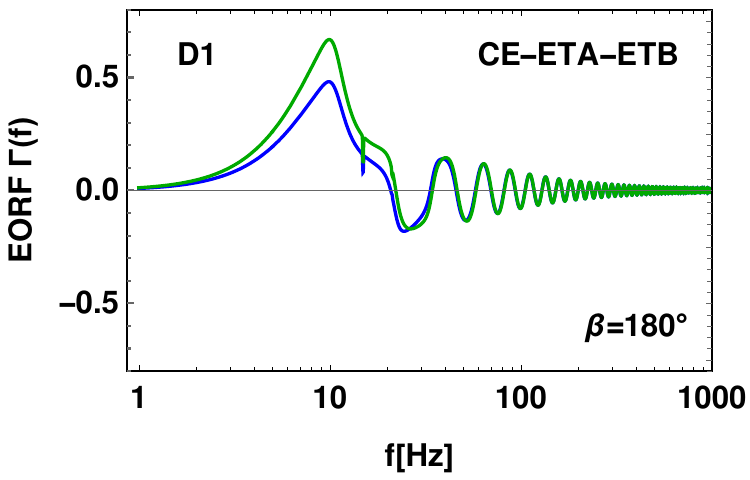}\\
\caption{Vector (blue) and scalar (green) noise-affected EORFs (see Eq.\eqref{eq:ETEORFX}) corresponding to optimal configurations listed in Tabs.\ref{tab:TVorientationsETET} and \ref{tab:TSorientationsETET} respectively. The ``dominant'' detector is represented by ETB and the configuration used is D1, which refers to the ETA and ETB interferometers in the xylophone configuration along with CE in Stage 1.}
\label{fig:ETETEORFVS}
\end{figure*}

We now retrace these steps considering the interferometer ETB as the ``dominant'' detector: once again, we begin with the expected SNR ($\Omega^X_{ref}=10^{-12}$ and $T$ = $5$ yrs). Keeping $\sigma_1$ and $\sigma_2$ as the ETB and CE bisector orientations respectively, our results are shown in Figs.\ref{fig:2TVD12ETET} (vector modes) and \ref{fig:2TSD12ETET} (scalar modes).
Interestingly, recalling we only need to focus on ORFs related to the ETB-CE detector pair (see Eq.\eqref{eq:ETEORFX}), we find we can describe all possible scenarios in terms of optimal Type I and Type II layouts introduced in Eq.\eqref{eq:type}: results for different values of the angular separation between ET and CE are shown in Tabs.\ref{tab:TVorientationsETET} (vector modes) and \ref{tab:TSorientationsETET} (scalar modes) using the D1 configuration.
\begin{table}
\begin{ruledtabular}
\begin{tabular}{cc}
Angular separation&\multicolumn{1}{c}{Optimal config.}\\
\hline
$\beta$ & Vector D1 \\ \hline
$30^{\circ}$ & Type I \\

$60^{\circ}$ & Type II \\

$90^{\circ}$ & Type II \\

$120^{\circ}$ & Type II \\

$150^{\circ}$ & $\approx\sigma_1=-\sigma_2$ mod$\pi/2$ \\

$180^{\circ}$ & $\sigma_1=-\sigma_2$ mod$\pi/2$ \\
\end{tabular}
\end{ruledtabular}
\caption{Possible optimal orientations for ETB (``dominant'' detector) and CE which maximize SNR$_V$ for a SGWB made of tensor and vector modes. The layout used is D1, while the Type I and II configurations are described in Eq.\eqref{eq:type}.}
\label{tab:TVorientationsETET}
\end{table}
\begin{table}
\begin{ruledtabular}
\begin{tabular}{cc}
Angular separation&\multicolumn{1}{c}{Optimal configuration}\\
\hline
$\beta$ & Scalar D1  \\ \hline
$30^{\circ}$ & Type II \\

$60^{\circ}$ & Type II \\

$90^{\circ}$ & Type II \\

$120^{\circ}$ & Type II \\

$150^{\circ}$ & $\approx \sigma_1=-\sigma_2$ mod $\pi/2$ \\

$180^{\circ}$ & $\sigma_1=-\sigma_2$ mod $\pi/2$ \\
\end{tabular}
\end{ruledtabular}
\caption{Possible optimal orientations for ETB (``dominant'' detector) and CE which maximize SNR$_S$ for a SGWB made of tensor and scalar modes. The layout used is D1, while the Type I and II configurations are described in Eq.\eqref{eq:type}.}
\label{tab:TSorientationsETET}
\end{table}
Additionally, the corresponding optimal noise-affected EORFs are shown in Fig.\ref{fig:ETETEORFVS}, with the peaks appearing shortly after $10$ Hz coming from the spikes in the ET-D sensitivity curve, as shown in the left panel of Fig.\ref{fig:NOISE}. In Tabs.\ref{tab:VECTORONLYETET} and \ref{tab:SCALARONLYETET} we list the results for both expected SNR ($\Omega^X_{ref}=10^{-12}$ and $T=5$ yrs) and detectable energy density (SNR$=5$ and $T=5$ yrs) using configurations B1 and D1 with optimal orientations for ET and CE.  With the exception of small spatial distances between the pair ($\beta$ = $\pi/6$) where B1 gives better forecasts, once again, we find that the D1 configuration is the best one, where the network greatly improves its sensitivity to both vector and scalar polarization modes in the SGWB. Moreover, better forecasts for the expected SNR and detectable energy density are related to larger angular separations between the two detectors.
\begin{table}
\begin{ruledtabular}
\begin{tabular}{ccccccccc}
 &\multicolumn{2}{c}{SNR$_V$}&\multicolumn{2}{c}{$h_0^2\Omega^V_{GW}$}\\
$\beta$ & B1 & D1 & B1 & D1 \\ \hline
$30^{\circ}$ & $0.58$ & $0.35$ & $8.58\times 10^{-12}$ & $1.44\times 10^{-11}$ \\

$60^{\circ}$ & $0.71$ & $1.13$ & $6.96\times 10^{-12}$ & $4.43\times 10^{-12}$ \\

$90^{\circ}$ & $0.77$ & $2.02$ & $6.43\times 10^{-12}$ & $2.47\times 10^{-12}$ \\

$120^{\circ}$ & $0.73$ & $2.39$ & $6.82\times 10^{-12}$ & $2.09\times 10^{-12}$ \\

$150^{\circ}$ & $0.67$ & $2.47$ & $7.45\times 10^{-12}$ & $2.02\times 10^{-12}$ \\

$180^{\circ}$ & $0.65$ & $2.47$ & $7.70\times 10^{-12}$ & $2.02\times 10^{-12}$ \\
\end{tabular}
\end{ruledtabular}
\caption{Expected SNR$_V$ ($\Omega^V_{ref}=10^{-12}$ and $T$ = $5$ yrs) and detectable energy density (SNR$=5$ and $T=5$ yrs) for a SGWB made of tensor and vector modes with ETB (``dominant'' detector) and CE with optimal orientations listed in Tab.\ref{tab:TVorientationsETET} for a fixed angular separation ($\beta$). Numerical results are shown using configurations B1 and D1 for the detector pair.}
\label{tab:VECTORONLYETET}
\end{table}
\begin{table}
\begin{ruledtabular}
\begin{tabular}{ccccc}
 &\multicolumn{2}{c}{SNR$_S$}&\multicolumn{2}{c}{$\xi h_0^2\Omega^S_{GW}$}\\
$\beta$ & B1 & D1 & B1 & D1\\ \hline
$30^{\circ}$ & $0.91$ & $0.55$ & $5.46\times 10^{-12}$ & $9.00\times 10^{-12}$ \\

$60^{\circ}$ & $1.51$ & $1.91$ & $3.16\times 10^{-12}$ & $2.62\times 10^{-12}$ \\

$90^{\circ}$ & $1.28$ & $3.20$  & $3.89\times 10^{-12}$ & $1.56\times 10^{-12}$ \\

$120^{\circ}$ & $0.97$ & $3.59$ & $5.17\times 10^{-12}$ & $1.39\times 10^{-12}$ \\

$150^{\circ}$ & $0.82$ & $3.55$ & $6.07\times 10^{-12}$ & $1.41\times 10^{-12}$ \\

$180^{\circ}$ & $0.79$ & $3.51$ & $6.33\times 10^{-12}$ & $1.42\times 10^{-12}$ \\
\end{tabular}
\end{ruledtabular}
\caption{Expected SNR$_S$ ($\Omega^S_{ref}=10^{-12}$ and $T$ = $5$ yrs) and detectable energy density (SNR$=5$ and $T=5$ yrs) for a SGWB made of tensor and scalar modes with ETB (``dominant'' detector) and CE with optimal orientations listed in Tab.\ref{tab:TSorientationsETET} for a fixed angular separation ($\beta$). Numerical results are shown using configurations B1 and D1 for the detector pair.}
\label{tab:SCALARONLYETET}
\end{table}
Finally, comparing results in Tabs.\ref{tab:VECTORONLY} and \ref{tab:SCALARONLY}, and in Tabs.\ref{tab:VECTORONLYETET} and \ref{tab:SCALARONLYETET} we find how the network is slightly more sensitive to scalar modes instead of vector modes, independently from the value of the angular variable $\beta$. Despite the arbitrary choice of the ``dominant'' detector, we showed that forecasts relative to both the expected SNR and detectable energy density for a SGWB made of tensor and $X$-polarization modes (with $X=V$ or $S$) can be directly compared to results obtained for the space-based LISA-Taiji network in \cite{omiya2020searching}.

\section{SGWB made of Tensor, Vector and Scalar modes \label{3GWB}}
In the most general scenario a SGWB might be made of tensor, vector and scalar polarization modes at the same time. In literature, it is a well-known result that in order to separate these polarization modes we need to consider detector networks involving at least three interferometers \cite{nishizawa2009probing}. Then, the SNR expression for tensor, vector and scalar polarization modes separately can be derived (see Appendix A) and we have
\begin{widetext}
\begin{eqnarray}
\text{SNR$_M$}=\frac{3H_0^2}{10\pi^2}\sqrt{T}\left\{2\int_{0}^{+\infty}df\frac{(\Pi(f)\Omega_{GW}^M(f))^2}{f^6\bigl[ (\alpha^M_1(f))^2P_{1}(f)P_{2}(f)+(\alpha^M_2(f))^2P_{2}(f)P_{3}(f)+(\alpha^M_3(f))^2P_{3}(f)P_{1}(f)\bigr]}\right\}^{1/2},\nonumber\\
 \label{eq:wide}
\end{eqnarray}
\end{widetext}
where $\alpha^M_{1,2,3}(f)$, with $M$=$T$, $V$, $S$, and $\Pi(f)$ are given by Eqs.\eqref{eq:aalphaT}, \eqref{eq:aalphaV}, \eqref{eq:aalphaS} and \eqref{eq:api} respectively. Clearly, tensor, vector and scalar polarization modes are allowed to be algebraically isolated as long as the condition $\Pi(f)\neq 0$ is satisfied: we anticipate that for an ET-CE network this is true only for finite frequency ranges with relative values much smaller than both ET and CE characteristic frequency $f_*$. Therefore, we are safe to work in the low-frequency limit once again and we proceed to discuss our results for ET and CE. In particular, in order to better understand when the isolation of tensor, vector and scalar polarization modes is allowed, in the following we investigate two different cases: first, we consider the network involving CE along with two ET interferometers (e.g. ETA and ETB), then we move on to the network involving three ET interferometers (i.e. ETA, ETB and ETC). We finally present our forecasts for tensor, vector and scalar polarization modes detectable energy density contributions to the SGWB and we compare them to the ones recently provided by \cite{abbott2021upper} for second-generation ground-based interferometers. 
\subsection{ETA-ETB-CE network}
\begin{figure*}
\includegraphics[width=0.27\textwidth,valign=m]{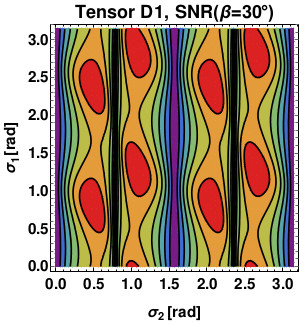}
\includegraphics[width=0.5cm,height=3.85cm,valign=m]{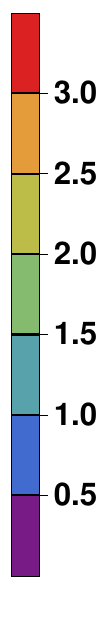}
\hspace{0.19cm}
\includegraphics[width=0.27\textwidth,valign=m]{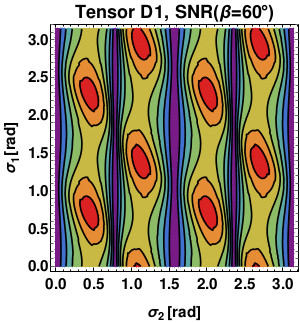}
\includegraphics[width=0.4cm,height=3.85cm,valign=m]{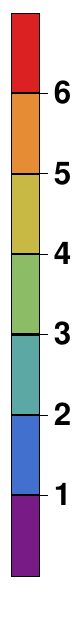}
\hspace{0.2cm}
\includegraphics[width=0.27\textwidth,valign=m]{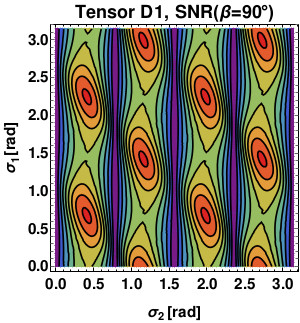}
\includegraphics[width=0.5cm,height=3.85cm,valign=m]{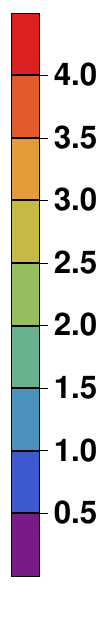}\\
\vspace{0.3cm}
\includegraphics[width=0.27\textwidth,valign=m]{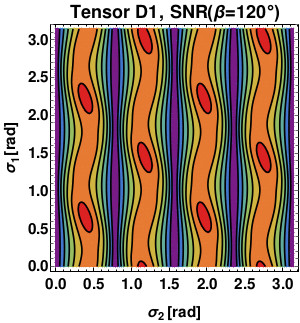}
\includegraphics[width=0.5cm,height=3.85cm,valign=m]{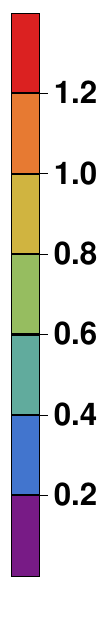}
\hspace{0.19cm}
\includegraphics[width=0.27\textwidth,valign=m]{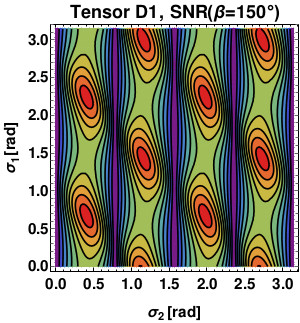}
\includegraphics[width=0.7cm,height=3.85cm,valign=m]{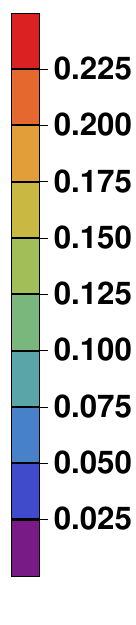}\\
\caption{Expected SNR$_T$ ($\Omega^T_{ref}$ = $10^{-11}$ and $T=5$ yrs) for a SGWB made of tensor, vector and scalar polarization modes as a function of both ETB ($\sigma_1$) and CE ($\sigma_2$) orientations for fixed angular separations ($\beta$) between the pair (ETA orientation is consequently fixed due to ET triangular topology).The configuration D1 refers to two Einstein Telescope interferometers (e.g. ETA and ETB) in the xylophone layout and the Stage 1 CE.}
\label{fig:3TD12}
\end{figure*}

\begin{figure*}
\includegraphics[width=0.27\textwidth,valign=m]{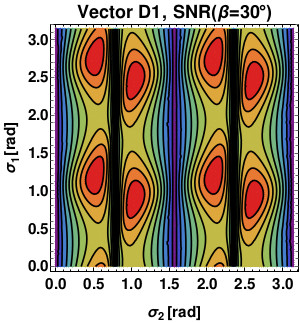}
\includegraphics[width=0.6cm,height=3.85cm,valign=m]{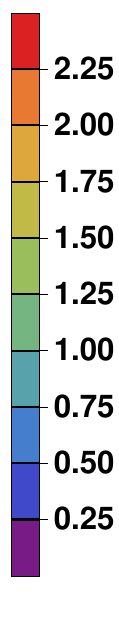}
\hspace{0.18cm}
\includegraphics[width=0.27\textwidth,valign=m]{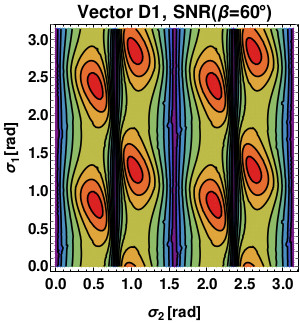}
\includegraphics[width=0.5cm,height=3.85cm,valign=m]{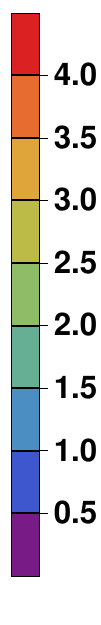}
\hspace{0.19cm}
\includegraphics[width=0.27\textwidth,valign=m]{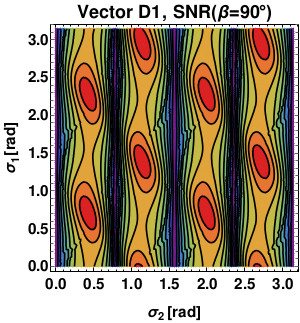}
\includegraphics[width=0.6cm,height=3.85cm,valign=m]{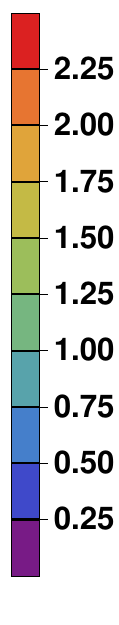}\\
\vspace{0.3cm}
\includegraphics[width=0.27\textwidth,valign=m]{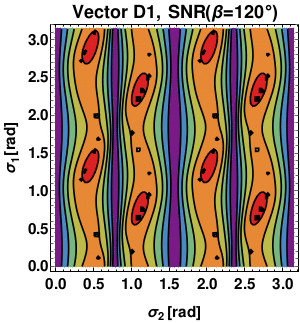}
\includegraphics[width=0.5cm,height=3.85cm,valign=m]{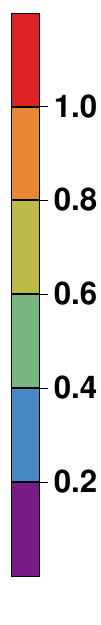}
\hspace{0.19cm}
\includegraphics[width=0.27\textwidth,valign=m]{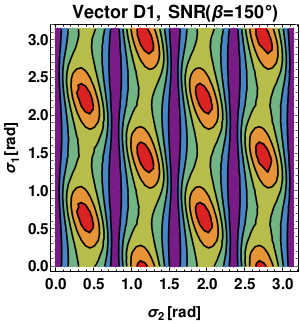}
\includegraphics[width=0.6cm,height=3.85cm,valign=m]{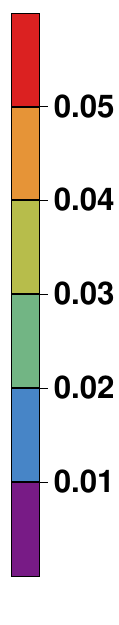}\\
\caption{Expected SNR$_V$ ($\Omega^V_{ref}$ = $10^{-11}$ and $T=5$ yrs) for a SGWB made of tensor, vector and scalar polarization modes as a function of both ETB ($\sigma_1$) and CE ($\sigma_2$) orientations for fixed angular separations ($\beta$) between the pair (ETA orientation is consequently fixed due to ET triangular topology).The configuration D1 refers to two Einstein Telescope interferometers (e.g. ETA and ETB) in the xylophone layout and the Stage 1 CE.}
\label{fig:3VD12}
\end{figure*}

\begin{figure*}
\includegraphics[width=0.27\textwidth,valign=m]{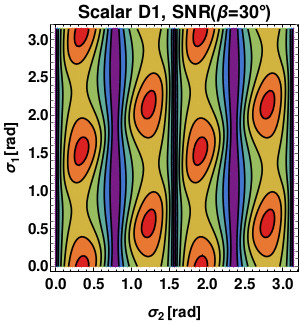}
\includegraphics[width=0.5cm,height=3.85cm,valign=m]{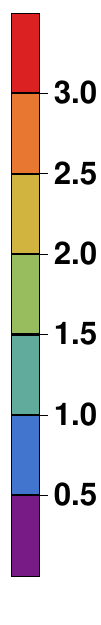}
\hspace{0.19cm}
\includegraphics[width=0.27\textwidth,valign=m]{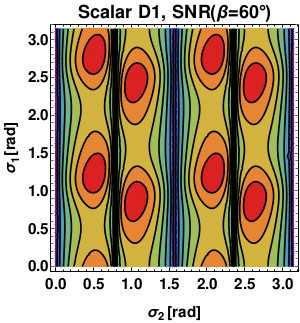}
\includegraphics[width=0.4cm,height=3.85cm,valign=m]{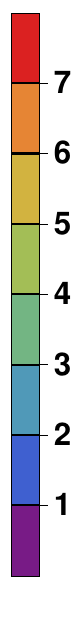}
\hspace{0.2cm}
\includegraphics[width=0.27\textwidth,valign=m]{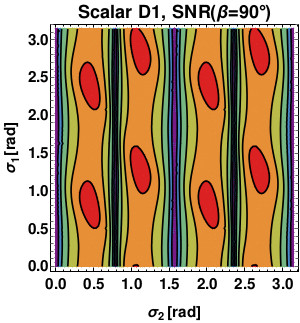}
\includegraphics[width=0.4cm,height=3.85cm,valign=m]{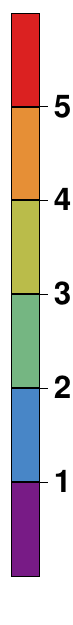}\\
\vspace{0.3cm}
\includegraphics[width=0.27\textwidth,valign=m]{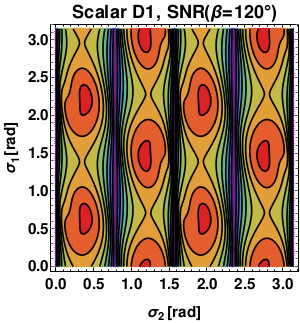}
\includegraphics[width=0.5cm,height=3.85cm,valign=m]{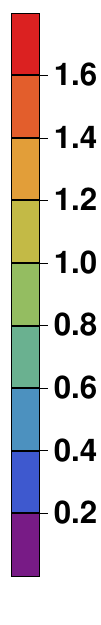}
\hspace{0.19cm}
\includegraphics[width=0.27\textwidth,valign=m]{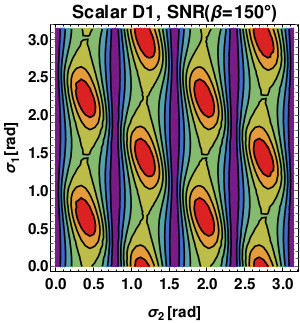}
\includegraphics[width=0.6cm,height=3.85cm,valign=m]{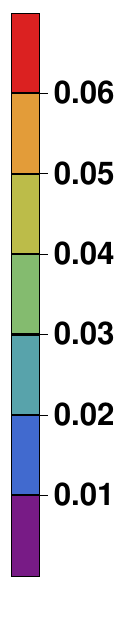}\\
\caption{Expected SNR$_S$ ($\Omega^S_{ref}$ = $10^{-11}$ and $T=5$ yrs) for a SGWB made of tensor, vector and scalar polarization modes as a function of both ETB ($\sigma_1$) and CE ($\sigma_2$) orientations for fixed angular separations ($\beta$) between the pair (ETA orientation is consequently fixed due to ET triangular topology).The configuration D1 refers to two Einstein Telescope interferometers (e.g. ETA and ETB) in the xylophone layout and the Stage 1 CE.}
\label{fig:3SD12}
\end{figure*}

\begin{figure*}
\includegraphics[width=0.45\textwidth]{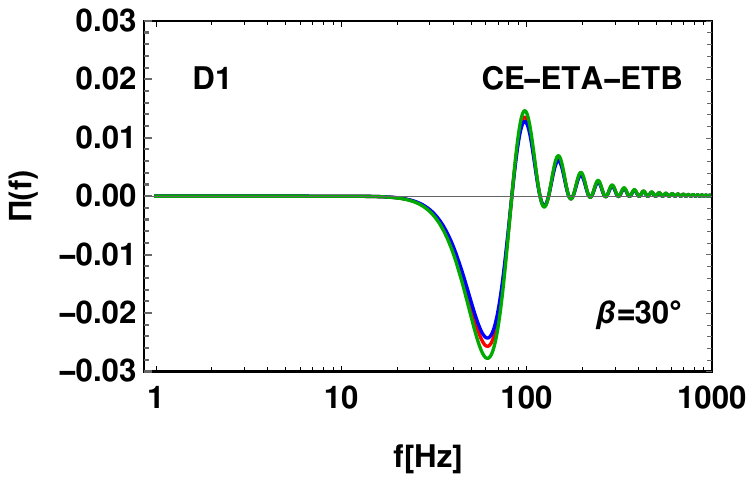}
\hspace{1cm}
\includegraphics[width=0.45\textwidth]{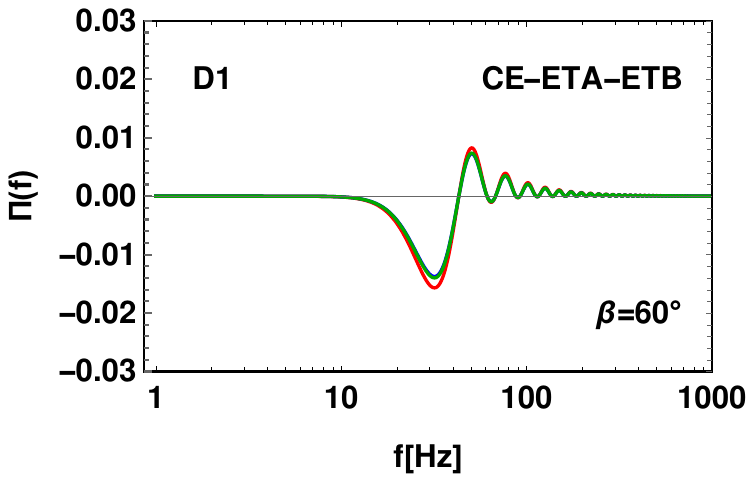}\\
\vspace{0.3cm}
\includegraphics[width=0.45\textwidth]{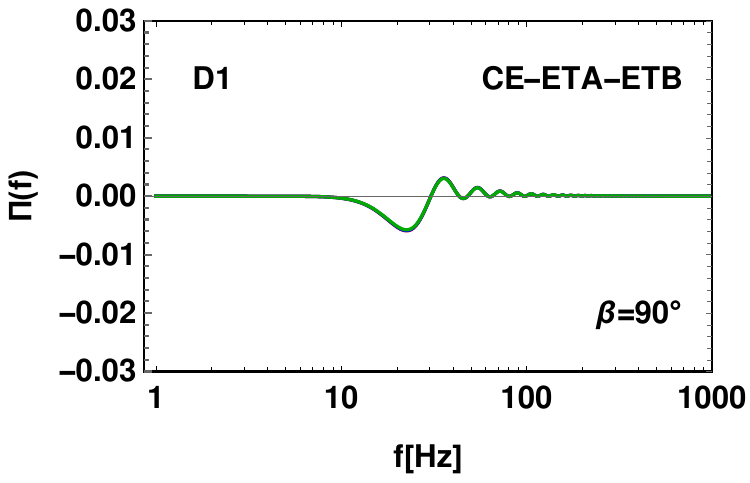}
\hspace{1cm}
\includegraphics[width=0.45\textwidth]{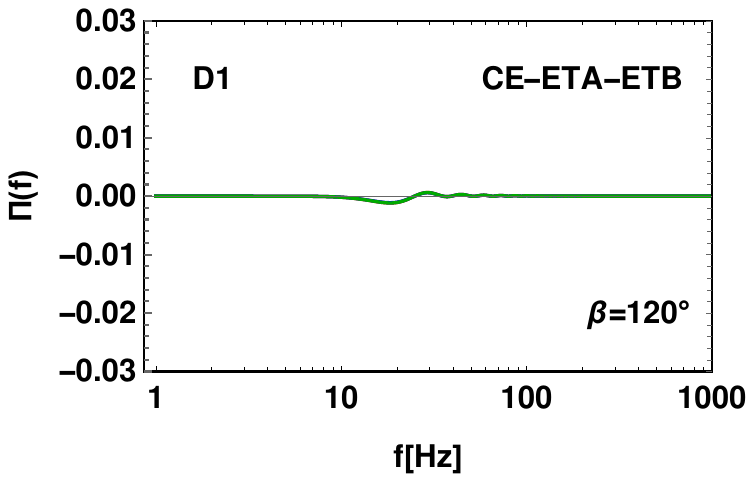}\\
\vspace{0.3cm}
\includegraphics[width=0.45\textwidth]{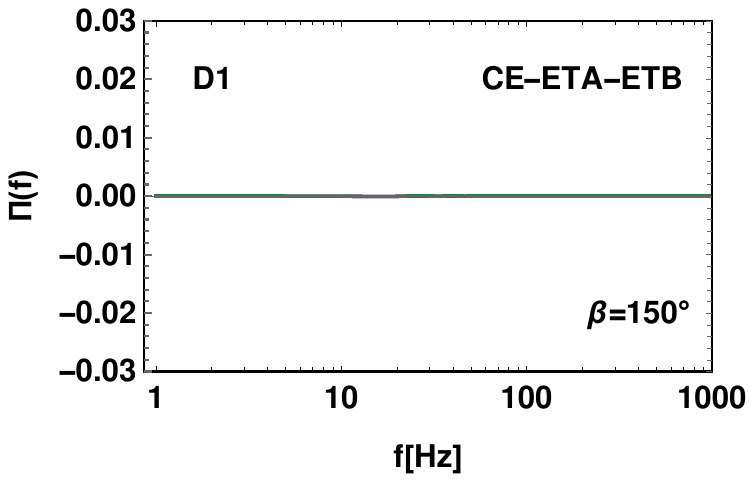}
\caption{Tensor (red), vector (blue) and scalar (green) $\Pi$ functions corresponding to optimal configurations shown in Figs.\ref{fig:3TD12}, \ref{fig:3VD12} and \ref{fig:3SD12} respectively. The configuration used is D1, which refers to the ETA and ETB interferometers in the xylophone configuration along with CE in Stage 1.}
\label{fig:DETTVS}
\end{figure*}

Let us first consider the case where interferometers $1$, $2$ and $3$ are given by ETA, ETB and CE respectively: this might be the first third-generation collaboration among ground-based detectors to be fully exploited in the following two decades, therefore we now start our analysis of the network. Let $\sigma_2$ be the orientation of the CE bisector, $\sigma_1$ the orientation of the ETB one and $\sigma_1-2\pi/3$ the one relative to ETA. Additionally, since we are working in the low-frequency limit, we further assume $\gamma^T_{12}=\gamma^V_{12}=\gamma^S_{12} \approx -3/8$ (see Fig.\ref{fig:ETETORF}). For a fixed value of $\beta$, it is straightforward to show that
\[\Pi(f)\propto \sin(4\sigma_2)(7+3\cos\beta)\cos^4\bigl(\frac{\beta}{2}\bigr)j_2(f)j_4(f).\]
Note that this result does not depend on ET orientation due to its triangular topology, while the CE orientation filters the sensitivity to GWs: indeed we find that whenever $\sin(4\sigma_2)=0$, then $\Pi=0$ and the signal is null for tensor, vector and scalar modes, no matter what the separation angle is. When $\beta=0$, then $j_2=j_4=0$ and the signal is null, while if $\beta=\pi$, then $\cos(\pi/2)=0$ and the signal is again null: this means that the detector plane orthogonal direction cannot be the same or opposite for both detectors, similarly to a SGWB made of tensor and $X$-polarization modes (with $X=V$ or $S$), with CE being the ``dominant'' detector. We now consider Eq.\eqref{eq:wide} in order to compute the expected SNR for an energy density reference value $\Omega^M_{ref}=10^{-11}$, while assuming a frequency independent energy density spectrum for tensor, vector and scalar polarization modes. In Fig.\ref{fig:3TD12}, Fig.\ref{fig:3VD12} and Fig.\ref{fig:3SD12} we show the results for the expected SNR using the D1 configuration for several fixed values of $\beta$ while leaving the orientation of both detectors free to vary.
We immediately recognize four ``bands'' separated by vertical straight lines of null SNR values, corresponding to the case where $\sin (4\sigma_2)=0$. With the exception of small angular separations ($\beta \approx \pi/3$), from Figs.\ref{fig:3TD12}, \ref{fig:3VD12} and \ref{fig:3SD12} we clearly see that for tensor, vector and scalar polarization modes the CE optimal orientation is once again approximately the same for each value of $\beta$. On the other hand, ET optimal orientations are not greatly shared among different polarization mode (i.e. they are not approximately the same for tensor, vector and scalar polarization modes), meaning that if we choose bisector orientations in order to maximize SNR$_T$, we lose some sensitivity in terms of scalar and vector modes or the other way around. Indeed, this behavior affects the corresponding optimal $\Pi$ functions, which are shown in Fig.\ref{fig:DETTVS}: given the angular separation, these look similar to each other being only proportional to the CE orientation and ignoring the ET one as we discussed. This means that the isolation of tensor, vector and scalar polarization modes is allowed in almost identical frequency ranges.
We show in Tabs.\ref{tab:ThreeTensor}, \ref{tab:ThreeVector} and \ref{tab:ThreeScalar} results for both expected SNR ($\Omega^M_{ref}=10^{-11}$ and $T=5$ yrs) and detectable energy density (SNR$=5$ and $T=5$ yrs) for tensor, vector and scalar modes respectively using configurations B1 and D1 with optimal orientations for ET and CE.
\begin{table}
\begin{ruledtabular}
\begin{tabular}{ccccc}
 &\multicolumn{2}{c}{SNR$_T$}&\multicolumn{2}{c}{$h_0^2\Omega^T_{GW}$}\\
$\beta$ & B1 & D1 & B1 & D1 \\ \hline
$30^{\circ}$ & $6.16$ & $3.48$ & $8.11\times 10^{-12}$ & $1.44\times 10^{-11}$ \\

$60^{\circ}$ & $5.51$ & $6.48$ & $9.07\times 10^{-12}$ & $7.71\times 10^{-12}$ \\

$90^{\circ}$ & $2.70$ & $4.11$ & $1.85\times 10^{-11}$ & $1.22\times 10^{-11}$ \\

$120^{\circ}$ & $1.36$ & $1.27$ & $3.69\times 10^{-11}$ & $3.92\times 10^{-11}$ \\

$150^{\circ}$ & $0.27$ & $0.24$ & $1.84\times 10^{-10}$ & $2.09\times 10^{-10}$ \\
\end{tabular}
\end{ruledtabular}
\caption{Expected SNR$_T$ ($\Omega^T_{ref}=10^{-11}$ and $T=5$ yrs) and detectable energy density (SNR$=5$ and $T=5$ yrs) for a SGWB made of tensor, vector and scalar modes for ETB and CE with optimal orientations shown in Fig.\ref{fig:3TD12}) for a fixed angular separation ($\beta$). Numerical results are shown using B1 and D1 configurations.}
\label{tab:ThreeTensor}
\end{table}
\begin{table}
\begin{ruledtabular}
\begin{tabular}{ccccc}
 &\multicolumn{2}{c}{SNR$_V$}&\multicolumn{2}{c}{$h_0^2\Omega^V_{GW}$}\\
$\beta$ & B1 & D1 & B1 & D1 \\ \hline
$30^{\circ}$ & $4.17$ & $2.47$ & $1.20\times 10^{-11}$ & $2.03\times 10^{-11}$ \\

$60^{\circ}$ & $4.71$ & $4.25$ & $1.06\times 10^{-11}$ & $1.18\times 10^{-11}$ \\

$90^{\circ}$ & $2.69$ & $2.43$ & $1.86\times 10^{-11}$ & $2.05\times 10^{-11}$ \\

$120^{\circ}$ & $1.08$ & $1.09$ & $4.63\times 10^{-11}$ & $4.65\times 10^{-11}$ \\

$150^{\circ}$ & $0.11$ & $0.06$ & $4.36\times 10^{-10}$ & $9.03\times 10^{-10}$ \\

\end{tabular}
\end{ruledtabular}
\caption{Expected SNR$_V$ ($\Omega^V_{ref}=10^{-11}$ and $T=5$ yrs) and detectable energy density (SNR$=5$ and $T=5$ yrs) for a SGWB made of tensor, vector and scalar modes for ETB and CE with optimal orientations shown in Fig.\ref{fig:3VD12}) for a fixed angular separation ($\beta$). Numerical results are shown using B1 and D1 configurations.}
\label{tab:ThreeVector}
\end{table}
\begin{table}
\begin{ruledtabular}
\begin{tabular}{ccccc}
 &\multicolumn{2}{c}{SNR$_S$}&\multicolumn{2}{c}{$\xi h_0^2\Omega^S_{GW}$}\\
$\beta$ & B1 & D1 & B1 & D1 \\ \hline
$30^{\circ}$ & $5.25$ & $3.17$ & $9.52\times 10^{-12}$ & $1.58\times 10^{-11}$  \\

$60^{\circ}$ & $8.70$ & $7.73$ & $5.75\times 10^{-12}$ & $6.47\times 10^{-12}$  \\

$90^{\circ}$ & $4.57$ & $5.46$ & $1.09\times 10^{-11}$ & $9.16\times 10^{-12}$  \\

$120^{\circ}$ & $1.26$ & $1.65$ & $3.98\times 10^{-11}$ & $3.03\times 10^{-11}$  \\

$150^{\circ}$ & $0.06$ & $0.07$ & $8.83\times 10^{-10}$ & $7.15\times 10^{-10}$  \\

\end{tabular}
\end{ruledtabular}
\caption{Expected SNR$_S$ ($\Omega^S_{ref}=10^{-11}$ and $T=5$ yrs) and detectable energy density (SNR$=5$ and $T=5$ yrs) for a SGWB made of tensor, vector and scalar modes for ETB and CE with optimal orientations shown in Fig.\ref{fig:3SD12}) for a fixed angular separation ($\beta$). Numerical results are shown using B1 and D1 configurations.}
\label{tab:ThreeScalar}
\end{table}
In terms of tensor modes, we find that the D1 configuration is the best one, with the usual exception of small spatial distances between the pair ($\beta$ = $\pi/6$) where B1 gives better forecasts. On the other hand, we find that B1 can be equally or more sensitive to vector modes in the SGWB with respect to D1. Finally, in terms of scalar modes, the best configuration depends on the angular separation between ET and CE: we have B1 more sensitive to scalar modes for $\beta=\pi/6$ and $\beta=\pi/3$, while the D1 arrangement gives better forecasts for all other suggested values. For each polarization class and configuration, as the angular separation increases, the expected SNR and detectable energy density first reach a maximum and minimum value respectively, then, the processes reverses, leading to worse expected forecasts (indeed in Fig.\ref{fig:DETTVS} we see that $\Pi(\beta=2\pi/3$, $5\pi/6) \approx 0$ and the expected SNR gets smaller). We now compare results in Tabs.\ref{tab:ThreeTensor}, \ref{tab:ThreeVector} and \ref{tab:ThreeScalar} to energy density upper limits for tensor and non-GR provided by \cite{abbott2021upper}. If we are allowed to choose the proper angular separation between ET and CE, we find that a network built with third-generation ground-based interferometers approximately improves its sensitivity to tensor and extra polarization modes in the SGWB by a factor $10^3$ in terms of corresponding detectable energy density contributions to the background energy density.

\subsection{ETA-ETB-ETC network}
Let us consider Fig.\ref{fig:3} once again: ET triangular configuration provides three ground-based interferometers. We mentioned that in order to separate tensor, vector and scalar polarization modes we first need to consider Eq.\eqref{eq:api}, in particular it must be $\Pi(f) \neq 0$ at least for a finite frequency range where the polarization modes are allowed to be separated. Moreover, we also stated that we can assume $\gamma^M_{AB}=\gamma^M_{BC}=\gamma^M_{AC}$ for $M$=$T$, $V$ and $S$ in the low-frequency limit. This configuration clearly presents a problem, since no matter what angular separation or frequency range we are considering, we always get $\Pi=0$, meaning the SNR is null. This issue only arises because of ET triangular topology \footnote{As shown in detail in \cite{philippoz2018gravitational}, small perturbations of the simmetry of ET are not sufficient to allow the isolation of tensor, vector and scalar polarization modes using the three interferometers provided by the observatory.} and it is not related to incoming GW properties: even when we exit the low-frequency limit, as long as ORFs are the same for each pair, ET alone is not able to distinguish different polarization modes in terms of their energy density contributions to the SGWB. 

\section{Cosmic Explorer replacing LIGO observatories \label{2CE}}

\begin{figure*}
\includegraphics[width=0.4\textwidth]{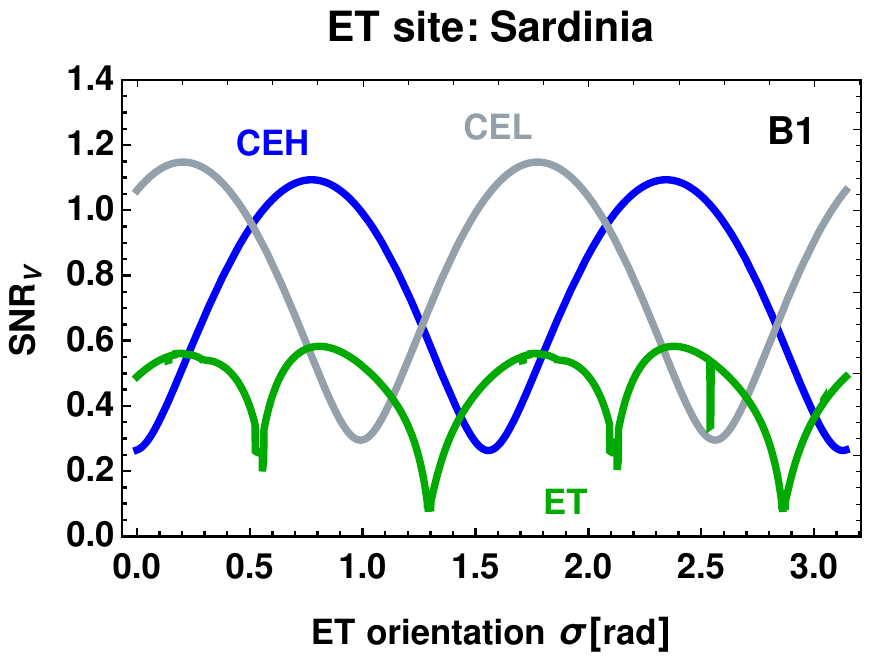}
\hspace{0.3cm}
\includegraphics[width=0.4\textwidth]{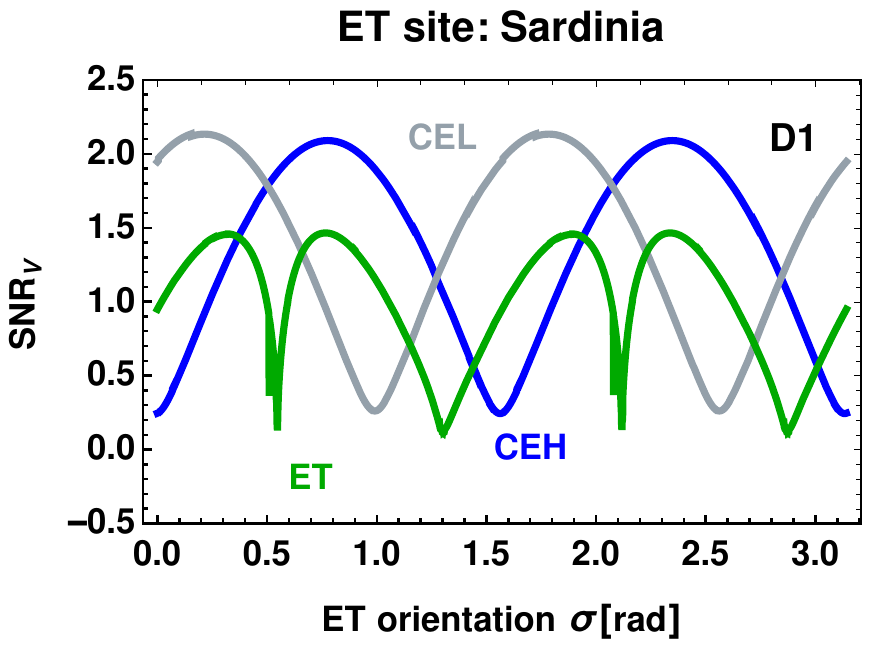}\\
\vspace{0.3cm}
\includegraphics[width=0.4\textwidth]{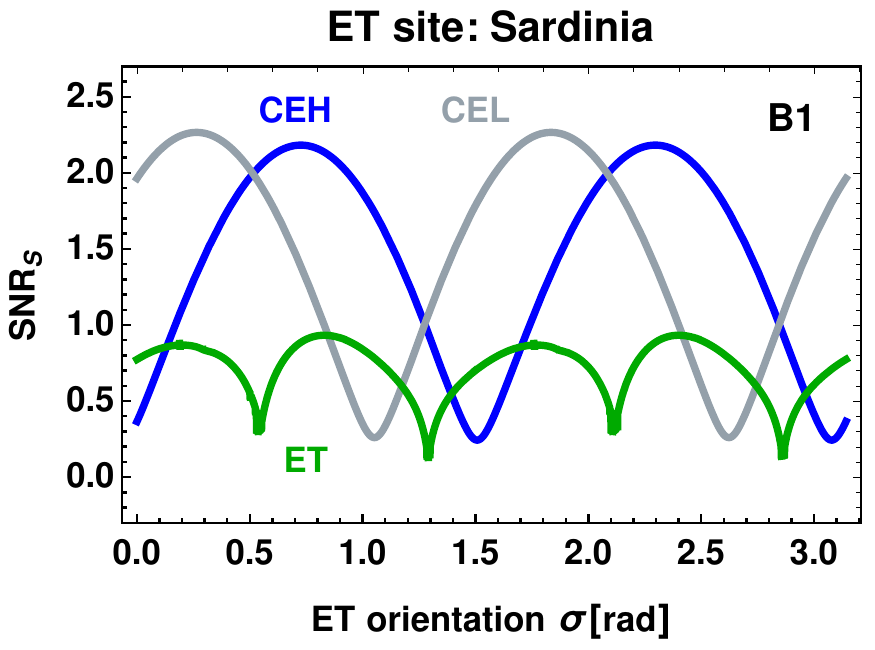}
\hspace{0.3cm}
\includegraphics[width=0.4\textwidth]{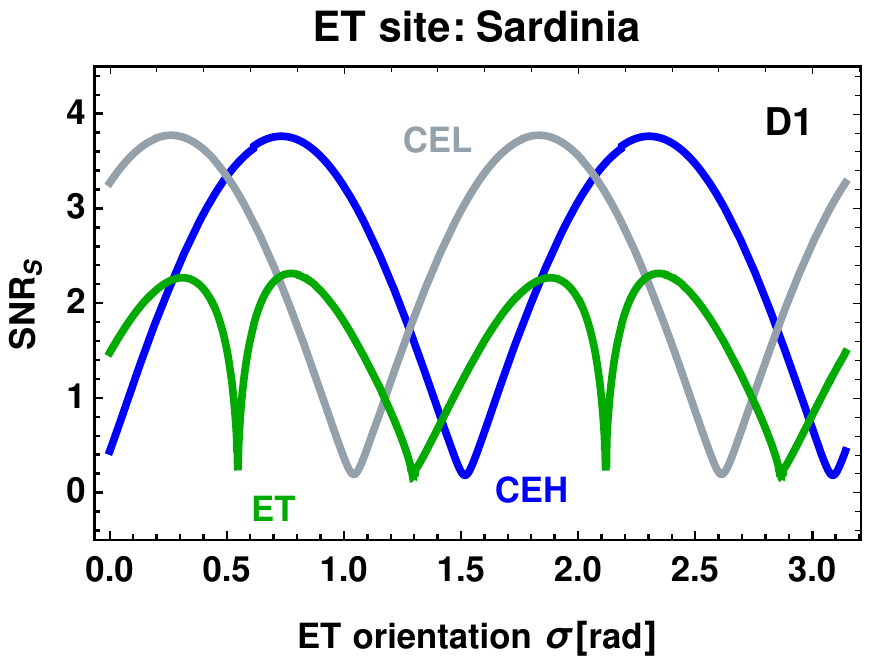}\\
\vspace{0.3cm}
\includegraphics[width=0.4\textwidth]{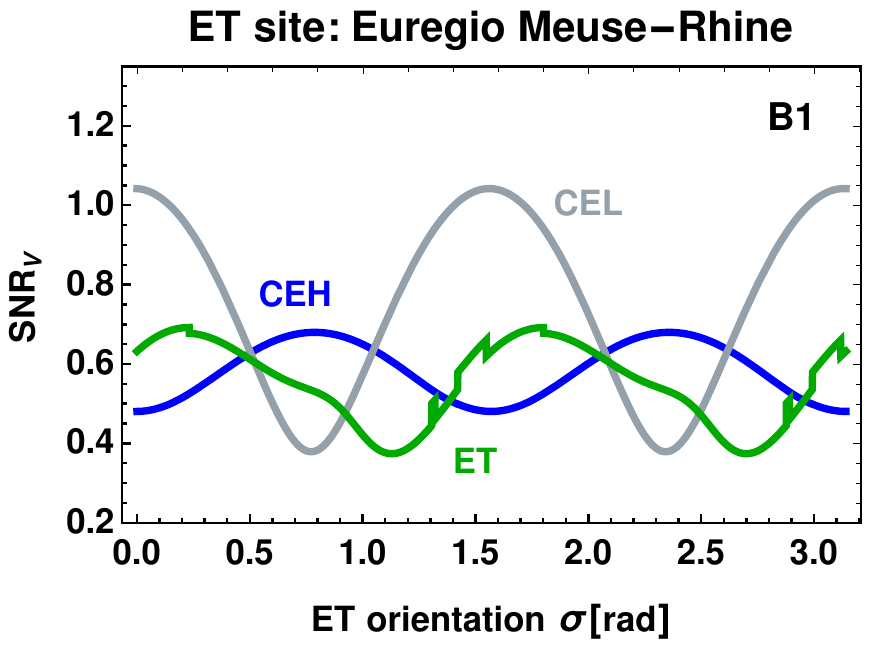}
\hspace{0.3cm}
\includegraphics[width=0.4\textwidth]{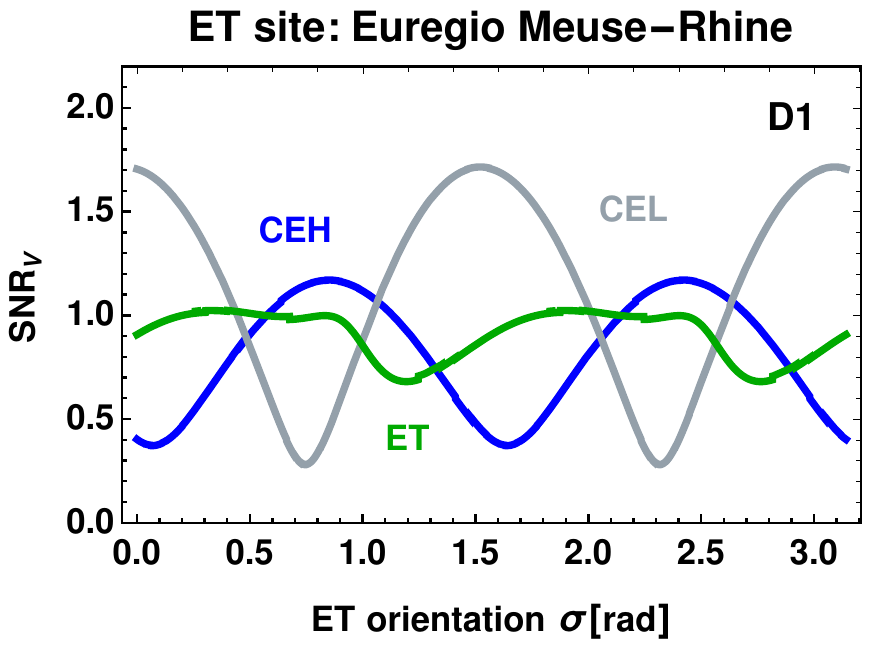}\\
\vspace{0.3cm}
\includegraphics[width=0.4\textwidth]{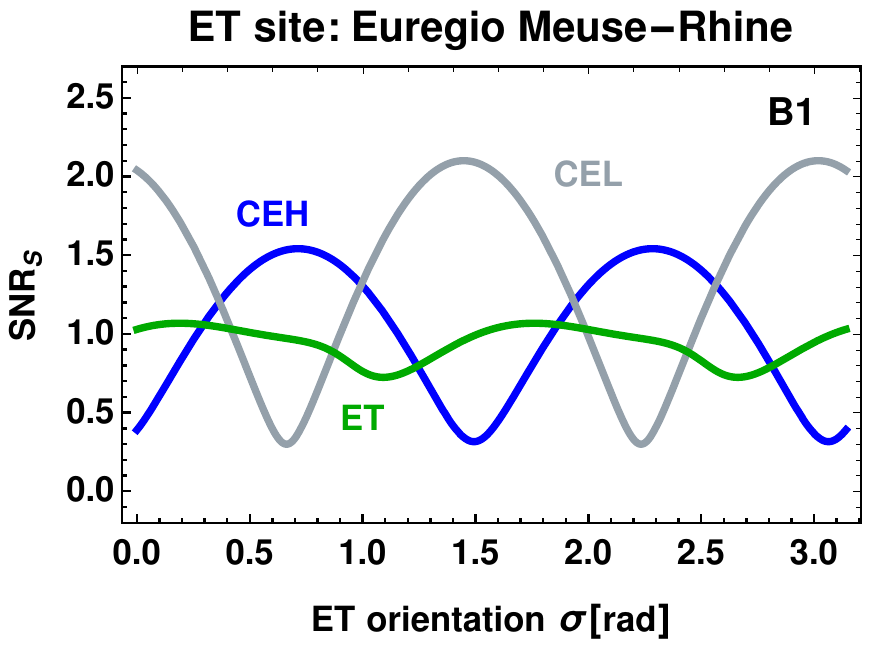}
\hspace{0.3cm}
\includegraphics[width=0.4\textwidth]{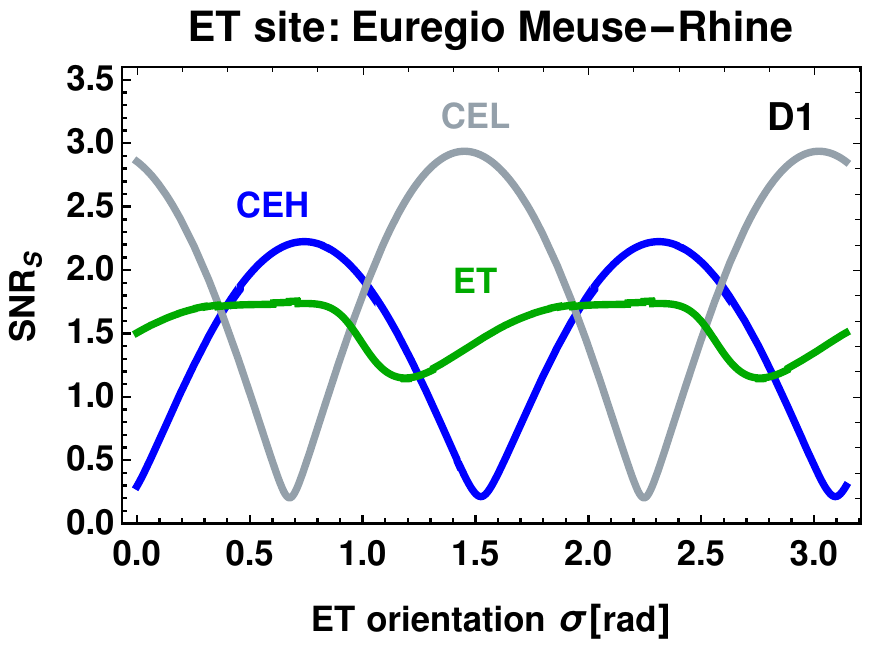}\\
\caption{Expected SNR$_X$, $X=V$ or $S$ ($\Omega^X_{ref}=10^{-12}$ and $T=5$ yrs), for a SGWB made of tensor and vector or scalar modes considering a network of two CE-like interferometers replacing the LIGO observatories in Livingston and Hanford (CEL and CEH) and one ET interferometer located either in Sardinia or Euregio Meuse-Rhine. Each curve corresponds to a particular choice of the ``dominant'' detector: CEH (blue), CEL (gray) and ET (green).}
\label{fig:2CE2SGWB}
\end{figure*}

In sections \ref{2GWB} and \ref{3GWB} we discussed that only two ET interferometers can be taken to work independently (see also e.g. \cite{philippoz2018gravitational}) along with CE in order to successfully isolate tensor, vector and scalar polarization modes energy density contributions to the SGWB. In this section, we extend our previous results to the particular case where we dispose of a network made of one ET interferometer and two CE-like interferometers replacing the two LIGO observatories in North America both in location and orientation: we refer to these detectors as ``CEL'' for the Livingston site and ``CEH'' for the Hanford site. Moreover, we assume ET to be located either in Italy (Sardinia site) or at the border region between the Netherlands, Belgium and Germany (Euregio Meuse-Rhine site), while we leave its orientation $\sigma$ (which we measure in a counterclokcwise manner with respect to the great circumference connecting CEL and ET) free to vary. Let us begin by considering a SGWB made of tensor and $X$-polarization modes, with $X$ = $V$ or $S$. In order to remove tensor modes contributions to the SNR for the SGWB, we can safely recover Eq.\eqref{eq:SNREXTRA} (since there are no colocated interferometers we assume detector noises to be uncorrelated): we know all detector locations along with CEH and CEL angular orientations, which means that the SNR can be seen as a function of the only ET orientation. In Fig.\ref{fig:2CE2SGWB} we show the expected SNR$_X$ ($\Omega^X_{ref}=10^{-12}$ and $T=5$ yrs) we computed assuming a frequency-independent energy density spectrum for both tensor and  $X$-polarization modes and considering the two possible ET sites and layouts (described by ET-B and ET-D sensitivity curves as shown in the left panel of Fig.\ref{fig:NOISE}) along with the Stage 1 CE.
We remind the reader that Fig.\ref{fig:2CE2SGWB} (and Fig.\ref{fig:2CE3SGWB} below) was obtained while performing some numerical integrations: due to some limits on the precision and accuracy of numerical calculations, some of the related subfigures show small distortions. However, the latter do not change the physics of the problem and the considerations we made in this paper. Moreover, each SNR curve refers to a different ``dominant'' detector: we find that better forecasts for the expected SNR are mostly given by ``dominants'' CEH and CEL depending on the value of $\sigma$ for all possible scenarios. Additionally, the D1 configuration generally improves the network sensitivity to the SGWB with respect to B1. This is true in particular for the Sardinia site, where the sensitivity to non-GR polarization modes is approximately doubled. Note also that peaks appearing in both SNR$_V$ and SNR$_S$ while considering the D1 configuration are higher for the Sardinia site, meaning that the latter could provide better detectable energy density values for vector or scalar polarization modes with respect to the Euregio Meuse-Rhine site: indeed SNR$_X$ is of order unity, therefore if we e.g. set SNR$_X=5$ to claim detection, $T=5$ yrs and we compute $\Omega^X_{GW}$, the latter can be directly compared to results listed in Tabs.\ref{tab:VECTORONLY}, \ref{tab:VECTORONLYETET}, \ref{tab:SCALARONLY}, \ref{tab:SCALARONLYETET} where two ET interferometers and one CE were considered, reconfirming the network gain in sensitivity to extra polarization modes with respect to second-generation ground-based interferometers. Let us now move on to the most general case where tensor, vector and scalar polarization modes are present at the same time: we further consider Eq.\eqref{eq:wide} and we investigate what is the network sensitivity to GWs of different polarizations. In Fig.\ref{fig:2CE3SGWB} we show the expected SNR$_M$, with $M=T$, $V$, and $S$, ($\Omega^M_{ref}=10^{-11}$ and $T=5$ yrs) which we again computed assuming a frequency-independent energy density spectrum for tensor, vector and scalar polarization modes and considering the two possible B1 and D1 layouts.
\begin{figure*}
\includegraphics[width=0.45\textwidth]{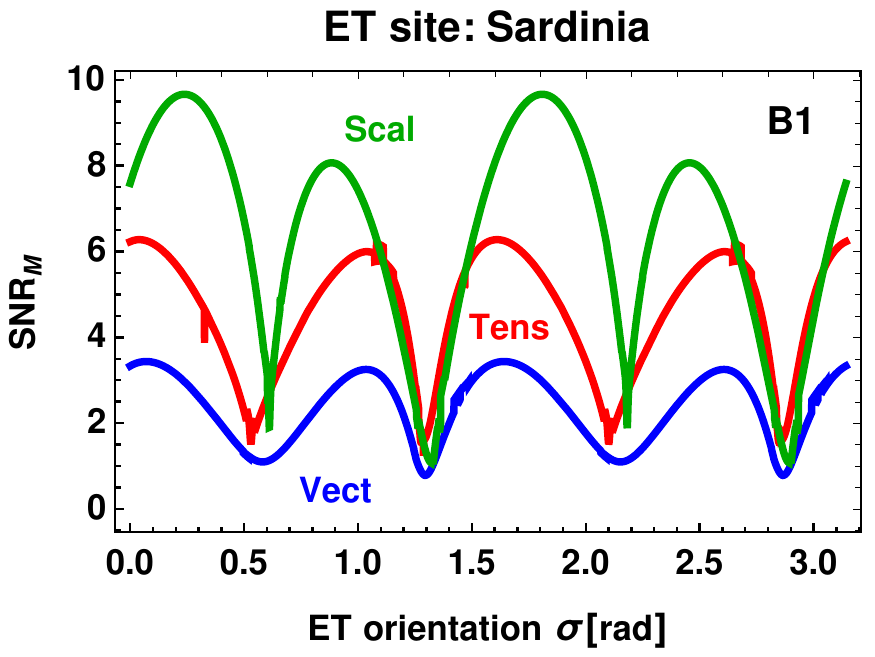}
\hspace{0.3cm}
\includegraphics[width=0.45\textwidth]{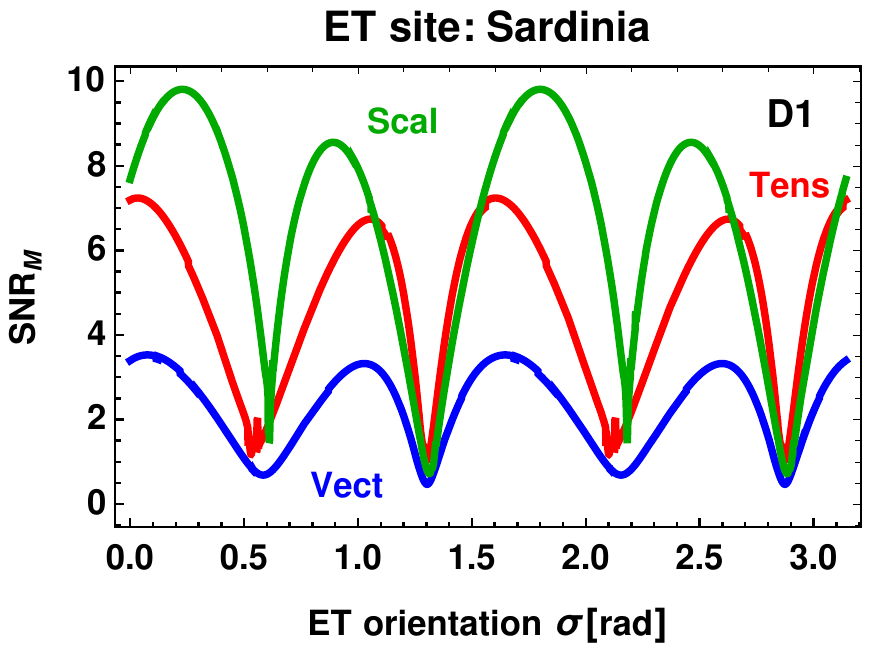}\\
\vspace{0.3cm}
\includegraphics[width=0.45\textwidth]{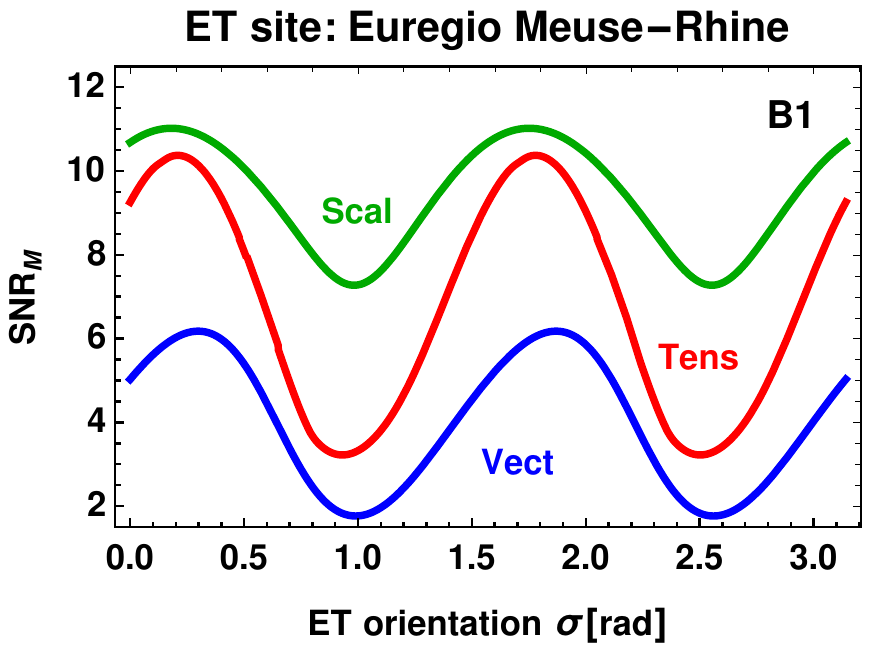}
\hspace{0.3cm}
\includegraphics[width=0.45\textwidth]{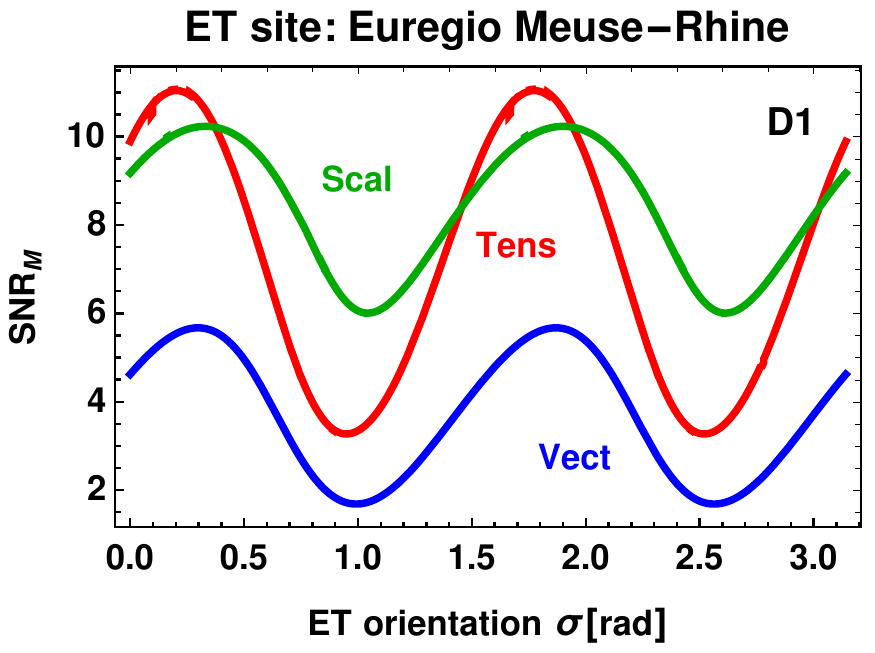}\\
\caption{Expected SNR$_M$, $M=T$, $V$ and $S$ ($\Omega^M_{ref}=10^{-11}$ and $T=5$ yrs), for a SGWB made of tensor, vector and scalar modes considering a network of two CE-like interferometers replacing the LIGO observatories in Livingston and Hanford (CEL and CEH) and one ET interferometer located either in Sardinia or Euregio Meuse-Rhine. Different curves refer to the expected SNR for each polarization class: tensor (red), vector (blue) and scalar (green).}
\label{fig:2CE3SGWB}
\end{figure*}
Once we choose the site for ET, we find that there are no significant changes for tensor and scalar modes while switching between configurations B1 and D1. An exception is made for SNR$_V$ curves, where corresponding peaks are slightly higher while considering ET in its xylophone configuration. Additionally, we also find that the network is approximately twice more sensitive to tensor and scalar polarization modes with respect to vector modes. Moreover, SNR$_M$ peaks approximately correspond to the same value of $\sigma$ for $M=T$, $V$ and $S$, meaning that if we optimally choose the ET orientation, we get close-to-ideal SNR$_M$ values for tensor, vector and scalar polarization modes at the same time. Although this applies to both ET sites, the best choice would be the Euregio Meuse-Rhine location, which would slightly improve the network sensitivity (SNR peaks) to all polarization modes. This can be better understood with a concrete example, therefore we choose $\sigma=\pi/2$ for the Sardinia site and $\sigma=7\pi/12$ for the Euregio Meuse-Rhine site. We further set SNR$_M=5$ and we compute the detectable energy density $\Omega^M_{GW}$ for both cases: results are shown in Tab.\ref{tab:2CE} and, once again, can be directly compared to some of the optimal ones listed in Tabs.\ref{tab:ThreeTensor}, \ref{tab:ThreeVector} and \ref{tab:ThreeScalar}: indeed, with respect to second-generation detectors \cite{abbott2021upper}, we find once again that a third-generation network of ground-based interferometers approximately improves its sensitivity to tensor and extra polarization modes in the SGWB by a factor $10^3$.
\\
\begin{table}
\begin{ruledtabular}
\begin{tabular}{ccc}
 &\multicolumn{1}{c}{Sardinia site}&\multicolumn{1}{c}{Eur. Meuse-Rhine site}\\
 &$(\sigma=90^{\circ})$ &$(\sigma=105^{\circ})$ \\ \hline
$h_0^2\Omega^T_{GW}$ & $6.95\times 10^{-12}$ & $4.55\times 10^{-12}$ \\

$h_0^2\Omega^V_{GW}$ & $1.47\times 10^{-11}$ & $8.84\times 10^{-12}$ \\

$\xi h_0^2\Omega^S_{GW}$ & $6.51\times 10^{-12}$ & $4.91\times 10^{-12}$ \\
\end{tabular}
\end{ruledtabular}
\caption{Detectable energy density (SNR$_M=5$, $M=T$, $V$ and $S$, and $T=5$ yrs) for tensor, vector and scalar polarization modes assuming ET to be located in the Sardinia site with orientation $\sigma=90^{\circ}$ or in the Euregion Meuse-Rhine site with orientation $\sigma=105^{\circ}$.}
\label{tab:2CE}
\end{table}

\section{Detector responses to high gw frequencies \label{GWhigh}}

\begin{figure*}
\includegraphics[width=0.4\textwidth]{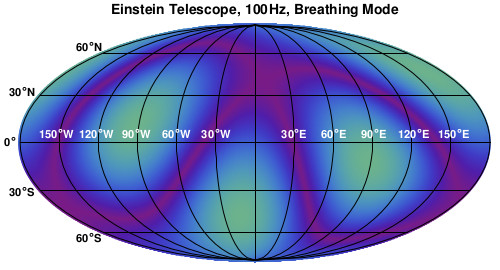}
\includegraphics[width=0.6cm,height=3.5cm]{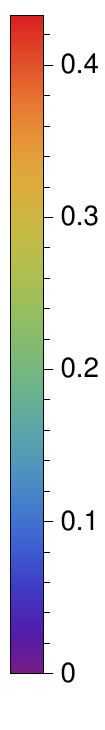}
\hspace{0.3cm}
\includegraphics[width=0.4\textwidth]{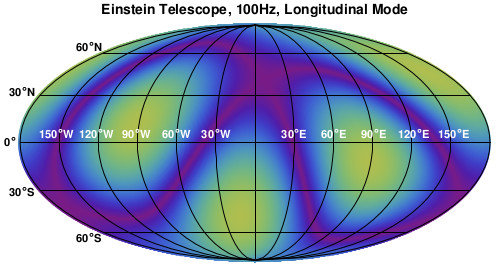}
\includegraphics[width=0.6cm,height=3.5cm]{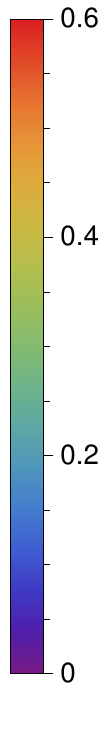}\\
\vspace{0.8cm}
\includegraphics[width=0.4\textwidth]{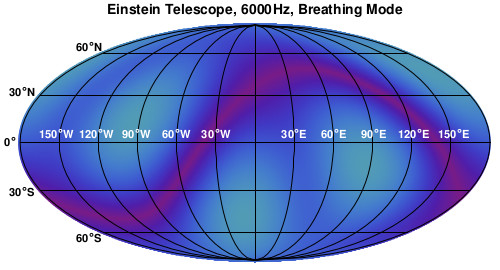}
\includegraphics[width=0.6cm,height=3.5cm]{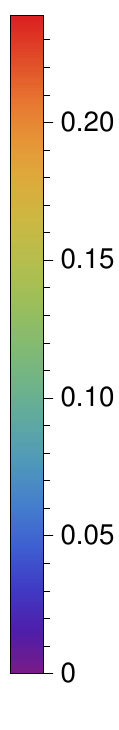}
\hspace{0.3cm}
\includegraphics[width=0.4\textwidth]{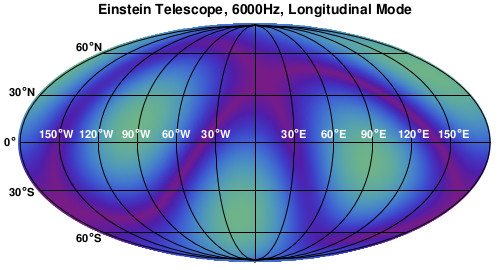}
\includegraphics[width=0.6cm,height=3.5cm]{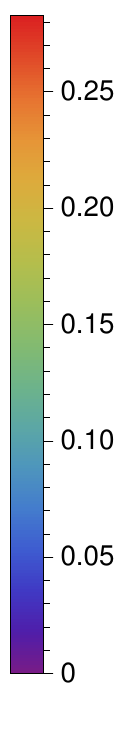}\\
\vspace{0.8cm}
\includegraphics[width=0.4\textwidth]{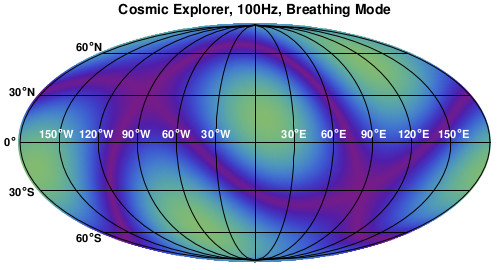}
\includegraphics[width=0.6cm,height=3.5cm]{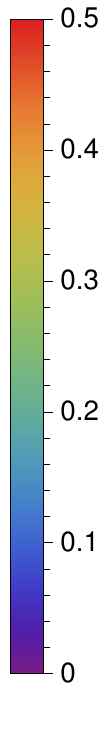}
\hspace{0.3cm}
\includegraphics[width=0.4\textwidth]{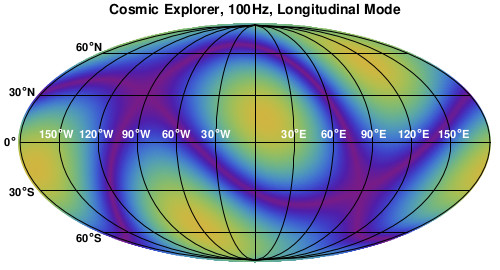}
\includegraphics[width=0.6cm,height=3.5cm]{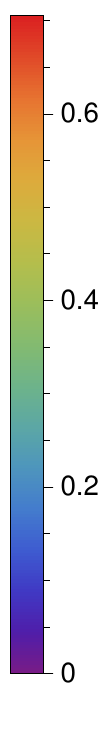}\\
\vspace{0.8cm}
\includegraphics[width=0.4\textwidth]{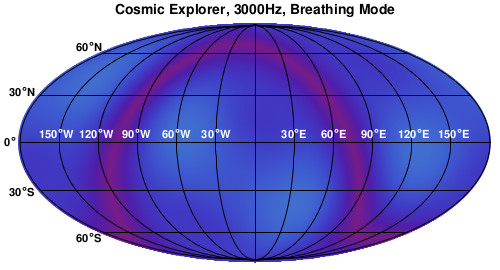}
\includegraphics[width=0.6cm,height=3.5cm]{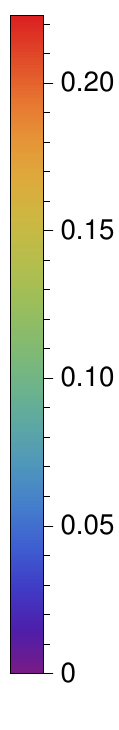}
\hspace{0.3cm}
\includegraphics[width=0.4\textwidth]{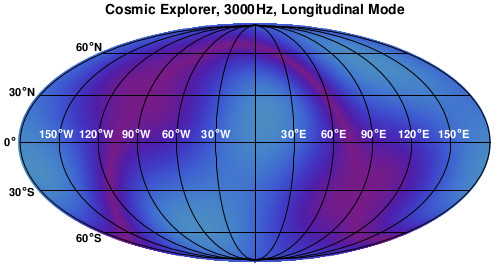}
\includegraphics[width=0.6cm,height=3.5cm]{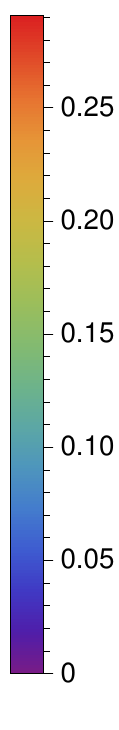}\\
\label{fig:breaking}
\caption{Mollweide projections of ET (single interferometer e.g. ETA) and CE angular response for breathing (left) and longitudinal (right) polarization modes considering GW frequencies lower and higher than the corresponding characteristic values $f_*$ ($f_*$ is approximately given by $4774$ Hz and $1194$ Hz for ET and CE respectively). CE was assumed to replace the LIGO Livingston observatory both in location and orientation, while ET was assumed to be located in the Sardinia site. Both detectors are assumed to be simple Michelson interferometers. Note that for sufficiently low GW frequencies the same detector is equally sensitive to the two scalar modes for different angular directions in the sky, but  when GW frequencies are higher than $f_*$, detector responses to breathing and longitudinal modes start to behave differently.}
\label{fig:25}
\end{figure*}

As long as one is working in the low-frequency limit, it is a well known result in literature that a single GW interferometer presents the same angular response to scalar-breathing and scalar-longitudinal polarization modes \cite{romano2017detection}, represented by the corresponding APFs which only differ for a constant factor, as shown in Eqs.\eqref{eq:breathET}, \eqref{eq:longET}, \eqref{eq:breathCE} and \eqref{eq:longCE}. We also mentioned that this prevents both ET and CE (and more generally all ground-based detectors) from distinguishing between the two scalar modes. However, the third-generation of ground-based interferometers is expected to be sensitive to GW frequencies equal to or even higher than the relative detector characteristic frequency $f_*$ (i.e. ET optimistically will be sensitive to frequencies in the $1$-$10000$ Hz range, though its characteristic frequency is $f_*\approx 4774$ Hz, while CE sensitivity curves CE1 and CE2 are defined between $5$-$5000$ Hz, though its characteristic frequency is $f_*\approx 1194$ Hz). Therefore, focusing on ET and CE, in this section we discuss how breathing and longitudinal APFs behave once we exit the low-frequency limit in order to study detector angular responses to scalar polarization modes considering higher GW frequencies. In the most general case, we dispose of an interferometer with opening angle $\nu$, with $0 < \nu \leq \pi/2$ and we define relative unit vectors directed along each detector arm as
\begin{center}
\[\mathbf{\hat{e}_{1}}=\biggl(\cos\biggl(\frac{\pi}{4}-\frac{\nu}{2}\biggr),\sin\biggl(\frac{\pi}{4}-\frac{\nu}{2}\biggr),0\biggr),\] \\
\[\mathbf{\hat{e}_{2}}=\biggl(\cos\biggl(\frac{\pi}{4}+\frac{\nu}{2}\biggr),\sin\biggl(\frac{\pi}{4}+\frac{\nu}{2}\biggr),0\biggr). \]
\end{center}
When we approach and/or surpass $f_*$, APFs are still given by the tensor contraction in Eq.\eqref{eq:5}, though the detector tensor is now frequency dependent because of the presence of the so-called transfer functions \cite{romano2017detection} defined for each detector arm, which are different depending on the nature of the interferometer considered (e.g. Michelson or Fabry-P\'erot interferometer \cite{schilling1997angular}). Given a general interferometer with relative transfer functions $\mathcal{T}_j$, with $j=1,2$, we get
\begin{eqnarray}
\mathbf{D}(\mathbf{\hat{\Omega}},f)=\frac{1}{2}\bigl\{(\mathbf{\hat{e}_1}\otimes\mathbf{\hat{e}_1})\mathcal{T}_1(\mathbf{\hat{\Omega}},f) - \mathbf{(\hat{e}_2}\otimes\mathbf{\hat{e}_2})\mathcal{T}_2(\mathbf{\hat{\Omega}},f) \bigr\},\nonumber\\
\end{eqnarray}
where
\[\mathcal{T}_j(\mathbf{\hat{\Omega}},f) \equiv \mathcal{T}(\mathbf{\hat{\Omega}} \cdot \mathbf{\hat{e}}_j,f).\]
Moreover, we are now able to compute breathing and longitudinal APFs while exiting the low-frequency limit, we get
\begin{widetext}
\begin{eqnarray}
F^b(f,\mathbf{\hat{\Omega}})=\frac{1}{4}(\mathcal{T}_1-\mathcal{T}_2)(1+\cos^{2}\theta)-\frac{1}{4}(\mathcal{T}_1-\mathcal{T}_2)\sin^{2}\theta\sin2\phi\cos\nu-\frac{1}{4}(\mathcal{T}_1+\mathcal{T}_2)\sin^{2}\theta\cos2\phi\sin\nu,
\label{dege1}
\end{eqnarray} 
\begin{eqnarray}
F^l(f,\mathbf{\hat{\Omega}})=\frac{\sqrt{2}}{4}(\mathcal{T}_1-\mathcal{T}_2)(1-\cos^{2}\theta)+\frac{\sqrt{2}}{4}(\mathcal{T}_1-\mathcal{T}_2)\sin^{2}\theta\sin2\phi\cos\nu+\frac{\sqrt{2}}{4}(\mathcal{T}_1+\mathcal{T}_2)\sin^{2}\theta\cos2\phi\sin\nu.
\label{dege2}
\end{eqnarray}
\end{widetext}
As a consistency check, if we assume $f\ll f_*$, then $\mathcal{T}_{1,2} \approx 1$ (considering normalized transfer functions) and if $\nu=\pi/2$ or $\nu=\pi/3$, APFs once again reduce to Eqs.\eqref{eq:breathCE} and \eqref{eq:longCE} or Eqs.\eqref{eq:breathET} and \eqref{eq:longET} respectively. If we take a closer look at Eqs.\eqref{dege1} and \eqref{dege2}, we see that the opening angle only affects the degenerate terms, meaning that the choice of $\nu$ does not break the scalar modes degeneracy, no matter which frequency range we are considering. However, if GW frequencies approach or surpass the characteristic frequency $f_*$, we find that the scalar modes degeneracy is always broken by the $(1\pm\cos^{2}\theta)$ factor and ground-based interferometers are able to distinguish between breathing and longitudinal polarization modes given their brand-new and different angular responses (see also \cite{liu2020constraining, liang2019frequency} for recent work on this topic with upcoming space-based interferometers). Note that this is no longer allowed while considering second-generation ground-based interferometers (e.g. LIGO, Virgo and KAGRA): due to their shorter arm lengths and worse sensitivities (with respect to ET and CE), these detectors are not sensitive to GW frequencies higher than their corresponding characteristic values $f_*$. We now consider an Earth-based coordinate and we assume a single ET interferometer (e.g. ETA) to be located in Sardinia and CE to replace the LIGO Livingston observatory both in location and orientation; moreover, we assume both detectors to be simple Michelson interferometers. We show in Fig.\ref{fig:25} plots of both detector angular responses to breathing and longitudinal polarization modes for the two cases where the low-frequency limit can or cannot be taken to be valid. We visibly recover what Eqs.\eqref{dege1} and \eqref{dege2} were suggesting: in terms of scalar modes, for sufficiently low GW frequencies the same detector is equally sensitive to different angular directions in the sky. However, when GW frequencies are higher than $f_*$, detector responses to breathing and longitudinal modes behave very differently.

\clearpage
\section{Conclusions \label{CONC}}

In this paper, we discussed in detail the detectability of non-GR polarization modes of a SGWB with third-generation ground-based detectors Einstein Telescope and Cosmic Explorer considering different possible orientations and locations on the Earth. The existence of such polarization modes is predicted by many alternative theories of gravitation and constraining these extra polarization modes would provide a useful tool to test such theoretical models (and at the same time a further test of GR), while their detection would be an inequivocable smoking gun of new physics beyond the standard model of GR.\\
We first investigated the ET network joint angular response to vector and scalar polarization modes. At the level of sky coverage, we found that in terms of extra polarization modes more isotropic joint responses are obtained with respect to a single L-shaped interferometer, though ET is still insensitive to GWs coming from the orthogonal direction to the detector plane, with respect to which joint responses present cylindrical symmetry.\\
We then considered possible detector networks with ET and CE in order to algebraically isolate \cite{nishizawa2009probing, omiya2020searching} tensor, vector and scalar energy density contributions to the SGWB, where we assumed frequency-independent energy density spectra (see Eq.\eqref{eq:spectrum}). It is worth  mentioning that ET alone is not able to exploit its three interferometers to achieve this result: this limitation comes from ET triangular topology \cite{philippoz2018gravitational}, therefore it is mandatory to also consider correlation with CE. We found results for several angular separations on Earth between ET and CE while using ideal orientations and considering five years of observation for each case. If only tensor modes are present, a SGWB with $h_0^2\Omega^T_{GW} \approx 10^{-12}$ could be detected by the ETA-CE detector pair in its D1 configuration, with ETA in its xylophone layout along with CE in Stage 1 (see Fig.\ref{fig:NOISE}). On the other hand, two ET interferometers (e.g. ETA-ETB) in their xylophone configuration could give equal or better forecasts with respect to the ETA-CE pair with an angular separation of $\beta \gtrsim \pi/6$. In the presence of additional $X$-polarization modes (with $X$ being one between vector or scalar modes) and despite the arbitrary choice of the ``dominant" detector (see Eq.\eqref{eq:dominant}), we found that by properly choosing the angular separation between ET and CE and their relative orientations the ETA-ETB-CE network in its D1 configuration could detect energy density contributions to the SGWB of order $h_0^2\Omega^X_{GW} \approx 10^{-12}$ for both vector and scalar polarization modes. Finally, when tensor, vector and scalar polarization modes are present at the same time, we have that the ETA-ETB-CE detector network loses some of its sensitivity to different polarization modes: using its D1 configuration, the network could isolate and detect tensor, vector and scalar modes energy density contributions to the SGWB of $h_0^2\Omega^M_{GW} \approx 10^{-11}$, for $M=T$, $V$ and $S$. Interestingly, in terms of vector modes, the same network in its B1 configuration would provide slightly better detection limits (in this case ET detectors are seen as single interferometers sensitive to the whole expected ET frequency range along with the Stage 1 CE).\\
Next, we investigated the possibility of having two CE-like interferometers replacing the LIGO observatories in North America both in location and orientation. We also considered the two proposed locations for ET, which are given by the Sardinia island site in Italy and the Euregio Meuse-Rhine site in northern Europe; we then focused on the CEL-CEH-ETA network in order to isolate and detect non-GR polarization modes in the SGWB. If only tensor and $X$-polarization modes (with $X=V$ or $S$) are present and choosing an optimal orientation for ETA, we found that the best configuration for the network is obtained with ET in a xylophone configuration located in the Sardinia site, with detection limits for the detectable energy density given by $h_0^2\Omega^X_{GW} \approx 10^{-12}$ for both vector and scalar polarization modes. If tensor, vector and scalar modes are present at the same time, the network sensitivity to all polarization modes (in particular to tensor and scalar modes) is slightly improved by assuming ET in the Euregio Meuse-Rhine site, while there are no significant changes in forecasts moving from the B1 to the D1 configuration and the other way around. Choosing the proper orientation for ETA, detection limits are approximately $h_0^2\Omega^{T,S}_{GW} \approx 5 \times 10^{-12}$ for tensor and scalar modes and $h_0^2\Omega^V_{GW} \approx 10^{-11}$ for vector modes.\\
In order to sum up our results, we can state that in terms of energy density contributions for tensor, vector and scalar polarization modes to the SGWB total energy density, all considered networks involving only ET and CE interferometers approximately improve their sensitivity to the SGWB by a factor $10^3$ with respect to current forecasts provided by \cite{abbott2021upper}; in particular, while considering valid locations and orientations for CE and ET, along with the latter in its proposed xylophone configuration, the corresponding network could detect energy density contributions to the SGWB in the range $\Omega^M_{GW} \approx 10^{-12}-10^{-11}$, with $M=T$, $V$ and $S$. We can further state that these detection limits can be directly compared to those found in \cite{omiya2020searching} using a network of space-based detectors (i.e. LISA and Taiji).\\
We finally investigated the possibility of breaking the scalar modes degeneracy with the new generation of ground-based interferometers. While considering GW frequencies $f \ll f_* = c/2\pi L$, with $L$ being the interferometer arm, detector angular responses to breathing and longitudinal polarization modes differ for a constant factor, making scalar modes indistinguishable for the interferometer. However, both ET and CE are expected to be sensitive to GWs with frequencies larger than their corresponding characteristic $f_*$: moving to this higher frequency-regime, we showed how different and new frequency-dependent angular responses can be obtained and how the degeneracy between scalar modes can be broken considering the new generation of ground-based interferometers (similarly to what is done in \cite{liu2020constraining, liang2019frequency} with the new generation of space-based interferometers). \\
The analysis performed in this paper can open the possibility to test many scalar-tensor and vector-tensor theories using  third-generation interferometers, constraining some modified gravity parameters~\cite{Bartolo:2020gsh}. We leave such an analysis for future works.

\section*{Acknowledgments}%
L.A.~would like to express his gratitude to A. Nishizawa for his useful comments given at the beginning of this project during his visit to the Department of Physics and Astronomy ``Galileo Galilei'', University of Padova.\\
A.R.~acknowledges funding from Italian Ministry of Education, University and Research (MIUR) through the ``Dipartimenti di
eccellenza'' project Science of the Universe.\\
N.B.~acknowledges partial financial support by ASI Grant No. 2016-24- H.0 and 2016-24-H.1-2018.\\
The work was supported by the International Helmholtz-Weizmann Research School for Multimessenger Astronomy, largely funded through the Initiative and Networking Fund of the Helmholtz Association.

\appendix

\section{Signal to Noise ratio}

Correlation analysis represents a useful and well-developed tool to detect a SGWB \cite{allen1999detecting, romano2017detection}. Here, we wish to retrace the steps needed to find the proper SNR expression for tensor and, when present in the SGWB, vector and scalar polarization modes. We draw heavily from \cite{nishizawa2009probing, omiya2020searching}, to which we refer the reader for more details.

\subsection{Tensor modes}

When only tensor modes are present, we begin by considering the output of two detectors $I$ and $J$ and we further introduce the cross-correlation signal
\begin{eqnarray}
Y_T=\int_{-\infty}^{+\infty}df\int_{-\infty}^{+\infty}df'\delta_T(f-f')s_I^*(f')s_J(f)  Q(f),\nonumber\\
\label{eq:a.1}
\end{eqnarray}
where $Q(f)$ is the so-called filter function used in the end to maximize the SNR and we introduced
\[\delta_T(f) = \int_{-T/2}^{T/2} dt e^{i 2\pi f t} =\frac{\sin (\pi f T)}{\pi f},\]
with $T$ the observation time; $\delta_T(f)$ coincides with the Dirac delta function in the limit $fT \rightarrow +\infty$, which we safely assume to be valid considering the sensitivity of ground-based detectors to GW-frequency ranges and an observation time of a few years. Taking the ensemble average of Eq.\eqref{eq:11} and also considering Eqs.\eqref{eq:a.1} and \eqref{eq:10}, we get
\begin{eqnarray}
S_T \equiv \langle Y_T \rangle &&=\int_{-\infty}^{+\infty}df\int_{-\infty}^{+\infty}df'\delta_T(f-f')G_{IJ}(f,f')Q(f).\nonumber\\
&&=\frac{3H_0^2}{20\pi^2}T\int_{-\infty}^{+\infty}dff^{-3}\bigl[\gamma^T(f)\Omega^T_{GW}(f) \bigr]Q(f). \nonumber\\
\label{eq:a.2}
\end{eqnarray}
The next step is to compute the corresponding variance in order to get the noise term: assuming GW-signals much smaller than the detector noise and recalling Eq.\eqref{eq:9} we get
\begin{eqnarray}
(N_T)^2&& \approx \bigl[\langle Y_T^2 \rangle - \langle Y_T \rangle^2\bigr]_{h=0} = \bigl[\langle Y_T^2 \rangle\bigr]_{h=0} \nonumber\\
&&=\frac{T}{4}\int_{-\infty}^{+\infty}dfP_{I}(f)P_{J}(f)|Q(f)|^2.
\label{eq:a.3}
\end{eqnarray}
We can express both Eqs.\eqref{eq:a.2} and \eqref{eq:a.3} in a simpler and compact way by introducing the following inner product
\begin{eqnarray}
\bigl(A(f)\cdot B(f)\bigr)\equiv && \int_{-\infty}^{+\infty}dfA^*(f)B(f)P_{I}(f)P_{J}(f),\nonumber\\
\end{eqnarray}
thus we have
\begin{eqnarray}
S_T=\frac{3H_0^2}{20\pi^2}T\left(Q(f) \cdot \frac{\gamma^T(f)\Omega^T_{GW}(f)}{f^{3}P_{I}(f)P_{J}(f)},\right)
\end{eqnarray}
\begin{eqnarray}
(N_T)^2=\frac{T}{4}\bigl(Q(f) \cdot Q(f)\bigr).
\end{eqnarray}
We can finally choose the proper expression for the filter function in order to maximize the SNR, which is given by
\[Q(f) \propto \frac{\gamma^T(f)\Omega^T_{GW}(f)}{f^{3}P_{I}(f)P_{J}(f)}.\]
In the end, we get the SNR expression for the SGWB in the presence of tensor modes only
\begin{eqnarray}
\text{SNR$_T$} &&= \biggl( \frac{S_T}{N_T} \biggr)\nonumber\\
&&=\frac{3H_0^2}{10\pi^2}\sqrt{T}\biggl[2\int_{0}^{+\infty}df\frac{(\gamma^T(f)\Omega^T_{GW}(f))^2}{f^3P_{I}(f)P_{J}(f)}\biggr]^{1/2}.\nonumber\\
\label{eq:a.4}
\end{eqnarray}

\subsection{Tensor and X-polarization modes}

We now extend the previous results to a SGWB made of tensor and $X$-polarization modes, where $X$ stands for vector or scalar: since we need (at least) three interferometers to work with, let us consider the following data combination
\begin{eqnarray}
\mu(f,f') =&& \gamma^T_{12}(f)s^*_{1}(f)s_{3}(f')-\gamma^T_{13}(f)s^*_{1}(f)s_{2}(f'),\nonumber
\label{eq:dominant}
\end{eqnarray}
where $1$, $2$ and $3$ denote the three detectors at our disposal. Analogously to what we did in Eq.\eqref{eq:a.1}, for the X-polarization modes we now define the cross correlation signal and its ensemble average as
\begin{eqnarray}
Y_X=\int_{-\infty}^{+\infty}df\int_{-\infty}^{+\infty}df'\delta_T(f-f')\mu(f)  Q(f),
\end{eqnarray}
\begin{eqnarray}
S_X \equiv \langle Y_X \rangle &&=\int_{-\infty}^{+\infty}df\int_{-\infty}^{+\infty}df'\delta_T(f-f')Q(f)\nonumber\\
&&\hspace{0.38cm}\times\bigl[ \gamma^T_{12}(f)G^{TX}_{13}(f,f')- \gamma^T_{13}(f,f')G^{TX}_{12}(f,f')\bigr],\nonumber\\
&&=\int_{-\infty}^{+\infty}df\int_{-\infty}^{+\infty}df'\delta_T(f-f')Q(f)\nonumber\\
&&\hspace{0.38cm}\times\bigl[ \gamma^T_{12}(f)G^{X}_{13}(f,f')- \gamma^T_{13}(f,f')G^{X}_{12}(f,f')\bigr],\nonumber\\
\label{eq:a12.0}
\end{eqnarray}
where we used Eqs.\eqref{eq:10} and \eqref{eq:11} and we defined
\[G^{TX}_{IJ}(f,f')=G^{T}_{IJ}(f,f')+G^{X}_{IJ}(f,f'),\]
and the relative variance
\begin{eqnarray}
(N_X)^2&& \approx \bigl[\langle Y_X^2 \rangle - \langle Y_X \rangle^2\bigr]_{h=0} = \bigl[\langle Y_X^2 \rangle\bigr]_{h=0} \nonumber\\
&&=\frac{T}{4}\int_{-\infty}^{+\infty}df|Q(f)|^2\nonumber\\
&&\hspace{0.38cm}\times P_1(f)\bigl[(\gamma^T_{12}(f))^2P_3(f)+(\gamma^T_{13}(f))^2P_2(f) \bigr].\nonumber\\
\label{eq:a12.01}
\end{eqnarray}
Given the data combination $\mu$, in the last equality of Eq.\eqref{eq:a12.0}, $G^T_{IJ}$ terms cancel out and we are left with the only contribution of $X$-extra polarization modes. We further introduce the following inner product in order to express Eqs.\eqref{eq:a12.0} and \eqref{eq:a12.01} in a more compact way
\begin{eqnarray}
\bigl(A(f)\cdot B(f)\bigr) \equiv && \int_{-\infty}^{+\infty}dfA^*(f)B(f)P_1(f)\nonumber\\
&&\times \bigl[(\gamma^T_{12}(f))^2P_3(f)+(\gamma^T_{13}(f))^2P_2(f) \bigr],\nonumber\\
\end{eqnarray}
thus both the signal and noise terms can be written as
\begin{eqnarray}
S_X=&&\left(Q(f) \cdot \frac{\bigl[\gamma^T_{12}(f)G^{X}_{13}(f)-\gamma^T_{13}(f)G^{X}_{12}(f)\bigr]}{P_1(f)\bigl[(\gamma^T_{12}(f))^2P_3(f)+(\gamma^T_{13}(f))^2P_2(f) \bigr]} \right),\nonumber\\
\label{eq:a15}
\end{eqnarray}
and
\begin{eqnarray}
(N_X)^2=\frac{T}{4}\bigl(Q(f)\cdot Q(f)\bigr).
\label{eq:a16}
\end{eqnarray}
The expression of the filter function that maximizes the SNR this time is given by
\[ Q(f) \propto \frac{\bigl[\gamma^T_{12}(f)G^{X}_{13}(f)-\gamma^T_{13}(f)G^{X}_{12}(f)\bigr]}{P_1(f)\bigl[(\gamma^T_{12}(f))^2P_3(f)+(\gamma^T_{13}(f))^2P_2(f) \bigr]}, \]
therefore we end up with the following SNR expression in the presence of a SGWB made of tensor and X-extra polarization modes
\begin{widetext}
\begin{eqnarray}
\text{SNR$_X$}&&=\frac{3 H_0^2}{10 \pi^2}\sqrt{T}\left\{2\int_{0}^{+\infty}df \frac{\bigl[\bigl(\gamma^T_{12}(f)\gamma^X_{13}(f)-\gamma^T_{13}(f)\gamma^X_{12}(f)\bigr)\Omega^X_{GW}(f)\bigr]^2}{f^6\bigl[(\gamma^T_{12}(f))^2P_1(f)P_3(f)+(\gamma^T_{13}(f))^2P_1(f)P_2(f) \bigr]} \right\}^{1/2}\, . 
\label{eq:aSNREXTRA}
\end{eqnarray}
\end{widetext}

\subsection{Tensor, Vector and Scalar polarization modes}

We finally consider a SGWB made of tensor, vector and scalar polarization modes in the general scenario where we dispose of three independent interferometers (e.g. ETA, ETB and CE): we introduce the following data combination defined for tensor, vector and scalar polarization modes as
\begin{eqnarray}
\mu^M(f,f')=&&\alpha^M_1(f)s^*_1(f')s_2(f)+\alpha^M_2(f)s^*_2(f')s_3(f)\nonumber\\
&&+\alpha^M_3(f)s^*_3(f')s_1(f),\nonumber
\end{eqnarray}
where $\alpha^M_{j}$, with $j=1$, $2$, $3$ and $M=T$, $V$ and $S$, are frequency-dependent coefficients. The idea is similar to what we did in the previous subsection: we need to find an expression for the frequency-dependent coefficients in order to remove the contribution of undesired polarization modes to the SNR for the SGWB. The simplest frequency-dependent coefficients needed to isolate one specific polarization class are given by
\begin{itemize}
\item Tensor modes
\begin{eqnarray}
\alpha^T_1=\gamma^S_{23}\gamma^V_{31}-\gamma^S_{31}\gamma^V_{23},\nonumber\\
\alpha^T_2=\gamma^S_{31}\gamma^V_{12}-\gamma^S_{12}\gamma^V_{31},\nonumber\\
\alpha^T_3=\gamma^S_{12}\gamma^V_{23}-\gamma^S_{23}\gamma^V_{12},
\label{eq:aalphaT}
\end{eqnarray}
\item Vector modes
\begin{eqnarray}
\alpha^V_1=\gamma^S_{23}\gamma^T_{31}-\gamma^S_{31}\gamma^T_{23},\nonumber\\
\alpha^V_2=\gamma^S_{31}\gamma^T_{12}-\gamma^S_{12}\gamma^T_{31},\nonumber\\
\alpha^V_3=\gamma^S_{12}\gamma^T_{23}-\gamma^S_{23}\gamma^T_{12},
\label{eq:aalphaV}
\end{eqnarray}
\item Scalar modes
\begin{eqnarray}
\alpha^S_1=\gamma^T_{23}\gamma^V_{31}-\gamma^T_{31}\gamma^V_{23},\nonumber\\
\alpha^S_2=\gamma^T_{31}\gamma^V_{12}-\gamma^T_{12}\gamma^V_{31},\nonumber\\
\alpha^S_3=\gamma^T_{12}\gamma^V_{23}-\gamma^T_{23}\gamma^V_{12}.
\label{eq:aalphaS}
\end{eqnarray}
\end{itemize}
We now recall Eqs.\eqref{eq:10} and \eqref{eq:11} in order to define the cross-correlation combination for each polarization mode (we already consider the ensemble average)
\begin{eqnarray}
\langle Y_M \rangle &&=\int_{-\infty}^{+\infty}df\int_{-\infty}^{+\infty}df'\delta_T(f-f')\bigl[\alpha^M_1(f)\langle s_1^*(f')s_2(f)\rangle \nonumber\\
&&\hspace{0.38cm}+\alpha^M_2(f)\langle s_2^*(f')s_3(f) \rangle +\alpha^M_3(f)\langle s_3^*(f')s_1(f) \rangle \bigr]Q(f)\nonumber\\
&&=\int_{-\infty}^{+\infty}df\int_{-\infty}^{+\infty}df'\delta_T(f-f')\bigl[\alpha^M_1(f)G_{12}(f,f')\nonumber\\
&&\hspace{0.38cm}+\alpha^M_2(f)G_{23}(f,f') +\alpha^M_3(f)G_{31}(f,f') \bigr]Q(f)\nonumber\\
\label{eq:a16.2}
\end{eqnarray}
which represents the signal $S_M\equiv \langle Y_M \rangle$ and we further introduce the corresponding variance
\begin{eqnarray}
(N_M)^2&&= \langle (Y_M)^2 \rangle-\langle Y_M \rangle^2\nonumber\\
&&=\frac{T}{4}\int_{-\infty}^{+\infty}df\bigl[ (\alpha^M_1(f))^2P_{1}(f)P_{2}(f)\nonumber\\
&&\hspace{0.38cm}+(\alpha^M_2(f))^2P_{2}(f)P_{3}(f)\nonumber\\
&&\hspace{0.38cm}+(\alpha^M_3(f))^2P_{3}(f)P_{1}(f)\bigr] |Q(f)|^2.
\end{eqnarray}
We still need to find an expression for the filter function, therefore we introduce
\begin{eqnarray}
W_s^M(f)=&&\alpha^M_1(f)G_{12}(f)+\alpha^M_2(f)G_{23}(f) \nonumber\\
&&+\alpha^M_3(f)G_{31}(f),\\
W_n^M(f)=&&\bigl[ (\alpha^M_1(f))^2P_{1}(f)P_{2}(f)+(\alpha^M_2(f))^2P_{2}(f)P_{3}(f)\nonumber\\
&&+(\alpha^M_3(f))^2P_{3}(f)P_{1}(f)\bigr].
\end{eqnarray}
Moreover, we can define the inner product 
\begin{eqnarray}
\bigl(A(f)\cdot B(f)\bigr)_M \equiv && \int_{-\infty}^{+\infty}dfA^*(f)B(f)W^M_n(f),
\end{eqnarray} 
so that we can express both signal and noise as
\begin{eqnarray}
S_M=\biggl(Q(f) \cdot \frac{W_s^M(f)}{W_n^M(f)}\biggr)\, ,\\
(N_M)^2=\frac{T}{4}\biggl(Q(f) \cdot Q(f)\biggr)\, .
\end{eqnarray}
In order to maximize the SNR, we now need to find the proper expression for the filter function. The latter is given by
\[Q(f)\propto\frac{W_s^M(f)}{W_n^M(f)}.\]
In the end, we finally get the SNR expression for tensor, vector and scalar polarization modes separately
\begin{widetext}
\begin{eqnarray}
\text{SNR$_M$}&&=\left[\int_{-\infty}^{+\infty}df\frac{(W_s^M(f))^2}{W_n^M(f)}\right]^{1/2}\nonumber\\
&&=\frac{3H_0^2}{10\pi^2}\sqrt{T}\left\{2\int_{0}^{+\infty}df\frac{(\Pi(f)\Omega_{GW}^M(f))^2}{f^6\bigl[ (\alpha^M_1(f))^2P_{1}(f)P_{2}(f)+(\alpha^M_2(f))^2P_{2}(f)P_{3}(f)+(\alpha^M_3(f))^2P_{3}(f)P_{1}(f)\bigr]}\right\}^{1/2},\nonumber\\
 \label{eq:awide}
\end{eqnarray}
where
\begin{eqnarray}
|\Pi(f)|^2=\bigl[\gamma_{12}^T(\gamma_{23}^S\gamma_{31}^V-\gamma_{31}^S\gamma_{23}^V)+\gamma_{23}^T(\gamma_{31}^S\gamma_{12}^V-\gamma_{12}^S\gamma_{31}^V)+\gamma^T_{31}(\gamma_{12}^S\gamma_{23}^V-\gamma_{23}^S\gamma_{12}^V)\bigr]^2.
\label{eq:api}
\end{eqnarray}
\end{widetext}

\nocite{*}

\bibliography{ArticoloCitazioni}

\end{document}